\documentclass[12pt]{article}

\usepackage{etex}
\usepackage{amsfonts}
\usepackage{latexsym}
\usepackage{natbib}
\usepackage{amsmath, pst-pdf, pst-all, amsthm, amssymb, latexsym, xfrac}
\usepackage{comment}
\usepackage{listings}
\usepackage{pstricks,pstricks-add} 
\usepackage{booktabs}
\usepackage{rotating}
\usepackage{multirow, tabularx}
\usepackage{lscape}
\usepackage{abstract}
\usepackage{lipsum} 
\usepackage{ctable}
\def\sym#1{\ifmmode^{#1}\else\(^{#1}\)\fi}
\usepackage{titlesec}
\usepackage{blindtext}
\usepackage{rotating}
\usepackage{epigraph}
\usepackage{geometry}
\usepackage{graphicx}
\usepackage{ragged2e}
\usepackage{graphicx, rotating, caption, lscape, threeparttable}

\usepackage{threeparttable}

\usepackage{fnpct}

 \newcolumntype{b}{>{\hsize=2.3\hsize}X}
\newcolumntype{s}{>{\hsize=.45\hsize}X}

\usepackage{chngcntr}

\usepackage[section]{placeins}
\usepackage{float}

\newcommand{\GG}[1]{}

\usepackage{afterpage}

\usepackage{relsize,etoolbox}
\AtBeginEnvironment{quote}{\smaller}

\usepackage{tabularx,booktabs}



\AtBeginDocument{%
  \addtolength\abovedisplayskip{-0.15\baselineskip}%
  \addtolength\belowdisplayskip{-0.15\baselineskip}%
  }

\newcolumntype{Y}{>{\centering\arraybackslash}X}

\usepackage{pb-diagram}

\makeatother
\usepackage{indentfirst}

\usepackage{enumerate}

\usepackage{hyperref}
\usepackage{cleveref}
\hypersetup{backref,  
pdfpagemode=FullScreen,  
colorlinks=true, citecolor=blue!50!black}

\usepackage{titlesec}

\renewenvironment{quote}{%
   \list{}{%
     \leftmargin0.45cm   
     \rightmargin\leftmargin
   }
   \item\relax
}
{\endlist}

\newcommand\Tstrut{\rule{0pt}{2.6ex}}      
\newcommand\Bstrut{\rule[-0.9ex]{0pt}{0pt}}
\newcommand{\TBstrut}{\Tstrut\Bstrut} 
 
\usepackage{pst-all}
\usepackage{fancyhdr}

\usepackage{caption}
\usepackage{subcaption}

  \titlespacing\section{0pt}{12pt plus 4pt minus 2pt}{11pt plus 2pt minus 2pt}
  
  \makeatletter
\def\@seccntformat#1{\@ifundefined{#1@cntformat}%
   {\csname the#1\endcsname\quad}  
   {\csname #1@cntformat\endcsname}
}
\let\oldappendix\appendix 
\renewcommand\appendix{%
    \oldappendix
    \newcommand{\section@cntformat}{\appendixname~\thesection\quad}
}
\makeatother
\usepackage{cleveref} 

\begin{document}

\title{Political Openness and Armed Conflict: \\ Evidence from Local Councils in Colombia}
\author{
 Hector Galindo-Silva\\
  \texttt{galindoh@javeriana.edu.co}\thanks{Pontificia Universidad Javeriana. An earlier draft of this paper was circulated under the title ``Political Representation and Armed Conflict: Evidence from Local Councils in Colombia''.  I thank seminar participants at  several seminars; the hospitality of the Barcelona Institute for Political Economy and Governance (IPEG), where part of this paper was written; and  Francesc Amat, Enriqueta Aragones, Laia Balcells, Carles Boix, Filipe Campante, Francesco Drago, Ruben Durante, Ruben Enikolopov, Donald Green, Jens Hainmueller, Raphael Godefroy, Gonzalo Hernandez, Paula Herrera, David Karp, Sebastian Lavezzolo, Marta Misas, Juan Morales, Suresh Naidu, Daniele Paserman, Maria Petrova, Vincent Pons, Didac Queralt, Marta Reynal-Querol, Alessandro Riboni, Andrea Saenz, Shanker Satyanath, James Snyder, Mathias Thoenig, Leonard Wantchekon Noam Yuchtman  and four anonymous referees for helpful comments and suggestions. Any remaining errors are my own. }
}
\date{First Draft: November 2015 \\This Draft: July 2020}
\maketitle

\begin{abstract}

In this paper, I empirically investigate how the openness of political institutions to diverse representation can impact conflict-related violence.  Regression discontinuity estimates that exploit plausibly exogenous variations in the number of councillors in Colombian municipalities show that political openness substantially decreases conflict-related violence, namely the killing of civilian non-combatants. Empirical evidence suggests that the lower level of political violence stems from parties with close links to armed groups having greater representation on larger municipal councils.  Using data about the types of violence employed by these groups, and government representation, I argue that armed violence has decreased not because of power-sharing arrangements involving armed groups linked to the parties with more political representation, but rather because armed groups with more political power deter other groups from initiating certain types of violence.

\bigskip
\emph{Journal of Economic Literature} Classification Numbers: H72, D72\\

\vspace{-0.3cm}
\emph{Keywords:} Political openness, armed conflict, municipal councils

\end{abstract}

\newpage

\section{Introduction}

In this paper, I empirically investigate the micro-level causal impact that political institutions being more open to direct representation from diverse groups (what I call ``political openness'') has on conflict-related violence.  Insofar as the level of openness of the political systems is an institutional component of democracy, studying its effects on armed conflict can shed light on the pacifying effects of democracy.\footnote{The term ``political openness'' is already used in the literature on democracy, within the theoretical framework of  ``political opportunity structure'' \citep{Kitschelt1986}. Kitschelt's notion of political openness differs from this study, where political openness is limited to the institutional structure (while the notion of political opportunity structure takes into account the whole political context). In this respect, political openness in this paper can be understood as a ``component of democracy'' \cite[see][]{coppedgegerringeta2011}.}  

The study focuses on Colombia, a country with serious deficiencies in the state's capacity to control violence, but which remains a democracy with regular free elections. This makes Colombia an ideal laboratory for studying the relationship between political openness and conflict.

The study begins by isolating plausibly exogenous variations in the level of openness of political institutions in medium and large Colombian municipalities. To do this, I exploit arbitrary discontinuities in the number of councillors for a municipality. By law, the number of municipal councillors is based on arbitrary population thresholds. Under certain conditions, municipalities with a population just below a given threshold can serve as a reasonable counterfactual for municipalities with a population just above the threshold.\footnote{See \cite{EggersFreierGrembiNannicini2018} for a review of the literature using regression discontinuity designs based on population thresholds and a discussion of the potential pitfalls.} 

Scholars have paid considerable attention to the size of councils, and of elected bodies in general,\footnote{See for instance \citeauthor[(PT, chap. VIII)]{Spinoza16762016}; \citeauthor[(SC, III)]{Rousseau1762};  \citeauthor[(Fed, 10 \& 55)]{Madison1055};  \citeauthor[(Letter No. 3)]{Brutus3}; \citet[chap. 15]{BuchananTullock1962} and \cite{Waldron1999}.} and have associated this factor with the degree of openness of such bodies to the entry of political parties: the greater the number of representatives, the greater the probability that more groups in the population will be directly represented. I provide regression discontinuity (RD) estimates suggesting that adding two additional seats to a municipal council leads, on average, to the election of 1.1 more political parties. The effect is meaningful, given that the average number of parties per council in the sample is 5.  

I use the exogenous discontinuity in the number of councillors to study violence and armed conflict. RD estimates show that the probability of a conflict-related homicide occurring during a council's term is 4.6 percentage points lower in municipalities with larger councils.\footnote{Conflict-related homicide is defined as the intentional killing of civilian non-combatants, which typically occurs when an armed group enters a village and executes one or more pre-identified inhabitants.} This is a large effect, since 14\% of municipality-periods in the sample experienced at least one conflict-related homicide during the period covered by the study.  These results are robust to alternative specifications, samples, and to the use of the conflict-related homicide rate rather than the extensive margin. 

Several mechanisms may explain these results. I focus on one that I believe to be the most plausible: larger municipal councils decrease conflict-related violence because they lead to more representation for parties with links with one non-state armed group, which increases its coercive power and deters its enemies from carrying out selective killings. This mechanism, which I call the  ``deterrence hypothesis,'' is related to the literature on the consolidation of power and monopoly on violence \cite[see][]{Olson1993, BatesGreifSingh2002, NorthWalliWeingast2009, AcemogluRobinsonSantos2013, Powell2013}.

I examine the plausibility of this hypothesis  in several  ways. First, I show that political parties that allegedly support the interests of right-wing paramilitaries have more direct representation on larger municipal councils. RD estimates document that adding two seats to a municipal council increases by 28.6 percentage points the probability that a party with paramilitary links will have at least one seat. Conversely, the estimates show no evidence of an effect on the political representation of left-wing parties.  Second, I show that the effect of a larger council on selective killings is larger for those municipalities with paramilitary-linked parties, and that in municipalities where paramilitaries do not have political representation, the effect is not statistically different from zero. Third, I show the decrease in conflict-related homicides is larger for very selective killings (four or fewer deaths per event), but almost nonexistent for massacres (the execution of many people at the same time).  Four, using data from an alternative source, I show that in municipalities with larger councils, left-wing guerrillas are the armed armed groups that reduce conflict-related violence the most. 

I conclude by analyzing the potential incentives for paramilitaries to increase their coercive power,  and the effect that the increase in paramilitary coercive power may have on the level of political openness of the municipal councils in other elections.  While I do not find evidence  of rent extraction, I find that when there are more councillors from paramilitary groups, the percentage of votes obtained by left-wing parties in the next election for the Colombian Senate is lower. Importantly, this result suggests that over time, political openness may have the somewhat paradoxical effect of making democracy less open by allowing right-wing paramilitary groups to have more political representation. 

This study contributes to the literature in several  ways. First, it adds to the vast conflict literature by providing new and well-identified evidence about the effects of democratic institutions on violence and armed conflict.  As previously mentioned, at the broad level of cross-country correlates, there is evidence that the openness of a political system can influence the probability of civil wars and political violence  \cite[see][]{ReynalQuerol2002jcr, ReynalQuerol2005ejpe}.  A smaller set of studies focuses on subnational variation, studying the role of  income shocks, climate shocks and the size of the franchise on conflict  \cite[see][]{DubeVargas2013, HarariLaFerrara2013, FergussonVargas2013}. To my knowledge, no study has examined the impact of increased political openness on conflict-related violence at the subnational level.\footnote{Perhaps the closest contribution to this paper is \cite{FergussonVargas2013}. They also study the effect of one aspect of democracy  (in their case, the size of the franchise) on conflict.  Using Colombian subnational data from 1821-1885, they find that municipalities where more voters were enfranchised experienced fewer violent political battles.  Three important differences are that i) they focus on the size of the franchise, which is different from political openness, ii) they focus on a different period and a different type of conflict, and iii) by using more recent and better conflict data (which allows me to identify types of violence and  perpetrators), I can shed more light on the mechanisms through which democracy affects violence.}  

My findings echo the logic of violence proposed by \cite{Kalyvas2006}: civil wars involve not just armed actors but also civilians, and the degree of control an armed group has over a territory is crucial in determining the intensity of violence directed toward the territory's inhabitants. I provide well-identified evidence in favor of one of \citeauthor{Kalyvas2006}'s key predictions: the higher the level of control, the less likely that selective violence will occur.

My empirical evidence comes from a single country. Therefore, I use caution in making claims about external validity. Nonetheless, I believe that the political mechanisms and empirical evidence presented in this paper are useful in understanding the effects of increasing the degree of political openness of elected bodies. At the very least, the empirical evidence showing that larger local councils are more open to the entry of political parties can be generalized to other countries and other elected bodies. Although the conclusions about the impact of greater participation of parties with paramilitary links in local government and about the reduction of conflict-related violence may be specific to Colombia, other countries appear to have or could have similar experiences with non-state armed groups in  politics.

The outline of the paper is as follows. Section \ref{background} provides a brief overview of the Colombian armed conflict and the country's local institutions. The data and empirical strategy are discussed in Section \ref{empiricalstrategy}.  The main results are presented in Section \ref{mainresults}.  Possible mechanisms are discussed in Section \ref{mechanisms}. Section \ref{conclusion} concludes. 


\section{Background}
\label{background}

\subsubsection*{Colombian Armed Conflict and Violence}

Colombia has suffered one of the world's longest-running internal conflicts. The conflict has its roots in struggles for land ownership rights, political exclusion, and weak institutions \citep{Sanchez2001}. Its persistence has been explained as the result of international influences and drug trafficking \citep{Deas2015}, as well as the decentralization of local politics and public spending \citep{SanchezPalau2006}. The start of the conflict coincided with the founding of the Revolutionary Armed Forces of Colombia (FARC), which was always Colombia's largest and best-equipped rebel group. The FARC was never affiliated with either of Colombia's main political parties, but was ideologically aligned with the Communist Party (see \citealt[p. 355]{PalaciosSafford2002} and \citealt[p. 62]{GMHFARC2013}). Other armed groups have participated in Colombia's conflict, including smaller left-wing insurgents and several right-wing paramilitary groups. Some authors have associated the origin of the paramilitary groups with local elites and drug cartels that faced threats of kidnapping and extortion from the guerrillas and felt betrayed by the central government's favorable view of political competition, agrarian reforms and peace talks \citep{Romero2005, GutierrezBaron2005, Lopez2010cap1}. In 1997,  paramilitary factions formed a national coalition called the United Self-Defense Groups of Colombia (AUC). Its creation considerably increased the effectiveness of the paramilitaries and, as a result, the guerrilla groups were driven out of large areas of the country. During this period of paramilitary expansion (1998 to 2002), violence associated with the conflict escalated dramatically. In 2002, with the arrival of a new president who eventually offered de-facto amnesty to paramilitaries, the level of violence began to decline, and by the end of the 2000s, the severity of the conflict decreased significantly. In July 2016, the  Colombian government and the FARC signed a historic peace deal,  which earned then-Colombian President Juan Manuel Santos the 2016 Nobel Peace Prize.

Colombian civilians have routinely been the target of massacres, selective assassinations, kidnappings and forced displacement. The Historical Memory Group  (\emph{Grupo de Memoria Hist\'orica}, or ``GMH''),  an independent group of academics created by the central government to record the history of the armed conflict, estimates that the conflict claimed at least 220,000 lives. Civilians accounted for about 81\% of this number \citep{GMH2013}.

Civilian victims died in different ways. The GMH classifies civilian deaths into two main categories: intentional killings in which an armed group enters a village and executes one or several pre-determined inhabitants, and unplanned deaths that occurred as a result of another action. The GMH documents at least 26,380 intentional killings from 1981 to 2012, affecting approximately 82\% of Colombian municipalities. Intentional killings escalated at the end of the 1990s and peaked in 2000.

\subsubsection*{Colombian Local Institutions}

Colombia has a long democratic tradition with almost no legacy of dictatorship. This contrasts with most other Latin American countries, which were led by dictators between the 1960s and 1990s. The armed conflict obviously damaged the nation's political institutions.  However, even when violence peaked at the end of the 1990s, democratic institutions did not collapse. Regular free elections (at least on paper) for political office, including local governments, has been the norm for more than two decades.\footnote{See \cite{Gutierrez2019}, who argues that electoral competition during this period was firmly established (p. 257-258), and served as a constraint on paramilitary groups (p. 370).}

At the local level, the fundamental administrative unit is the municipality, of which there were 1,102 as of June 2015. National laws that apply equally to all municipalities regulate elections and the duties of elected officials. Municipalities are governed by a mayor and a council elected by popular vote for a four-year term.\footnote{From 1993 to 2004, mayors and councillors were elected for a three-year term.} Municipal governments are responsible for providing certain public  goods related to education, health and infrastructure.\footnote{Municipal budgets do not include police or military expenditures.}

A key function of a municipal council is to approve proposals brought forward by the mayor.\footnote{According to Article 313 of the Colombian Constitution, the key responsibilities of municipal councils are the regulation of the delivery of public services; the supervision of contracts made by the mayor; the approval of local taxes and expenditures; the determination of the structure of the municipal administration, including salary scales; and the regulation of land use.} In practice, however, councils have a limited role in policymaking, with mayors being the key players. Despite these limitations, the municipal council is an important mechanism for the interplay of significant political forces, and an instrument used by the \emph{de facto} local powers to gain visibility and increase their control over municipalities.\footnote{One example is how the municipal council facilitates a mechanism called \emph{Cabildo abierto} (consultative public assembly): a public council meeting where citizens can participate directly in the discussion of community affairs (see Law 134 of 1994).}  

Councillors are elected using a multi-member single-district system.\footnote{The seats are allocated using the D'Hondt formula, with a minimum threshold of 3\%.} According to the national electoral law, the size of municipal councils is determined by population. As shown in columns (1) and (2) of Table \ref{t1obsallyears}, council sizes range from seven members for municipalities with 5,000 or fewer people, to 21 members for municipalities with more than 1 million residents. The population data used to determine the number of councillors is based on the central government's administrative records. Two or three months before an election, the National Department of Statistics sends the latest census data to electoral authorities, and these officials designate the number of people to be elected to each municipal council. 

It is important to note that, to the best of my knowledge, the population thresholds used to determine municipal council size are not used for any other relevant policy purpose.  The only relevant mechanism that might use the thresholds as an input is categorization (\emph{categorizaci\'on}), a process where municipalities are divided into seven groups, primarily according to their freely disposable revenue (essentially, current revenue excluding transfers and earmarked revenues).\footnote{\label{threshold5background}See Law 617 of 2000. Other policies that in some way use population thresholds --- in particular the 100,000 threshold --- are: (i) eligibility to have local fiscal oversight institutions (also in Law 617 of 2000), (ii) eligibility for independent education managing institutions (Decree 3940 of 2007), and (iii) a requirement to have more elaborate land-use plans (Law 388 of 1997).  I thank a referee for helping me to identify two of these policies. The most relevant of the three policies to this paper is likely the first. However, and importantly, the only scenario in which the 100,000 threshold (and not the municipal category) affects eligibility for local fiscal oversight institutions is if the municipality falls into the second category and has a population of at least 100,000 (see  Article 156 of the Law 617 of 2000). During the period studied, 23 municipalities satisfy these two conditions, of which only two have a population close to one of the thresholds also used to determine the size of a council (i.e.  the 100,000 threshold). In Web Appendix \ref{appomittedrobust}, I show that all the main results are robust to excluding not only these 2 (or 23) municipalities, but all the municipalities for which the council size is determined by the 100,000 threshold.} At the municipal level, categorization determines the salaries of the mayor, councillors and administrative staff; sets general administrative expenditure limits; and regulates entitlements to special transfers from the central government. The law on categorization (Law 617 of 2000) specifies population thresholds, some of which align with those used to designate the size of municipal councils. However, according to Law 617,  these thresholds are second-tier conditions and irrelevant in practice.\footnote{According to paragraph 1 of Article 6, if a municipality falls into one category based on population and a different category based on freely disposable revenue, the municipality must be classified according to its revenue. Thus, for 100\% of municipalities, the category is based on freely disposable revenue. In fact, during the period studied, approximately  60\% of municipalities would fall into a different category if only population were taken into account.}
 In Section \ref{empiricalstrategy}, I verify that this is indeed the case by showing that municipalities' categories and spending (which are positively correlated with municipal salaries and administrative expenditures) vary smoothly around population thresholds.
 
\subsubsection*{Representative Institutions and Non-State Armed Actors}

During the armed conflict in Colombia, the relationship between non-state armed actors and representative institutions progressed through different phases. The most relevant phase for this paper occurred during the 1990s and coincided with the creation of the AUC. This unification of paramilitary groups signaled an important change. An uncoordinated strategy of influencing politics by sponsoring particular local politicians from the two traditional parties was replaced by a strategic decision to influence politics directly and in a coordinated way at all levels of government  (see \citealp[pp. 245-246]{Romero2005} and \citealp[p. 43]{Lopez2010cap1}). This strategy resulted in a series of secret cooperation agreements between the AUC and a large number of politicians calling for the ``refounding of the nation." Tellingly, these agreements implied the capture, co-optation or creation of smaller parties. These agreements came to light in 2006, when the recovered laptop of a paramilitary leader was found to contain details of the  ``para-political" connections. This discovery launched an investigation known as \emph{parapolitics}, which by 2012 had resulted in the prosecution of at least 470 municipal mayors and councillors, and the imprisonment of 51 congressmen \cite[see][]{COLfiscaliaOCT232012, COLverdadabiertaOCT232012, indepa2012, MOE2013}. 

The  AUC  strategy  of  capturing  the  political  system  contrasted  with  that  of  the  FARC, which changed its approach around the same time: instead of sponsoring  specific  candidates from a few  left-wing parties, the FARC started a campaign to  sabotage local elections.   From its founding until 1997,  the  FARC supported  and  collaborated  with  a  small number of ideologically similar political  parties.   Initially,   collaboration  was  limited  to  the  Colombian  Communist  Party \cite[see][pp. 90-95]{GMHFARC2013}.  In 1985, collaboration expanded to include the Patriotic Union (UP), a national leftist party founded by the FARC as part of a first attempt at peace  negotiations  with  the  Colombian  government \cite[see][pp. 157-162]{GMHFARC2013}.   After numerous  violent  attacks  and  assassinations  of UP members,  the  FARC changed its strategy and began a hostile  campaign against elected representatives \cite[see][pp. 257-267]{GMHFARC2013}.  The hostilities, which peaked at  the  end  of  the 1990s, not only entailed promoting  voter abstention (a strategy that had been used previously) but also actively obstructing  the  electoral process. The hostilities  were accompanied by a series of mechanisms through which the FARC, although without a ``proper" party, tried to co-opt local political institutions  \cite[see][pp. 256-257]{GMHFARC2013}.  In the mid-2000s, the FARC moderated its attacks on the electoral process, reverting to social mobilization and political campaigning (see \citealp[p. 275]{GMHFARC2013} and \citealp[p. 385]{AvilaVelasco2012}).


\section{Data and Empirical Strategy}
\label{empiricalstrategy}

\subsection{Data}

My analysis uses data on armed conflict, council size and electoral outcomes in Colombian municipalities. The data on armed conflict comes from two main sources. The first is the GMH, which, as previously mentioned, is an independent group of academics created by the Colombian government to produce a historical account of the armed conflict.
In its final report, the GMH presented a series of datasets on the extent of violence during much of the conflict period (from the early 1980s to early 2010s).\footnote{This data can be found at \url{http://www.centrodememoriahistorica.gov.co/micrositios/informeGeneral/basesDatos.html}.} Covering several types of conflict-related violence and focusing on civilian victims, the data is based on reports from a network of Catholic non-governmental organizations. These reports describe incidents of political violence in nearly every municipality in the country (including those in remote regions). These Catholic organizations are regarded as neutral actors in the conflict, which minimizes concerns about potential over-reporting of violence perpetrated by a particular faction.\footnote{Their data collection framework includes internationally accepted definitions derived from human rights law and  international humanitarian law. For more about this source, see \cite{CINEP2008}.}

The GMH  classifies the killings of civilians in two main categories: intentional and unplanned killings. Intentional killings occur, for example, when an armed group enters a village and executes a particular villager (or villagers) perceived to be sympathetic to the opposing side. Unplanned killings are civilian deaths that occur as a result of another action (such as the bombing of an infrastructure or military target). As previously mentioned, in this paper I focus on intentional killings. To be classified by the GMH as intentional, a killing has to satisfy the following conditions: the victim(s) was killed in a state of helplessness, the perpetrator was an identified armed group or an identified group using weapons and wearing uniforms, and the victim was a social leader or an activist identified as a target by a non-state armed group. It is important to note that this category excludes deaths caused by mines, terrorist attacks, murders associated with the drug trade, murders suspected to have been carried out for personal reasons, and murders committed by vigilantes, social cleansing groups or gangs.\footnote{The GMH includes an additional criterion to describe how selective a killing was. A killing with three or fewer victims is labeled as a ``selective killing,'' and an event with four or more victims is a ``massacre" (this definition of a massacre is also used by the Colombian National Police Department).  Since selective killings and massacres share key characteristics (e.g. they are both selective and conflict-related), in some specifications I combine the two as ``conflict-related killings.'' In other specifications, I distinguish between selective killings and massacres.}

The GMH data provides a very thorough account of conflict-related violence in Colombia. However, two drawbacks are that information on the perpetrators of the killings is very imprecise, and that observations have not been systematically cross-checked against other sources. To mitigate concerns about the quality of GMH data, I also use data from the Conflict Analysis Resource Center (CERAC). CERAC is a private research organization that specializes in data-intensive studies of conflict and criminal violence. Like the GMH data, the CERAC data includes information about violent episodes in almost all Colombian municipalities over a lengthy period. Importantly, this data also includes information from reports prepared by the network of Catholic organizations mentioned above. However, the CERAC data also use media reports from several major newspapers, and events are cross-checked with several other sources.\footnote{For more information about the collection procedure, see \cite{RestrepoSpagatVargas2004}; see also \cite{DubeVargas2013}, who extensively use this dataset. I use public data, which is available for 1988-2009.} The CERAC data focuses on attacks and clashes between groups, which includes the bombing of pipelines, bridges and other infrastructure targets; the destruction of police stations and military bases; and ambushes of military convoys.\footnote{\label{dataNCHM}After the publication of GMH's final report, another entity with the same objective was created (called the National Centre for Historical Memory, NCHM). The NCHM also produced a series of datasets on conflict-related violence, but uses slightly different definitions. This data can be found at  \url{http://centrodememoriahistorica.gov.co/observatorio/bases-de-datos/}. Since the NCHM's profile is less academic (relative to  GMH's), quality concerns are even greater for the NCHM data. Thus, for baseline results, I focus on the GMH and CERAC data, and use the NCHM data only for robustness checks.}  

The data on the size of municipal councils and electoral outcomes  comes from the Colombian Electoral Agency. For council size, I have created a new dataset in which I integrate information from Electoral Agency resolutions and from Colombian Official Journals (\emph{Diario Oficiales}). As previously mentioned, by law the municipal council size is determined using population thresholds. Population data for this purpose is based on estimates from the National Administrative Department of Statistics (DANE). The DANE produces these estimates centrally, using census data on birth, mortality and immigration rates, and predicting the structure of the population of the municipalities between censuses using provincial data.\footnote{To produce these estimates, the DANE uses a version of the cohort-component method that predicts the age and sex structure of the population in small areas assuming that the demographic trends in these areas are consistent with the trends in larger areas \cite[see][]{DANE2009proypop}.} The fact that this process is carried out centrally, by public officials in Bogota, minimizes the potential for strategic manipulation of the data by local governments, particularly in medium and
large municipalities.\footnote{\label{footnotemanipulation}Since the population estimates use local data on birth, mortality and immigration rates, I cannot rule out the possibility that the data is manipulated by local authorities.  As I will discuss below, the analysis focuses on municipalities with at least 15,000 inhabitants, for which a manipulation the population data is less likely. For these municipalities, a battery of data-driven falsification tests provides empirical evidence about the validity of the design.}  Both the data on population size reported by the DANE and the number of seats assigned to each municipality are reported in Electoral Agency resolutions.\footnote{I use Official Journals 43.176, 44.056, 45.265, 46.639, 48.109 and 48.128, which can be consulted at \url{http://svrpubindc.imprenta.gov.co/diario/}; and Electoral Agency Resolutions 2823 to 2852 of 1997.} My analysis starts in 1994, and covers all local elections until the last and decisive cease-fire with the FARC in 2014 (i.e. it includes elections in 1997, 2000, 2003, 2007 and 2011).\footnote{\label{dataNCHM1}The legislation requiring that municipal council size be determined by population was passed in 1994 (Article 22 of Law 136 of 1994).  Since the GMH data stops at 2012, and because the government decided not to use new population estimates to determine the size of municipal councils for the 2011 and 2015 elections (see Official Journals 48.109, 48.128 and 49.327), for the baseline results I do not use the data on council size for the councils elected in 2011 and 2015. However, the main results are robust to the inclusion of this data, for which I use the NCHM conflict-related data mentioned in footnote \ref{dataNCHM}.}  Columns 1 and 2 of Table \ref{t1obsallyears} summarize the mapping between population and council size; Columns 3 to 8 show the number of observations within 5\%, 10\% and 15\% of each of the thresholds.

I merge GMH, CERAC, electoral outcomes and population data to create a dataset that spans from 1997 to 2011 and includes 2,355 local elections. These elections occurred in 564 municipalities, which corresponds to 51.5\% of the total number of Colombian municipalities.\footnote{\label{footnotesplitorcreated}Between 1997 and 2011, among these municipalities,  4 municipalities were created, 28 ceded territory to a new municipality, and no municipality disappeared. Since the creation of a municipality is partially decided by the local authorities, and because when a municipality is created its population and the population of other municipalities abruptly change, the inclusion of these municipalities in the baseline sample may be a concern. In Web Appendix \ref{appcreated}, I show that the main results are robust to excluding municipalities where the population changed due to the creation of another municipality. I thank a referee for suggesting this robustness check.}  The sources of other variables used for the evaluation of the sample balance are listed in the note to Table \ref{precharacteristics10}.

\subsection{Empirical Strategy}

I use a sharp RD design to study the impact of council size on armed conflict and political representation. This design addresses the potential endogeneity between political institutions, representation and conflict. An RD design relies on the existence of a dichotomous treatment variable that is a deterministic function of a single continuous covariate. If subjects pass some threshold level of the variable, they are assigned to the treatment group; otherwise, they are assigned to the control group. A law that bases the size of municipal councils on population is ideal for an RD design because council sizes increase deterministically and discontinuously at certain population thresholds. Thus, under certain conditions, municipalities with a population just below a given threshold can serve as a plausible counterfactual for municipalities with a population just above the threshold.

The baseline analysis estimates a regression model within a narrow window around a single discontinuity; that is, I pool all the thresholds together by normalizing population size according to the distance of each municipality's population from the threshold.\footnote{Intervals around each threshold are symmetric and constructed so that no municipality appears in more than one interval. As illustrated in Table \ref{t1obsallyears}, for municipalities with a population of at least 15,000, there are four thresholds, corresponding to the following critical population sizes at which discrete changes in council size occur: 20,000, 50,000, 100,000 and 250,000.} Specifically, my running variable is the normalized population $\tilde{N}_{mt}=(N_{mt}-N_{T})/N_{T}$, where $N_{mt}$ is the population of municipality $m$ in electoral year  $t$ and $N_{T}$ is the closest population threshold $T$.\footnote{This pooling-and-normalizing approach is widespread in empirical work employing RD designs (for a list of over 30 empirical papers where RD designs with multiple cutoffs are present and that use a pooling-and-normalizing approach, see \citealp[][Table SA-1]{CattaneoKeeleTitiunikVazquezBare2020}; see also \citealp{Bertanha2020}). This normalization procedure estimates an average of local treatment effects weighted by the relative density of individuals near each of the cutoffs \cite[][ Proposition 3]{CattaneoKeeleTitiunikVazquezBare2016}. A limitation of this approach is that  the estimated average is a meaningful summary measure  only if we are interested in the average effect near the existing cutoffs. For extensions of this approach, see   \cite{Bertanha2020} and \cite{CattaneoKeeleTitiunikVazquezBare2020}.}  

By definition, a municipality is treated (has a bigger council size) if $N_{mt}>0$.
The RD estimand for the effect of a bigger council size is defined as:
\begin{equation}
\tau^{RD}=\lim_{\tilde{N}_{mt} \downarrow 0}E[Y_{mt+1}|\tilde{N}_{mt}>0]-\lim_{\tilde{N}_{mt} \uparrow 0}E[Y_{mt+1}|\tilde{N}_{mt}<0]
\label{rdbaseline}
\end{equation}
where $Y_{mt+1}$ is the outcome of interest in municipality  $m$ in the term immediately following the election in year $t$. 

To estimate $\tau^{RD}$, I use local polynomial regressions. I implement this procedure using the robust bias-corrected estimator with a data-driven bandwidth selector proposed by \cite{CalonicoCattaneoTitiunik2014a}. Following \cite{CalonicoCattaneoTitiunik2014b}, I report a conventional estimate of $\tau^{RD}$ and conventional standard errors, as well as robust bias-corrected p-values. In all specifications, I include fixed effects of the thresholds and a full set of fixed effects based on the region and period. Additionally, I cluster the standard errors at the municipal level to account for any dependence over time within municipalities.

\begin{figure}[H]
\begin{center}
\caption{Manipulation tests: pooled thresholds}\label{McCraryall}
\vspace{-0.3cm}
\resizebox{10cm}{6cm}{\includegraphics[width=1cm]{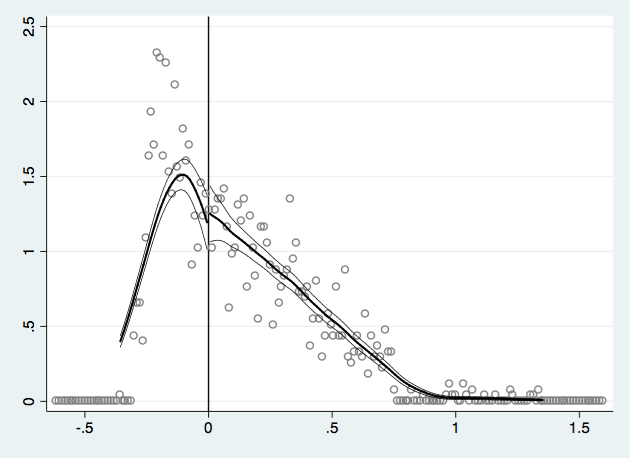}}
\end{center}
\vspace{-0.9cm}
\begin{center}
\begin{minipage}{10cm} \scriptsize This figure shows a finely-gridded histogram of  the population smoothed using local linear regressions, separately on either side of the cutoff  of the density function of the population. The figure  pools all years, uses data only around the normalized population threshold, and includes municipalities with a population of at least 15,000 inhabitants.  The estimate of the difference in the height at the threshold and  robust manipulation p-value (as implemented in Stata by the command rddensity.ado, \citep[see][]{CattaneoJanssonMa2018a} is 0.025 (s.e.  0.142) and p-value 0.577. 
\end{minipage}
\end{center}
\end{figure}

Identification requires that municipalities be unable to systematically manipulate population estimates. As previously mentioned, I focus on municipalities with a population of at least 15,000, for which a manipulation of the data by local governments is less plausible.  For these municipalities, I check for such manipulation by running kernel local linear regressions of the density separately on both sides of the relevant thresholds (in the spirit of \citealt{McCrary2008} and \citealt{CattaneoJanssonMa2018a, CattaneoJanssonMa2018b}).\footnote{For the population distribution, see Figure \ref{popdistribution} in Web Appendix \ref{appsuppfigtab}.}   Figure \ref{McCrarypopthres} in Web Appendix \ref{appsuppfigtab} reports the results of \citeauthor{McCrary2008}'s and \citeauthor{CattaneoJanssonMa2018a}'s tests for each population threshold. All the panels show no discontinuities.\footnote{The estimates of the difference in the height at the threshold (McCrary's test) and robust manipulation p-value (Cattaneo et al.'s test) for each population threshold are reported in the note to Figure \ref{McCrarypopthres}. Figure \ref{McCraryyear} in Web Appendix \ref{appsuppfigtab} repeats this analysis for each election year. No estimate is statistically significant. In addition, no municipality has a population identical to a threshold value, which minimizes the concern of a bias toward finding a jump in population, as noted by  \cite{EggersFreierGrembiNannicini2018}.}  For the pooled sample, both  \citeauthor{McCrary2008}'s and \citeauthor{CattaneoJanssonMa2018a}'s tests confirm that there is no discontinuity in the density at the normalized threshold (see Figure \ref{McCraryall}).\footnote{In Figure \ref{McCrarypopthressmall} in Web Appendix \ref{appsuppfigtab}, I check for systematic manipulation of population estimates in municipalities with a population of less than 15,000. The figure shows small jumps around the two smallest thresholds (5,000 and 10,000). Even though the jumps are small and weakly statistically significant, the use of data from these municipalities could reduce the trustworthiness of the estimations, in particular if, as discussed in footnote \ref{footnotemanipulation}, the small jumps  around the two smallest thresholds are related to the fact that the population estimates are based on data reported by local authorities. This is why, in a trade-off between thoroughness and transparency, I restrict the analysis to municipalities with a population of at least 15,000 inhabitants. However, all the results are robust to including smaller municipalities.}

Identification also requires that all relevant factors other than the treatment vary smoothly at the population thresholds. The idea of this falsification test is that if municipalities lack the ability to precisely manipulate their population size, there should be no systematic differences between municipalities with similar population size. Thus, this test complements the density test described in the last paragraph, and insofar as it implies that near the cutoffs, treated units are similar to control units in terms of observable (pre-treatment) characteristics, it is crucial for the validity of the design \cite[on the relevance of this test, see][]{CattaneoIdroboTitiunik2020}. To implement this formal falsification test, Table \ref{precharacteristics10} examines whether more than 50 pre-treatment characteristics are balanced across the normalized population threshold. Column (7) reports the RD estimand ($\tau^{RD}$) from Eq. (\ref{rdbaseline}) with bandwidths chosen using \cite{CalonicoCattaneoTitiunik2014a}'s procedure when the respective characteristic is used as the dependent variable. Column (8) reports the corresponding RD standard errors. There are no cases where the coefficients are statistically different from zero. This evidence strongly suggests that municipalities with populations just below a given population threshold are a valid control group for municipalities with populations just above that threshold.\footnote{Tables \ref{precharacteristics10dpopth3to4} and \ref{precharacteristics10dpopth5to7} in Web Appendix \ref{appsuppfigtab} repeat the analysis in Table \ref{precharacteristics10}, but distinguish between two groups of thresholds. Table \ref{precharacteristics10dpopth3to4} shows a statistically significant effect for some key pre-treatment variables. Tables \ref{precharacteristics10popth3} and \ref{precharacteristics10popth4} show that the effect is specific to municipalities for which the closest population threshold is 20,000. In Web Appendix \ref{appomittedrobust}, I show that all the main results are robust to excluding these municipalities.}

Finally, identification requires that no other relevant policies are based on a population discontinuity at the same thresholds. I argued in Section \ref{background} that there are no such policies. I now provide evidence that, to my knowledge, the only potential case for which there could be some doubt does not apply. As previously mentioned, this examination is important because the population affects the salary of the mayor, council members, and administrative staff, among other things. First,  note that Table  \ref{precharacteristics10} includes this variable as one of the pre-treatment characteristics; specifically, it includes the municipal category applied to the first year of a council's term, and determined the previous year. Table \ref{tableRDcategoria} in Web Appendix \ref{appsuppfigtab} extends this analysis to include the effect of a larger council size on the municipal category applied prior to the beginning of the council's term. Column (1) repeats the results from columns (7) and (8) in Table  \ref{precharacteristics10}, showing no statistically significant effect. Columns (2) and (3) in Table \ref{tableRDcategoria} focus on the category implemented one (column (2)) and two years (column (3)) prior to the first year of the council's term. None of the estimates are significant. These results provide strong evidence that municipal category does not determine population thresholds.\footnote{Figure \ref{catdistribution} in Web Appendix \ref{appsuppfigtab} maps municipal category against population estimates. Panel A  displays a scatter plot of the municipal category in the first year of a council's term; the vertical lines represent the population thresholds. Panel B depicts the same relationship, but with a scatter plot produced by averaging the municipal category over cells of 500 inhabitants. Additionally, a solid line plot predicts values from local regressions. No jump is visible.  Figure \ref{currspenddistribution}  in Web Appendix \ref{appsuppfigtab} shows a similar result when current spending is used instead of categorization (also included in Table \ref{precharacteristics10}), an additional proxy for the salaries of municipal councillors and administrative staff.}


\section{Main Results}\label{mainresults}


Panels (a)-(f) in Figure \ref{allselectiveRDfig}  examine how the size of a municipal council affects political violence.   Panels (a)-(b) show the probability that at least one conflict-related killing (either a selective killing or a massacre) occurs during the term that the council elected in $t$ is in office, plotted against the normalized population. Panel (b) shows that a larger municipal council significantly decreases the average probability of a  selective killing event. Panel (a) indicates that this probability is similar during the preceding term of a municipal council, regardless of whether the size of the succeeding council increases or not.

Panels (c)-(f) in Figure \ref{allselectiveRDfig}  repeat the previous graphical analysis, distinguishing between selective killings and massacres, which differ in the number of people killed per event. These results show that the presence of a larger municipal council has a significant effect on selective killings but almost no effect on massacres. 

Columns (1) to (3) of Panel A in Table \ref{tableRDconflictbaseline} report the estimates resulting from Eq. (\ref{rdbaseline})  for the outcomes analyzed in Figure \ref{allselectiveRDfig}, but including fixed effects for population threshold, electoral term and region. The estimates confirm that in municipalities with larger councils, the probability of a conflict-related killing is significantly lower. Estimates in column (1) show that the probability of a conflict-related  killing during a given period is 4.6 percentage points lower in municipalities with more councillors. This is a large effect, given that in the sample a conflict-related selective killing occurs in approximately 14\% of the periods. In addition, columns (2) and (3) in Table \ref{tableRDconflictbaseline} confirm that the effect of a larger council is concentrated on killings that are very selective. Columns (1) to (3) in Panels B and C in Table \ref{tableRDconflictbaseline} consider the effect of a larger council size on the number of conflict-related killings, as a total and per 100,000 residents, respectively. The results are consistent with those in Panel A:  the size of the municipal council affects conflict-related killings, but the effect is concentrated on very selective killings. Column (4) in Table \ref{tableRDconflictbaseline} confirms that a larger council has no impact on overall homicides, which includes non-conflict-related killings.\footnote{\label{footnotenothresholds3o5b}In Web Appendix \ref{appomittedrobust}, I show that the results in Table \ref{tableRDconflictbaseline} are robust to i) using as a dependent variable the ratio of conflict-related killings to total killings, ii) controlling for previous crime rates, iii) using yearly averages instead of averages over the electoral period, iv) excluding the thresholds 20,000 and 100,000, v) including the 2011 and 2015 elections (for which I use the NCHM data described in the footnote \ref{dataNCHM}). In Web Appendix \ref{appomittedrobust}, I also show that the effect of larger councils is specific to killings that deliberately target civilians (for which I use as a dependent variable other types of conflict-related violence).}


\section{Mechanisms}\label{mechanisms}

The results presented in the previous section can be explained in several ways. In this section, I focus on the explanation I believe to be the most plausible: larger municipal councils decrease conflict-related violence because they lead to more representation for parties with paramilitary links. This increases the coercive power of paramilitaries and deters guerrillas from carrying out selective killings.

This  ``deterrence'' hypothesis is based on two main assumptions. First, in municipalities with larger councils, there is a higher probability that more groups will be directly represented (I have called this ``political openness''); in Colombia, this phenomenon particularly benefited parties linked to paramilitary groups. Second, when politicians with paramilitary links enjoy greater influence, the coercive power of paramilitaries is higher, and this translates into a reduction in selective violence by guerrillas.  In the rest of this section, I provide evidence to support these assumptions. 
 
 
 \subsection{Political openness}\label{politicalopenness}
 
\begin{figure}[H]
\begin{center}
\caption{ RD figures for baseline electoral outcomes}\label{electoralRDfig}
\vspace{-0.3cm}
\begin{subfigure}[b]{0.45\textwidth}	
\resizebox{7cm}{3.5cm}{\includegraphics[width=6cm]{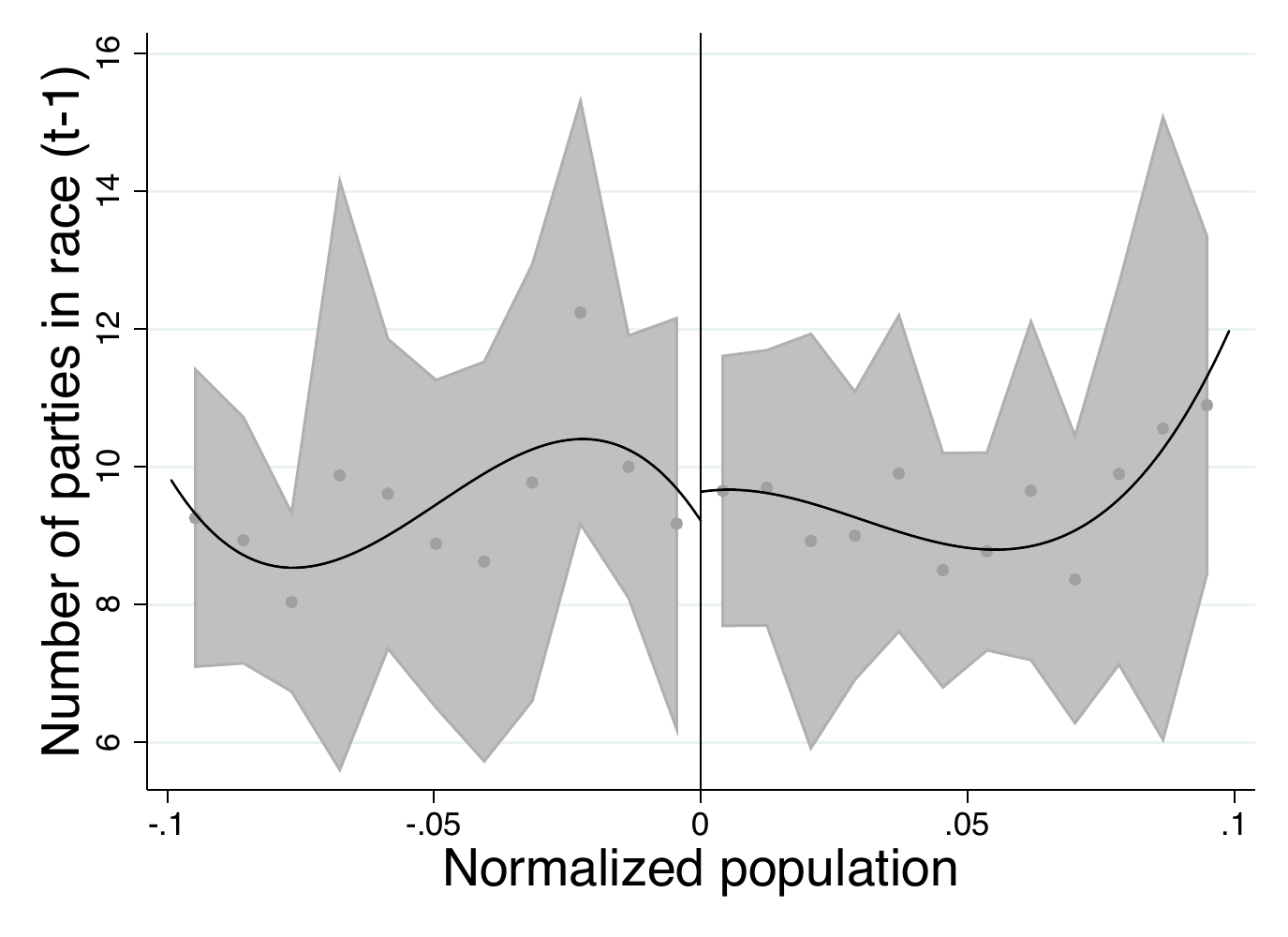}}
\vspace{-0.2cm}
\caption{Previous election}
 \end{subfigure}
\begin{subfigure}[b]{0.45\textwidth}	
\resizebox{7cm}{3.5cm}{\includegraphics[width=6cm]{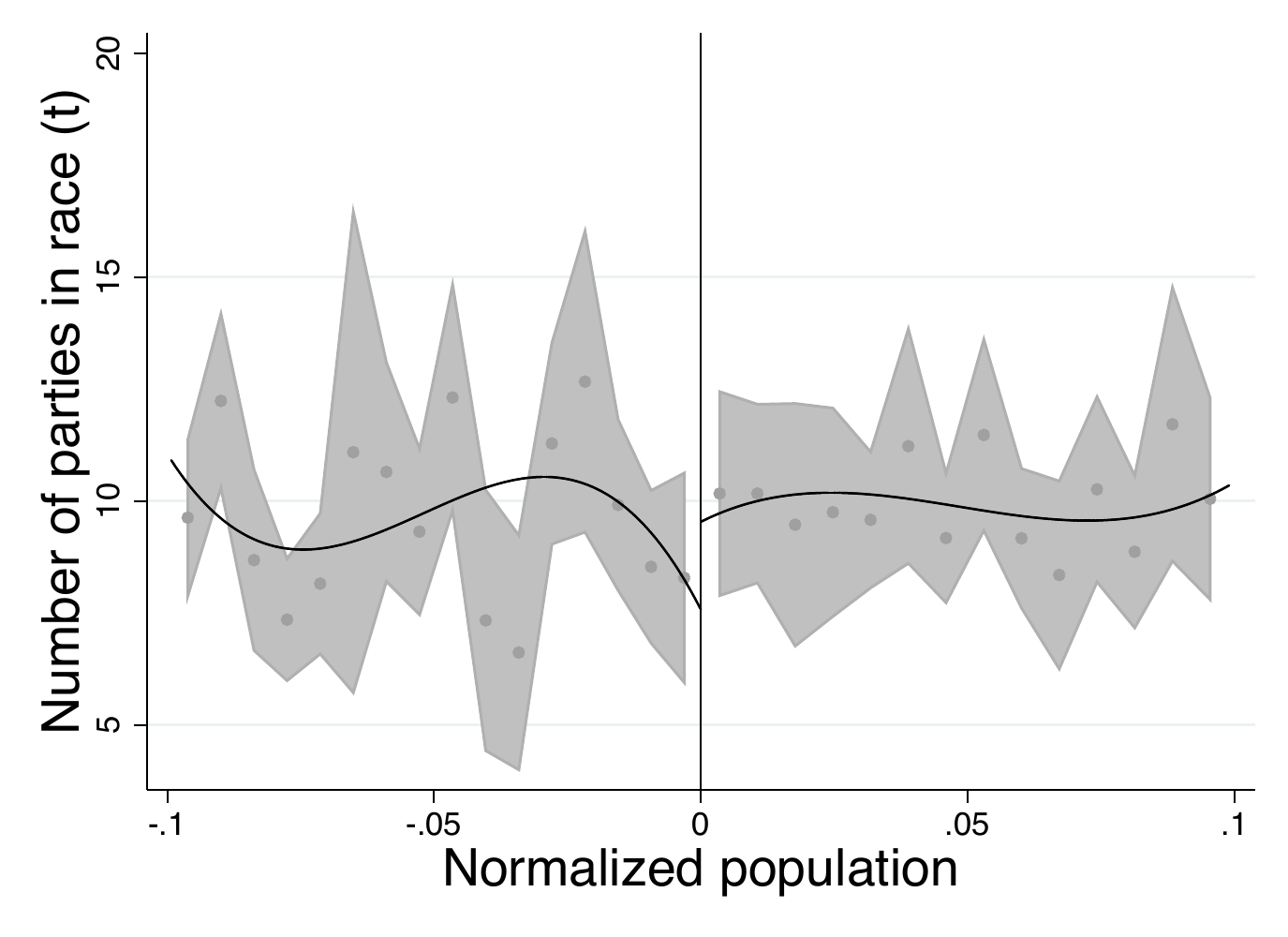}}
\vspace{-0.2cm}
\caption{Current election}
 \end{subfigure}
\begin{subfigure}[b]{0.45\textwidth}	
\resizebox{7cm}{3.5cm}{\includegraphics[width=6cm]{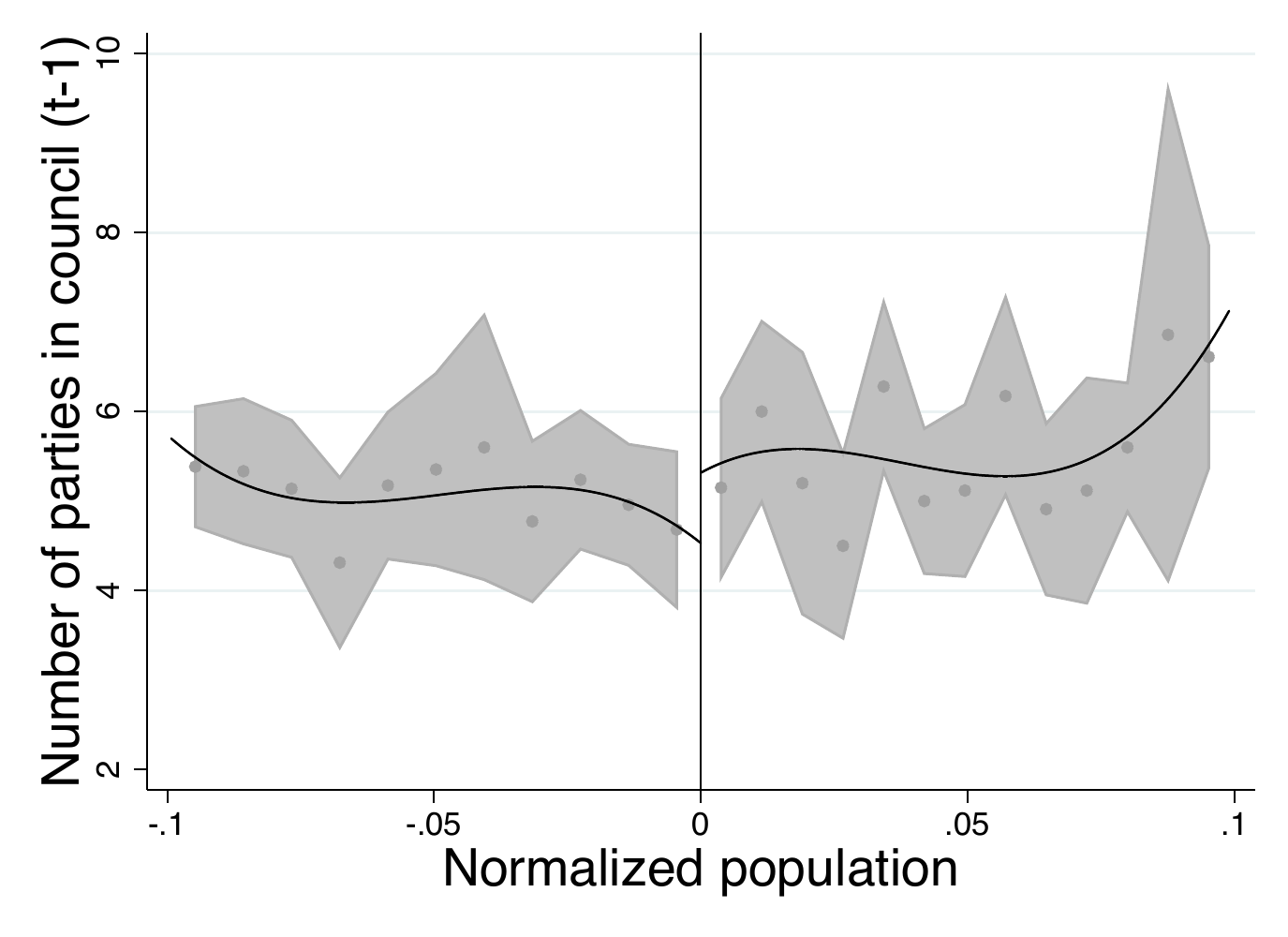}}
\vspace{-0.2cm}
\caption{Previous election}
 \end{subfigure}
\begin{subfigure}[b]{0.45\textwidth}	
\resizebox{7cm}{3.5cm}{\includegraphics[width=6cm]{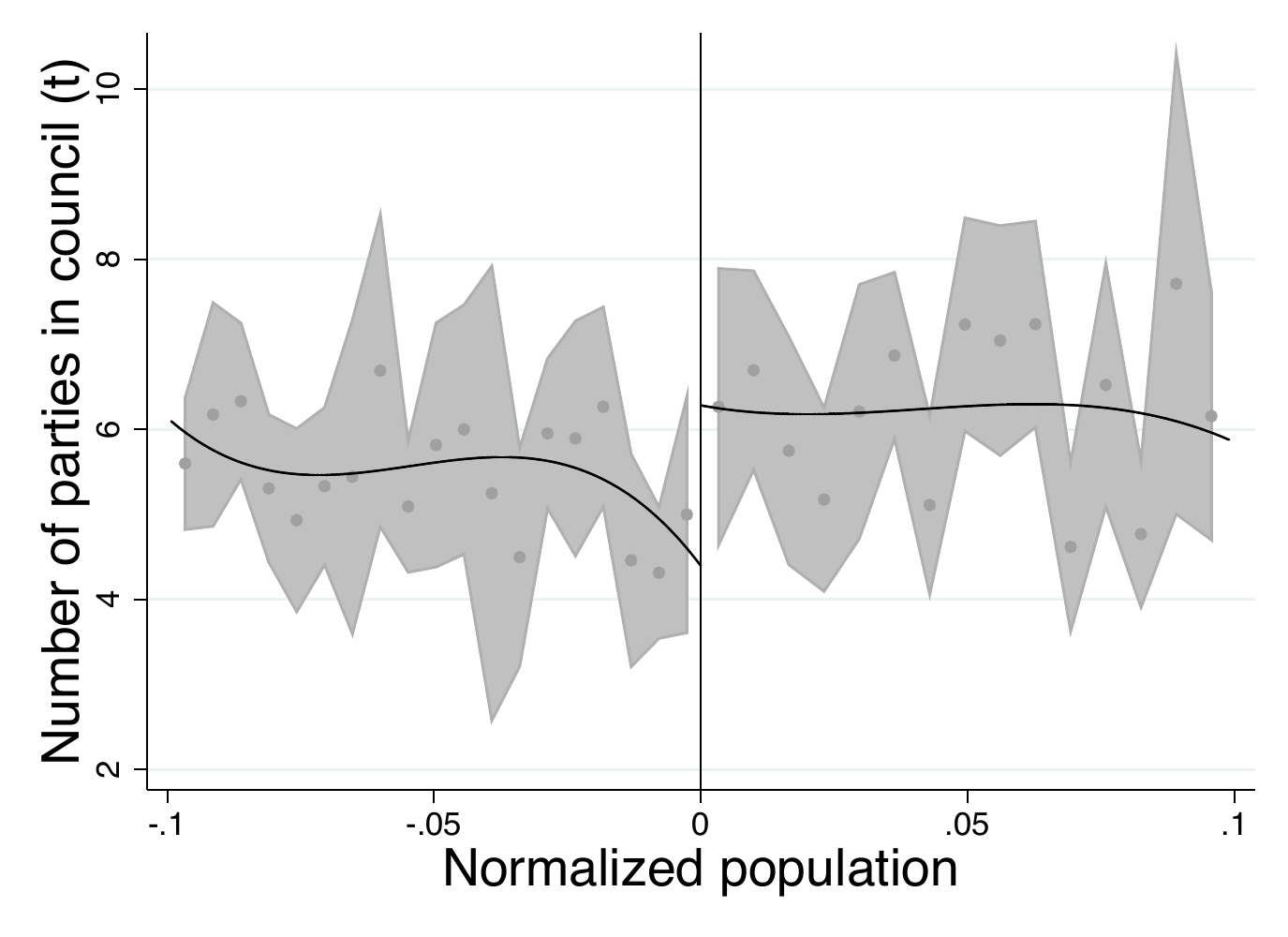}}
\vspace{-0.2cm}
\caption{Current election}
 \end{subfigure}
\begin{minipage}{15cm} \scriptsize These figures show electoral outcomes by normalized population size, with a negative value indicating smaller councils.  The bandwidth (h) is  chosen optimally using the algorithm by \cite{CalonicoCattaneoTitiunik2014a} as  implemented in Stata by the command rdplot.ado. No controls or fixed effects are included. The shaded regions represent the 95\% confidence interval.
\end{minipage}
\end{center}
\end{figure}

Panels (a)-(d) in Figure \ref{electoralRDfig}  examine how the size of a municipal council affects  the number of parties participating in local elections, and the number of political parties winning at least one seat. The solid line plots predicted values based on a local polynomial RD model with a kernel-weighted polynomial fit of order 3 and no covariates. The shaded areas indicate 95\% confidence intervals. While panel (b) shows no effect for the number of parties participating in an election, panel (d) shows a different pattern: being in the treatment group (i.e. having a larger municipal council) results in approximately one more party on the council. Panels (a) and (c) show no effect for the outcomes of the preceding election, which provides additional evidence that the RD sample is balanced.

Table \ref{electoralbasicRDtab} examines the previous results  in more detail  using my preferred non-parametric approach. All columns report estimates over bandwidths chosen using the procedure from \cite{CalonicoCattaneoTitiunik2014a}. Columns (2)-(3)  confirm the results in Figure \ref{electoralRDfig}, and column (1) includes estimates for voter turnout. While showing no effect for turnout or participation of political parties, these estimates confirm that the number of parties on the council is significantly higher in municipalities with larger councils, increasing by 1.086 (s.e. = 0.448). The effect is large, given that the average number of political parties per council in the sample is approximately 5, and that the increase is one-half of a standard deviation.

The results in Figure \ref{electoralRDfig} and Table \ref{electoralbasicRDtab} provide important evidence that a close relationship exists between council size and political openness.\footnote{\label{footnotenontraditional}An additional result is obtained if a peculiarity of the Colombian party system is taken into account: two political parties, the Liberal Party (center-left) and the Conservative Party (center-right), dominated the political landscape for over a century. Although this situation has changed in the last decade, Liberals and Conservatives continue to win a high percentage of elections at all levels of government (in my sample, Liberals and Conservatives are represented on 81\% and 69\% of councils, respectively). Column (4) in Table \ref{electoralbasicRDtab} repeats the analysis in column (3)  for parties other than Liberals and Conservatives, which I call ``non-traditional'' parties. The estimate shows that the number of non-traditional parties with at least one seat is significantly larger in municipalities with a larger council. These results provide additional and strong evidence that in municipalities with larger councils, more political parties have direct representation.} Although my motivation for examining this relationship is its potential impact on armed conflict, these findings are interesting in and of themselves. The results are related to a question widely discussed in the political science literature about the effect of electoral rules on electoral outcomes. According to this  literature, electoral district magnitude --- the number of seats elected in a given district  in a given election --- should increase the number of parties with representation \cite[see][]{Rae1967, TaageperaShugart1989, Lijphart1994}. The results in Figure \ref{electoralRDfig} and Table \ref{electoralbasicRDtab} are consistent with this prediction. The explanation is purely mechanical: given a fixed number of parties running,  in municipalities with a larger council more political parties should have direct representation.\footnote{By 	``mechanical'', I mean non-strategic, or to use Duverger's terminology, ``non-psychological'' \cite[see][p. 315]{Duverger1951}. An alternative --- and less mechanical --- explanation could be that in electoral systems with larger districts, voters are more  likely to cast a ballot for small parties; this would increase incentives for small parties to field candidates, which would cause their vote shares to increase, thereby also increasing the chances that they gain representation \cite[see][]{Duverger1951, TaageperaShugart1989, Lijphart1994, AmorimCox1997, ClarkGolder2006}. However, this explanation is inconsistent with the results in columns (1) and (2) in Table \ref{electoralbasicRDtab} the number of parties running and the voter turnout is not higher in larger districts. In addition, this explanation should only be revelant when voters would like to support small parties but are wary of ``wasting'' their votes on candidates who are unlikely to win. As the previously mentioned literature also emphasizes, if voters prefer large parties to begin with (which may occur in polities characterized by a lack of politically relevant social heterogeneity), increasing a district's magnitude should not affect the number of parties running, nor turnout rates, and a purely mechanical explanation for an increase the number of parties gaining representation may be more likely.}

 
 \medskip
 
The previous results provide strong evidence that in municipalities with larger councils, more political parties have direct representation. Given the background described in Section \ref{background}, a natural question that arises is whether the groups that gain representation as a result of a larger council are linked to armed actors. 

To examine this issue, I first classify Colombian political parties based on the historical characteristics of the party system and also according to their connections to the armed conflict. A first category includes parties with paramilitary links.  To identify such parties, I exploit the ``parapolitics" scandal described in Section \ref{background}, which revealed that certain non-traditional parties with representation across the country maintained closed links with paramilitary groups. Specifically, I focus on parties that are large enough to be represented in the Colombian Senate, and whose senators have been arrested (and in some cases found guilty) for having established a series of accords --- which at the time were secret --- with paramilitary groups.\footnote{Table \ref{tableparamilitaryparties} in the Web Appendix \ref{appsuppfigtab} provides some details about these parties, which have direct representation on approximately 45\% of municipal councils. It is noteworthy that most of these senators were the leaders of their parties. Among the parties with at least one senator connected to paramilitaries, I select those whose percentage of senators connected to the paramilitaries is more than 60\%; \cite{Lopez2010cap1} describes these parties as ``born captured''  by the paramilitaries \cite[see][p. 51]{Lopez2010cap1}. Even though 60\% is a reasonable threshold (as shown in Table \ref{tableparamilitaryparties}), the main results are robust to using a lower threshold.}

A second category includes left-wing parties. As mentioned in Section \ref{background}, only a few such parties have been explicitly linked to guerrilla groups. However, it is reasonable to expect that threats and violence targeted elected officials who were more ideologically distant, and that the guerrillas may have expected less resistance, and perhaps even collaboration or direct representation, from elected officials who were ideologically close.\footnote{Table \ref{tableleftwingparties} in Web Appendix \ref{appsuppfigtab} provides details about these parties, which are represented on 18\% of municipal councils. Even though in some cases the relationship between certain left-wing parties and the guerrillas may have been ambivalent \cite[][]{Duncan2015}, during the period I focus on, the  biggest guerrilla group supported Colombia's most important left-wing party  \cite[see][]{ElclarinAUG272007, Basset2008}.} 

A third category consists of Colombia's most traditional parties:  the Liberal Party and the Conservative Party.\footnote{See footnote \ref{footnotenontraditional} for details about these parties.}

Table  \ref{tableRDsuccessparties}  shows the success of paramilitary-linked, left-wing and Liberal or Conservative parties. It considers the probability of at least one party within each group winning at least one seat on a municipal council. Interestingly and importantly, while column (1) shows a  statistically significant  increase in the probability that a party with paramilitary links wins at least one seat, column (2) shows that parties that may be ideologically close to guerrilla groups (those that I classify as ``left-wing'') do not increase their representation when the municipal council is larger. These results provide evidence that paramilitary groups benefited the most from the greater political openness of larger councils in Colombian municipalities during the period I focus on.\footnote{In Web Appendix \ref{appomittedrobust}, I show that the results in Table  \ref{tableRDsuccessparties} are robust to using the alternative definition of paramilitary-linked and left-wing parties mentioned in the notes to Tables \ref{tableparamilitaryparties} and \ref{tableleftwingparties}.}  

\medskip

The previous results are important for the rest of this paper.\footnote{The results in Table  \ref{tableRDsuccessparties}  can also be understood as providing evidence against a possible concern related to the interpretation of the results in Section \ref{politicalopenness} as ``political openness,'' insofar as a political system more open to groups closely linked to local elites (i.e. the paramilitaries) may also be less open to other groups traditionally excluded from power (i.e. left-wing parties). In this respect, the estimates in columns (2) and (4)  of Table  \ref{tableRDsuccessparties} show that an increase in representation for paramilitary-linked parties is not accompanied by a simultaneous decrease in representation for left-wing parties.}   However, before examining this relationship, it is convenient to briefly consider the plausibility of an alternative and intuitive explanation, based on the notion of power-sharing \cite[e.g. as it is presented in][]{Lijphart1977,  ReynalQuerol2002jcr, ReynalQuerol2002DPE, HartzellHoddie2003, FrancoisRainerTrebbi2015}.  According to this explanation, greater representation of parties linked to the armed groups could make these groups see formal political power as a substitute for violence or, alternatively, may make peaceful interactions between these groups more attractive.  Importantly, the results in Table  \ref{tableRDsuccessparties}  show that left-wing parties do not gain more political representation on larger councils, which contradicts the hypothesis that conflict-related killings are lower because left-wing insurgents are substituting violence for formal political power.  As for the paramilitaries, their close links with local elites and their consequent lack of incentives to use violence to challenge the \emph{status quo} mean it is unlikely that they see formal political power as a substitute for violence.  Further, the estimates in Table  \ref{tableRDsuccessparties} are also inconsistent with the hypothesis that larger councils facilitate peaceful interactions between armed groups: column (4) shows that larger councils do not substantially increase the likelihood that a council will have both a party with paramilitary links and a left-wing party.\footnote{Even though column (4) shows a statistically significant increase, the increase is not  different from the effect on the success of paramilitary-linked parties found in column (1).} 

 
 \subsection{Paramilitary-linked representation and political violence}\label{paramilitarylinkedrepresentationpoliticalviolence}
 
Although the results in the previous subsection are consistent with the deterrence hypothesis, they only provide indirect evidence of a causal relationship between political representation with paramilitary links and a lower prevalence of conflict-related violence. In this subsection, I provide evidence that it is not coincidental that larger municipal councils are associated both with more parties with paramilitary links and with fewer selective killings. In addition, I propose and empirically examine a mechanism through which more paramilitary-linked representation translates into less political violence.

First, I examine the effect of a larger council on selective killings, distinguishing between municipalities where paramilitary-linked parties are represented on council (columns (1) to (3) of panel A in Table \ref{tableRDconflictbytypeparty}) and municipalities without paramilitary representation (columns (4) to (6) of panel A in Table \ref{tableRDconflictbytypeparty}). The results confirm that the effect is larger for those municipalities with paramilitary-linked parties. In municipalities where paramilitaries do not have political representation, the effect is not statistically different from zero. 

The results in panel A of Table \ref{tableRDconflictbytypeparty} are problematic given that the increased representation of political parties linked to paramilitaries clearly results from council size. As an alternative, I explore paramilitary political representation in mayoral elections. If the deterrence hypothesis is true, a mayor from a paramilitary-linked party should reinforce the effect of councillors with paramilitary links on selective violence. Although the outcomes of mayoral elections are plausibly related to those for a municipal council, mayors are directly and independently elected, which alleviates concerns about the election directly depending on the size of the council. Panel B in Table \ref{tableRDconflictbytypeparty} examines the effect of a larger council on selective killings, distinguishing between municipalities with and without a mayor from a paramilitary-linked party (columns (1) to (3) and (4) to (6), respectively).  Even though there is a statistically significant decrease in selective killings in municipalities with a mayor from a party that does not have paramilitary links, the effect is substantially greater in municipalities with a mayor from a paramilitary-linked party.\footnote{In Web Appendix \ref{appomittedrobust}, I show that a larger council does not affect the probability of electing a mayor from any group that I have already defined. However, since endogeneity concerns about the mayor's party may remain, in Panel C in Table \ref{tableRDconflictbytypeparty}, I examine robustness by looking at close mayoral elections. Thus, in addition to municipalities with populations just above or below the population thresholds, I further restrict the sample to close mayoral elections where either the first- or second-place candidate is from a party with paramilitary links (Figure \ref{McCraryparamayorpopthresl} in Web Appendix \ref{appsuppfigtab} reports the RD density tests for manipulation). This panel shows results that are consistent with those in panels A and B in Table  \ref{tableRDconflictbytypeparty} (although the results are imprecise given the significant loss of observations). }

\medskip

Now I examine how  more paramilitary-linked representation translates into less political violence. As previously mentioned, my preferred mechanism is based on the idea that when politicians with paramilitary links enjoy greater influence, the coercive power of paramilitaries is higher, and in such a scenario, it is likely that guerrillas are deterred from carrying out selective killings. In the rest this section, I provide more details and empirical evidence in favor of this specific channel.

An extensive literature has documented the increase in coercive power of Colombian paramilitaries stemming from the co-opting of politicians and political parties. There is now a consensus among scholars who study the Colombian conflict that a key objective of paramilitaries was the capture of institutions and the imposition of a new social order. To attain this objective, it was crucial to gain the support of local officials through co-optation, threats or a combination of the two \cite[see][]{GutierrezBaron2005, Valencia2007, Garay2008, Avila2010, AcemogluRobinsonSantos2013, GafaroIbanezJustino2014, Gutierrez2019}.\footnote{For instance, \cite{AcemogluRobinsonSantos2013} show that when a senator receives a greater proportion of votes in areas with a significant paramilitary presence, the senator is more likely to subsequently be arrested for illegal connections with paramilitaries and to have supported legislation viewed as lenient towards such organizations. According to  \cite{GafaroIbanezJustino2014}, the presence of armed groups in Colombian municipalities is associated with increased participation in political organizations, and this greater participation is driven by the capture of these organizations by the non-state armed actors, who then create networks that impose stronger controls over the population.}

Among the qualitative evidence gathered by the GMH, it is common to find testimonies showing that in municipalities governed by politicians with paramilitary links, the coercive power of these armed groups is expressed through the absence of denunciations. The following is one example:
 
\begin{quote}
\footnotesize
We did not denounce it [some irregularities associated with the presence in power of  politicians linked to paramilitaries] because, on the one hand, there were pressures; on the other hand, we were afraid to do it ... We have been asked: ``Hey, why did you not denounce it?" But we never seriously considered that ... Look,  in these regions ... you have to be quiet, because if you start talking or something ... you have to pay \cite[see][p. 350]{GMH2013}
\end{quote}

How does greater paramilitary coercive power cause a reduction in selective violence? A possible explanation, consistent with the anecdotal evidence documented above, is the following. In order for an armed group to be able to carry out selective killings, collaboration from the population is required (for example, by providing information about key targets). In areas controlled by paramilitary groups, collaboration with the guerrillas is riskier. In such a scenario, it is reasonable to expect i) less selective killings by the guerrillas, and ii) no significant difference in selective killings by the paramilitaries.\footnote{The effect of paramilitaries on selective killings would be ambiguous because on the one hand they could obtain more collaboration from the population to carry out killings, but on the other hand there would be fewer targets (if their deterrence strategy is successful). This mechanism is consistent with \citealt[][capt. 7]{Kalyvas2006}.} 

An ideal empirical evaluation of this mechanism requires a direct measure of paramilitary coercive power. Unfortunately, such a measure does not exist. Therefore, in addition to examining whether this mechanism is consistent with the evidence from previous sections, I conduct several exercises to test certain key implications. 

I start by re-examining the results of the impact of municipal council size on armed conflict. Table \ref{tableRDarmedconflicgroup} looks at this again using an alternative dataset from the CERAC. Like the GMH data, the CERAC data provides time-varying information about violent crimes against persons in a municipality. However, it includes other kinds of actions (e.g. infrastructure attacks) and, importantly, provides precise information about the main armed group behind each action.\footnote{For some killings, the GMH identifies a group that may have carried out the killing. Unfortunately, this data is very fragmentary: the perpetrator is explicitly identified in only 9.8\% of killing events.} In addition to confirming the robustness of the main results in Section \ref{mainresults}, Table \ref{tableRDarmedconflicgroup} sheds more light on the deterrence hypothesis. Column (1) shows that that larger councils significantly decrease violent actions by any non-state armed group. Columns (2) to (4) distinguish between violent actions by guerrillas, paramilitaries and the Colombian army. Column (2) shows that municipalities with larger councils have a significantly lower probability of violent action by a guerrilla group. Column (3) also shows a negative effect for violent actions caused by paramilitaries, but it is not statistically significant.\footnote{In Web Appendix \ref{appomittedrobust}, I show that the results in Table \ref{tableRDarmedconflicgroup} are robust to using yearly averages instead of averages over the electoral period.} 

The deterrence mechanism may also help explain a result in Tables \ref{tableRDconflictbaseline} and \ref{tableRDconflictbytypeparty} that I have not discussed yet: the impact of greater political representation of paramilitary-linked parties on selective violence is significantly lower for less selective killings (i.e. massacres). How is this result consistent with the deterrence hypothesis? A possible explanation is that in order to carry out a highly selective homicide, an armed group requires high cooperation from the population. The massacres, insofar as  they are less selective killings, require less collaboration from the population to be carried out. Therefore, it is reasonable to expect that the coercive power  that an armed group may have over a certain population has a lower (or null) effect on the occurrence of massacres.\footnote{This mechanism is also consistent with \citealt[][capt. 7]{Kalyvas2006}.}\footnote{In Web Appendix \ref{appmechanism}, I provide additional (indirect) evidence on the deterrence hypothesis. Specifically, I show that the effect of an exogenous increase in the influence of politicians with paramilitary links on selective killings is amplified in areas that have been more  contested.} 


\subsection{Rent Extraction and Power Consolidation}\label{rentextractionandpowerconsolidation}

The results in the last sections provide evidence consistent with the hypothesis that the decrease in conflict-related homicides in municipalities with larger councils is explained by greater coercive power of paramilitaries, which results in less violence because guerrilla groups are deterred from certain kinds of killings.  I call this the ``deterrence'' hypothesis, which I relate to the literature on the consolidation of power and the establishment of a monopoly on violence (of paramilitaries). In this last section, I provide additional evidence consistent with this hypothesis, particularly the incentives of paramilitaries to increase their coercive power and consolidate their power. 

First, recall that Table \ref{tableRDarmedconflicgroup} shows that paramilitary groups do not significantly reduce their level of violence in municipalities where they have more political representation. This is counter to the alternative hypothesis that paramilitaries replace their own violence with rent extraction. However, this does not necessarily mean that paramilitaries' opportunities for rent extraction are not elevated for other reasons. In fact, greater paramilitary influence may imply more instruments to impose order and organize violence. Additionally, as some literature on stationary banditry shows \cite[see][]{Olson1993, BatesGreifSingh2002, SanchezdelaSierra2015}, as paramilitaries' power consolidates, their time horizon may be longer, which may give them incentives to invest in rent-seeking activities. 

Empirically assessing the extent to which paramilitary-linked parties capture more rents is a daunting task. I examine the plausibility of this hypothesis indirectly, by looking at the effect of a larger municipal council on fiscal outcomes, public goods provision and coca cultivation.  Table \ref{tableRDfiscaloutcomes} shows the RD estimates of the effect of a larger municipal council on key fiscal outcomes: capital expenditures (i.e. investments in urban infrastructure, education, health and housing), current expenditures (i.e. supplies and government employee salaries), revenue from local taxes, and capital revenue (i.e. transfers from the central government and royalties from natural resource extraction). Greater political power for parties with paramilitary links may imply a higher capacity for extracting rents through friendly means (e.g. from natural resource royalties, as suggested by \citealp{DubeVargas2013}). However, the results in Table \ref{tableRDfiscaloutcomes} show that this does not seem to be the case: the estimated effect for all fiscal outcomes is not statistically significant.\footnote{Although the results in Table \ref{tableRDfiscaloutcomes} do not provide evidence of rent-seeking, it may be that resources are extracted through providing fewer public goods or from more illegal activities. Tables \ref{tableRDpublicgoods} and  \ref{tableRDcoca} in the Web Appendix \label{appA} complement these results by looking at certain key indicators of local public goods provision and coca cultivation. The tables show no evidence of less local public goods provision nor more coca cultivation or aerial spraying  in municipalities with larger councils.}

Tables \ref{tableRDfiscaloutcomes} shows no evidence of rent extraction when there are councillors from paramilitary groups. An additional explanation is based on the idea that the paramilitaries' main objective is not short-term rent extraction at the local level, but political influence at the national level. According to this explanation, political representation at the local level should help paramilitaries get direct political representation at the national level, or at least reduce the representation of their main competitors (e.g. the left).\footnote{Leaders of paramilitaries have explicitly recognized this objective \cite[see][]{Gutierrez2015}. As mentioned in Section \ref{background}, paramilitaries had direct political representation on the Colombia Senate from 2002-2010. Scholars have identified the following potential motivations for the paramilitaries to penetrate national political institutions: i)  helping to approve laws giving \emph{de facto} amnesty to paramilitaries, and ii) helping to block a series of laws related to  redistributing land titles \cite[see][]{Garay2008, Lopez2010cap1, AcemogluRobinsonSantos2013, Gutierrez2019}.}  Table \ref{tableNTRDsuccesspartiessenate}  considers this possibility by examining the percentage of votes obtained by paramilitary-linked and left-wing parties in the next election for the Colombian Senate. Consistent with this hypothesis, column (2) in panel A shows that municipalities with larger councils voted less for left-wing parties. Importantly, these results suggest that over time, political openness may have the somewhat paradoxical effect of making democracy less open by allowing paramilitary groups to have more political representation. As for the effect on the vote for paramilitary-linked parties, the estimate is positive, but statistically insignificant.

Since it is clear that paramilitary-linked parties obtained representation on the Colombian Senate, as a final exercise I examine whether there is an effect of a larger council size on the percentage of votes obtained by these parties in municipalities where they obtained at least one seat on the local council. In addition, I examine whether the effect is stronger in the 2002 and 2006 Senate elections, which correspond to the period of maximum penetration of paramilitaries in national politics. Columns (4) and (5) in panel A of Table \ref{tableNTRDsuccesspartiessenate} show the estimates of paramilitary-linked parties' vote shares in municipalities with and without paramilitary representation on the local council. Panel B in Table  \ref{tableNTRDsuccesspartiessenate} focuses on the 2002 and 2006 Senate elections. Importantly, the results are consistent with paramilitary-linked parties getting more votes in the 2002 and 2006 Senate elections in municipalities in which larger councils allowed them to get direct political representation.


\section{Conclusion}\label{conclusion}

This study examines how Colombian political institutions that are exogenously more open to the entry of political parties into local government (which I call ``political openness'') affects violence and armed conflict. RD estimates show that non-traditional parties are represented more frequently in municipalities with larger councils, and that this phenomenon particularly benefits parties linked to paramilitary groups. RD estimates suggest that the probability of a conflict-related homicide is significantly lower in municipalities with larger councils and with more councillors linked to paramilitary groups. Further analysis suggests that the lower level of conflict-related violence is associated with a rise in the political power of paramilitary groups, and a consequent decrease in violent action by guerrillas. 

Several opportunities exist for future research. While I have established that more political openness results in less conflict-related violence, it is not clear how permanent this effect is. One could also examine whether the decreases in conflict-related violence within municipalities reflects an overall reduction in violence or merely a shifting of crime across municipalities.  Finally, there is the question of how increased political openness, and the consequent political participation of extremist and paramilitary-linked groups, affects the capacity of the Colombian state to monopolize violence.


\newpage


\section*{Tables and Figures}


\renewcommand{\arraystretch}{0.8} 
\begin{table}[H]
\begin{center}
\small
\caption{Local councils in Colombia from 1997-2007}\label{t1obsallyears}
\begin{tabular}{cccccccc}
\hline\hline
\TBstrut
 &    &  \multicolumn{6}{c}{Obs. in bins of (relative to upper cut-off)}\\
  &   Council &  \multicolumn{2}{c}{0.05} &  \multicolumn{2}{c}{0.10} &  \multicolumn{2}{c}{0.15} \\\cmidrule[0.2pt](l){3-4}\cmidrule[0.2pt](l){5-6}\cmidrule[0.2pt](l){7-8}
Municipal population  &   members & Below &    Above &  Below &    Above &  Below &    Above \\
(1)   & (2) & (3) & (4) & (5) & (6) &  (7) &  (8)\\
\midrule
0-5000      &           7&          62&          91&         125&         152&         178&         211\\
5001-10000  &           9&         104&         126&         231&         243&         350&         351\\
10001-20000 &          11&          84&          91&         197&         174&         322&         268\\
20001-50000 &          13&          40&          34&          79&          64&         121&          89\\
50001-100000&          15&          20&          15&          35&          26&          50&          40\\
100001-250000&          17&           2&           7&           9&          12&          14&          18\\
250001-1000000&          19&           1&           0&           2&           0&           4&           2\\

More than 1 million    & 21 &  0 &   0 &   0 &   0 &  0  &  0 \\
\hline\hline
\multicolumn{8}{p{14.3cm}}{\scriptsize \textbf{Notes}: Columns 1 and 2  summarize the mapping of municipal populations to council sizes as prescribed by Law 136 of 1994: if a municipality's population is less than or equal to 5,000, the council must consist of seven members; if the population is greater than 5,000 but less than or equal to 10,000, the council size must be nine members, and so on. Columns 3-8 show the number of observations for different bandwidths (the widths of the ``window'' of observations around the respective population thresholds). As is discussed in Section \ref{empiricalstrategy}, the analysis focuses on municipalities with a population of at least 15,000 inhabitants.}
\end{tabular}
\end{center}
\end{table}


\newgeometry{left=2cm,bottom=1.5cm,top=1.2cm}
{ 
\renewcommand{\arraystretch}{0.4} 
\setlength{\tabcolsep}{2pt}
\begin{table}[H]
\footnotesize
\caption{Pre-treatment characteristics}\label{precharacteristics10}
\vspace{-0.4cm}
\begin{tabular}{lcccccccc}
\hline\hline\addlinespace[0.15cm]
   &\multicolumn{3}{c}{Entire sample} &\multicolumn{3}{c}{10\% population spread}  \\
&&&&& & &RD& SE on \\
   & Obs.  & Mean & St. dev.& Obs.  & Mean& St. dev. &estimate &estimate \\\cmidrule[0.2pt](l){2-4}\cmidrule[0.2pt](l){5-7}\cmidrule[0.2pt](l){8-9}
            &         (1)&         (2)&         (3)&         (4)&         (5)&         (6)&         (7)&         (8)\\
\midrule
\emph{\textbf{Conflict-related violence (pre-election term)}}&            &            &            &            &            &            &            &            \\
Killings (overall)&        2355&       2.619&       6.646&         598&       2.229&       5.869&      -0.653&       0.545\\
Very selective killings&        2355&       1.746&       4.237&         598&       1.455&       3.665&      -0.499&       0.382\\
Massacres   &        2355&       0.873&       3.063&         598&       0.774&       2.654&      -0.246&       0.287\\
Occurrence of a killing (overall)&        2355&       0.076&       0.120&         598&       0.065&       0.109&      -0.018&       0.016\\
Occurrence of a very selective killing&        2355&       0.069&       0.112&         598&       0.059&       0.101&      -0.018&       0.015\\
Occurrence of a massacre&        2355&       0.011&       0.034&         598&       0.011&       0.032&      -0.002&       0.003\\
\emph{\textbf{Armed conflict (pre-election term)}}&            &            &            &            &            &            &            &            \\
Action by guerrilla&        1900&       0.480&       0.403&         468&       0.469&       0.402&      -0.075&       0.073\\
Action by paramilitaries&        1900&       0.206&       0.292&         468&       0.201&       0.277&      -0.046&       0.063\\
Action by Nal. army&        1900&       0.436&       0.385&         468&       0.442&       0.387&      -0.123&       0.071\\
Encounter paramilitaries vs guerrilla&        1900&       0.138&       0.256&         468&       0.122&       0.240&      -0.017&       0.058\\
Encounter Nal. army vs guerrilla&        1900&       0.363&       0.381&         468&       0.356&       0.379&      -0.086&       0.067\\
Encounter Nal. army vs paramilitaries&        1900&       0.116&       0.233&         468&       0.110&       0.218&      -0.011&       0.041\\
Event of massive expulsion&        2355&       0.085&       0.199&         598&       0.088&       0.210&      -0.057&       0.038\\
\emph{\textbf{Crime (pre-election term)}}&            &            &            &            &            &            &            &            \\
Overall homicide rate&        2315&      58.840&      56.840&         583&      58.897&      59.844&      -1.246&       8.074\\
Log of hectares of coca&        1931&       1.147&       2.294&         498&       1.244&       2.399&      -0.554&       0.456\\
Log of hectares of aerial spraying on coca&        1931&       0.745&       2.050&         498&       0.858&       2.164&      -0.663&       0.448\\
\emph{\textbf{Elections (previous election)}}&            &            &            &            &            &            &            &            \\
Turnout rate&        1423&       0.572&       0.105&         376&       0.578&       0.103&      -0.013&       0.020\\
Number of parties in council&        1905&       5.221&       2.110&         489&       5.327&       2.113&       0.643&       0.332\\
Council fractionalization&        2349&       0.975&       0.100&         596&       0.985&       0.078&      -0.017&       0.013\\
Liberal party in council&        1905&       0.898&       0.303&         489&       0.857&       0.351&       0.094&       0.050\\
Conservative party in council&        1905&       0.708&       0.455&         489&       0.681&       0.467&       0.175&       0.078\\
Left-wing party in council&        1905&       0.256&       0.437&         489&       0.254&       0.436&       0.017&       0.068\\
Party with paramilitary links in council&        1905&       0.510&       0.500&         489&       0.560&       0.497&       0.057&       0.078\\
Mayor from Liberal party&        2349&       0.359&       0.480&         596&       0.344&       0.475&      -0.041&       0.074\\
Mayor from Conservative party&        2349&       0.174&       0.379&         596&       0.181&       0.386&       0.081&       0.056\\
Mayor from left-wing party&        2349&       0.027&       0.162&         596&       0.020&       0.141&       0.022&       0.032\\
Mayor from party with paramilitary links&        2349&       0.076&       0.265&         596&       0.096&       0.294&       0.063&       0.048\\
\emph{\textbf{Economy and institutions}}&            &            &            &            &            &            &            &            \\
Municipal category (first year of term)&        1424&       5.396&       1.347&         377&       5.414&       1.271&       0.158&       0.186\\
\% unsatisfied basic needs (1993 or 2005)&        2355&      48.479&      22.310&         598&      48.100&      22.708&      -4.549&       2.964\\
Schools per 1000 inhab. (1997)&        2213&      34.779&      19.559&         554&      37.190&      20.639&      -3.234&       4.450\\
Hospitals per 1000 inhab. (1997)&        2213&       2.842&       2.561&         554&       2.941&       2.716&      -0.464&       0.387\\
Bank branches per 1000 inhab. in 1997&        2028&       7.364&       3.928&         502&       7.458&       3.756&       0.479&       0.690\\
Courts per 1000 inhab. (1997)&        2213&      10.030&       7.906&         555&      10.212&       8.190&       1.289&       1.384\\
Police stations per 1000 inhab. (1997)&        2208&       4.306&       2.123&         552&       4.432&       2.139&      -0.213&       0.274\\
\emph{\textbf{Fiscal outcomes (pre-election term)}}&            &            &            &            &            &            &            &            \\
Log current spending per capita&        2331&      -2.526&       0.537&         593&      -2.519&       0.557&       0.043&       0.094\\
Log fixed capital spending per capita&        2331&      -2.222&       0.630&         593&      -2.167&       0.674&       0.059&       0.117\\
Log other capital spending per capita&        2326&      -2.014&       0.932&         593&      -1.962&       0.903&       0.022&       0.165\\
Log tax revenue per capita&        2330&      -3.291&       1.153&         592&      -3.268&       1.156&       0.071&       0.170\\
Log royalties per capita&        1661&      -5.075&       2.591&         419&      -5.031&       2.665&       0.113&       0.596\\
Log transfers per capita&        2308&      -1.728&       0.546&         590&      -1.708&       0.542&       0.119&       0.084\\
Total deficit per capita&        2334&      -0.018&       0.234&         593&      -0.033&       0.414&      -0.066&       0.084\\
\emph{\textbf{Geographic characteristics}}&            &            &            &            &            &            &            &            \\
Surface area (km2)&        2355&    1223.097&    3634.450&         598&    1367.573&    4335.841&    -919.130&     658.769\\
Mean altitude (m)&        2294&     991.990&    1424.019&         576&     985.476&     918.371&     154.226&     141.588\\
Distance to Bogota (km)&        2355&     353.962&     192.983&         598&     354.501&     195.120&       1.050&      22.075\\
Distance to the capital of department (km)&        2355&      77.396&      60.821&         598&      75.774&      60.309&      -5.536&       9.593\\
\% Municipalities on the Atlantic coast&        2355&       0.248&       0.432&         598&       0.241&       0.428&       0.026&       0.095\\
\% Municipalities in the eastern region&        2355&       0.186&       0.390&         598&       0.224&       0.417&       0.071&       0.076\\
\% Municipalities in the central region&        2355&       0.172&       0.377&         598&       0.154&       0.361&       0.000&       0.060\\
\% Municipalities on the Pacific coast&        2355&       0.199&       0.399&         598&       0.197&       0.398&      -0.083&       0.067\\
\% Municipalities in Antioquia&        2355&       0.144&       0.352&         598&       0.129&       0.335&      -0.026&       0.068\\
\% Municipalities in the Amazon region&        2355&       0.050&       0.218&         598&       0.055&       0.229&       0.029&       0.037\\

\addlinespace[0.15cm]
\hline\hline\addlinespace[0.1cm]
\multicolumn{9}{p{18cm}}{\scriptsize \textbf{Notes:} 
The sample in all columns includes municipalities with a population of at least 15,000 inhabitants (i.e. it excludes municipalities with a population where the closest threshold is one of the two smallest thresholds). Pre-election term means that the variable is averaged for the previous election term. Data on conflict-related killings are from the  Historical Memory Group (GMH). Data on armed conflict are from the Conflict Analysis Resource Center (CERAC); this data is available for 1988-2009. Data on crime and geographic characteristics are from the Center of Studies on Economic Development (CEDE). Electoral data are from the Electoral Agency. Data on municipal public finance and municipal categories are from the National Planning Department (DNP).  Data on population and the proportion of people with unsatisfied basic needs are from the National Administrative Department of Statistics (DANE). Data on the number of courts, bank branches, hospitals, schools and community organizations are from the Social Foundation, a non-profit foundation. Column (7) reports the RD estimate from equation (\ref{rdbaseline}) when the respective characteristic is used as the dependent variable, with a data-driven bandwidth chosen optimally using the algorithm by  \cite{CalonicoCattaneoTitiunik2014a} as implemented in Stata by the command rdrobust.ado, and includes  fixed effects for population threshold, electoral term and region.  Column (7) reports the standard errors clustered by municipality.} \\
\end{tabular}
\end{table}
}

\newgeometry{left=4cm,bottom=1.5cm,top=4cm}
{ 
\begin{figure}[H]
\begin{center}
\caption{ RD figures for conflict-related violence}\label{allselectiveRDfig}
\vspace{-0.3cm}
\begin{subfigure}[b]{0.45\textwidth}
\resizebox{7cm}{3.5cm}{\includegraphics[width=6cm]{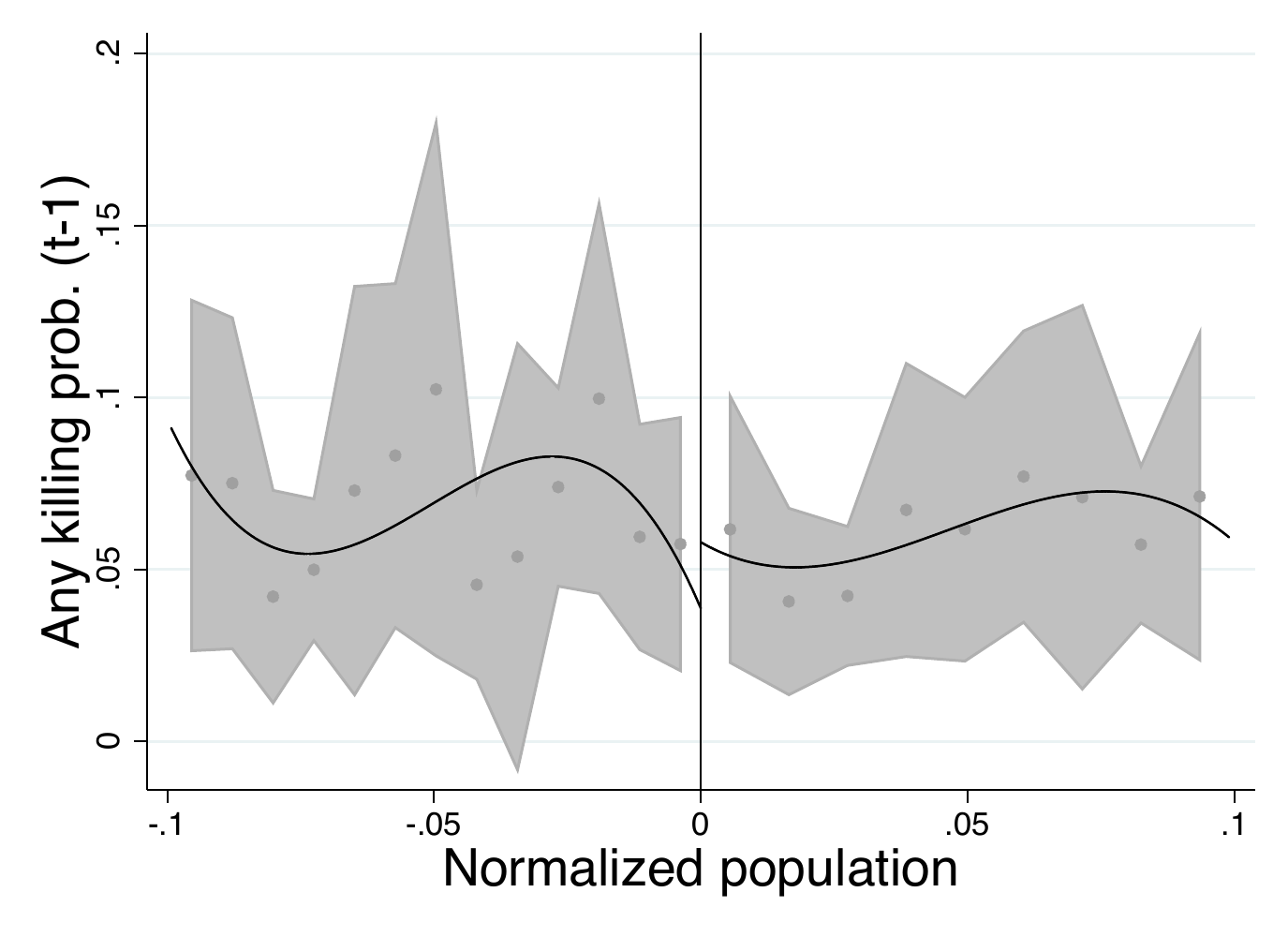}}
\vspace{-0.2cm}
\caption{Any killing prob. (pre-election)}
 \end{subfigure}
\begin{subfigure}[b]{0.45\textwidth}	
\resizebox{7cm}{3.5cm}{\includegraphics[width=6cm]{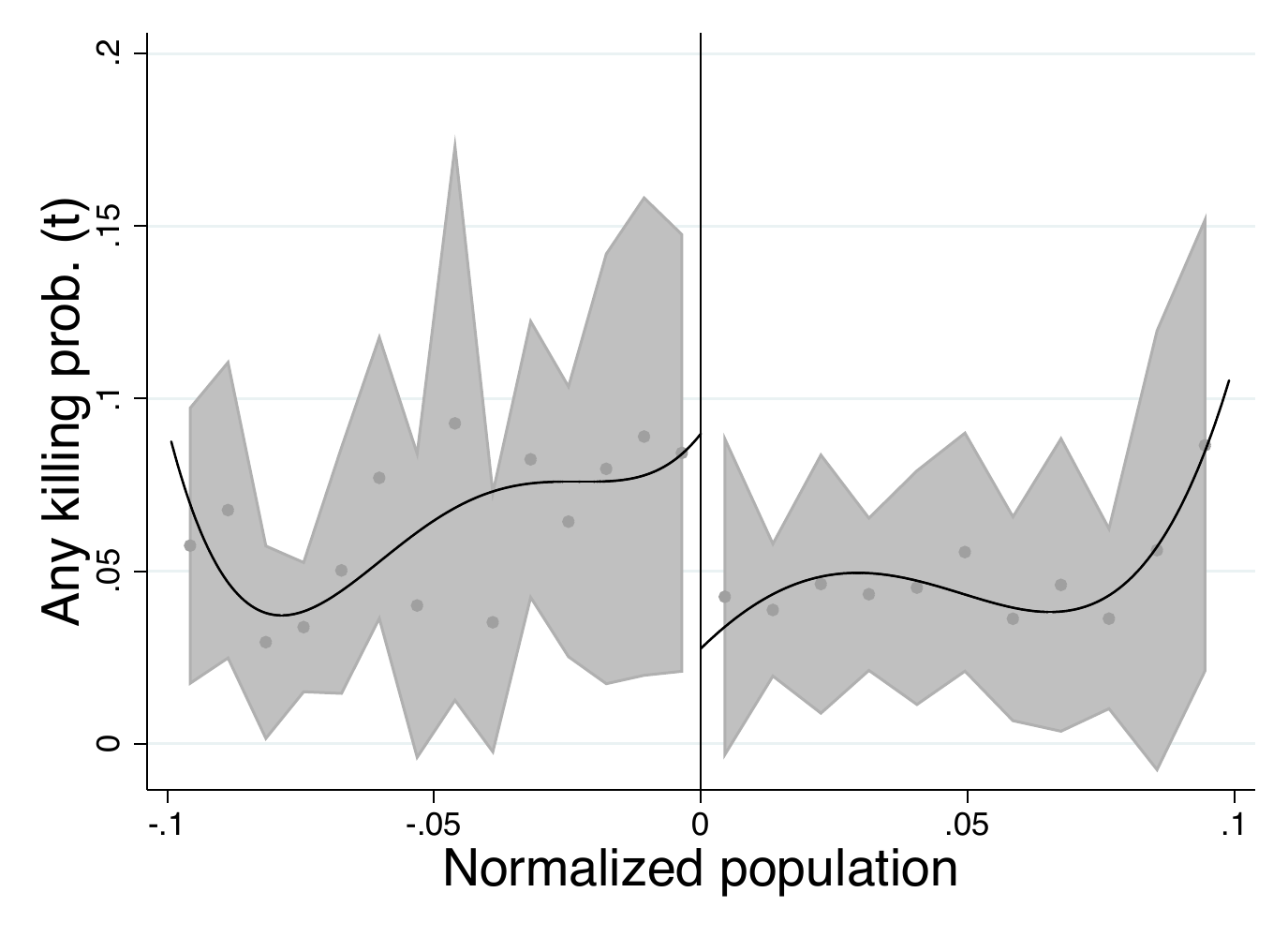}}
\vspace{-0.2cm}
\caption{Any killing prob. (post-election)}
 \end{subfigure}\\
\begin{subfigure}[b]{0.45\textwidth}
\resizebox{7cm}{3.5cm}{\includegraphics[width=6cm]{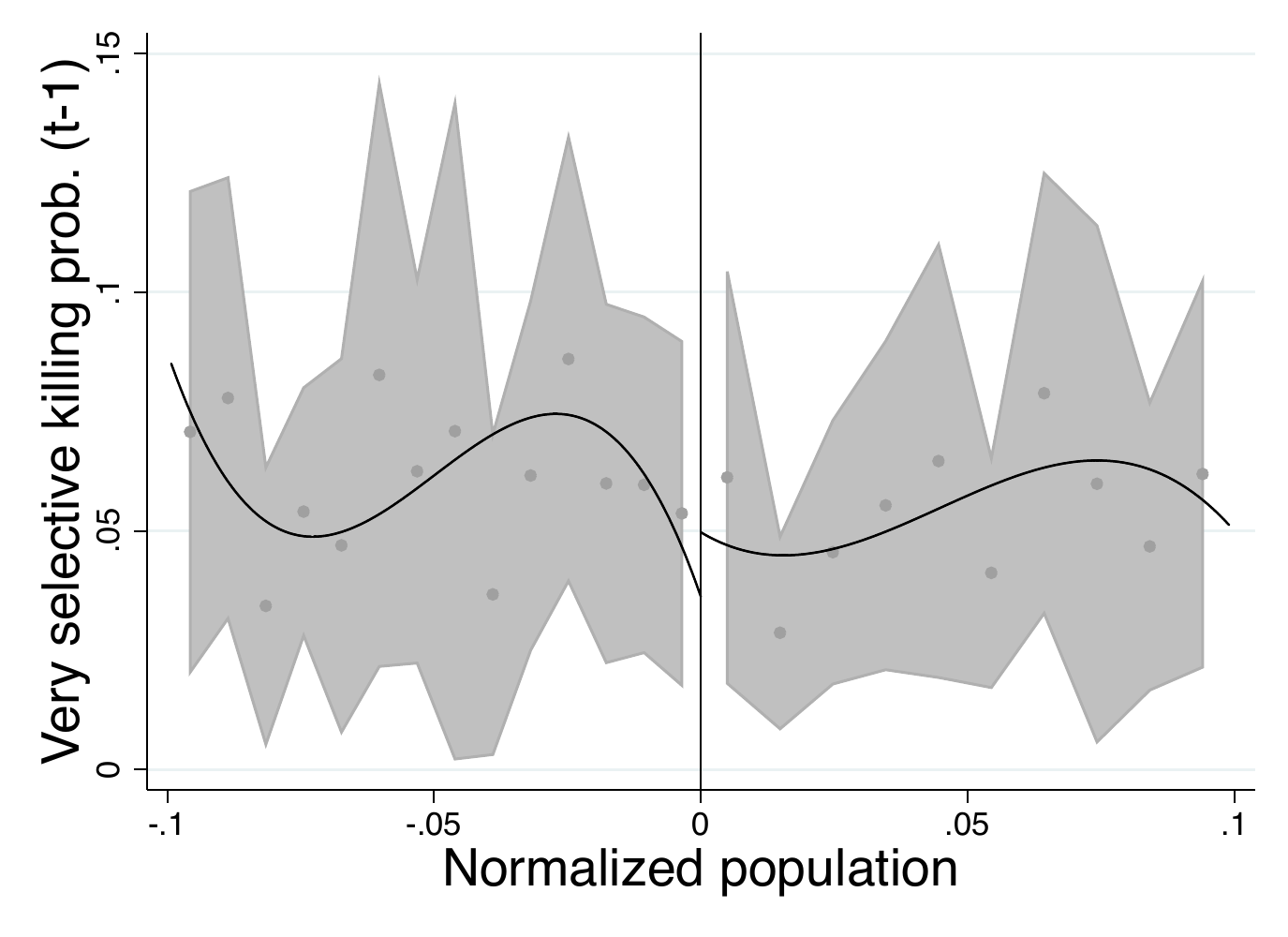}}
\vspace{-0.2cm}
\caption{Selective killing prob. (pre-election)}
 \end{subfigure}
\begin{subfigure}[b]{0.45\textwidth}	
\resizebox{7cm}{3.5cm}{\includegraphics[width=6cm]{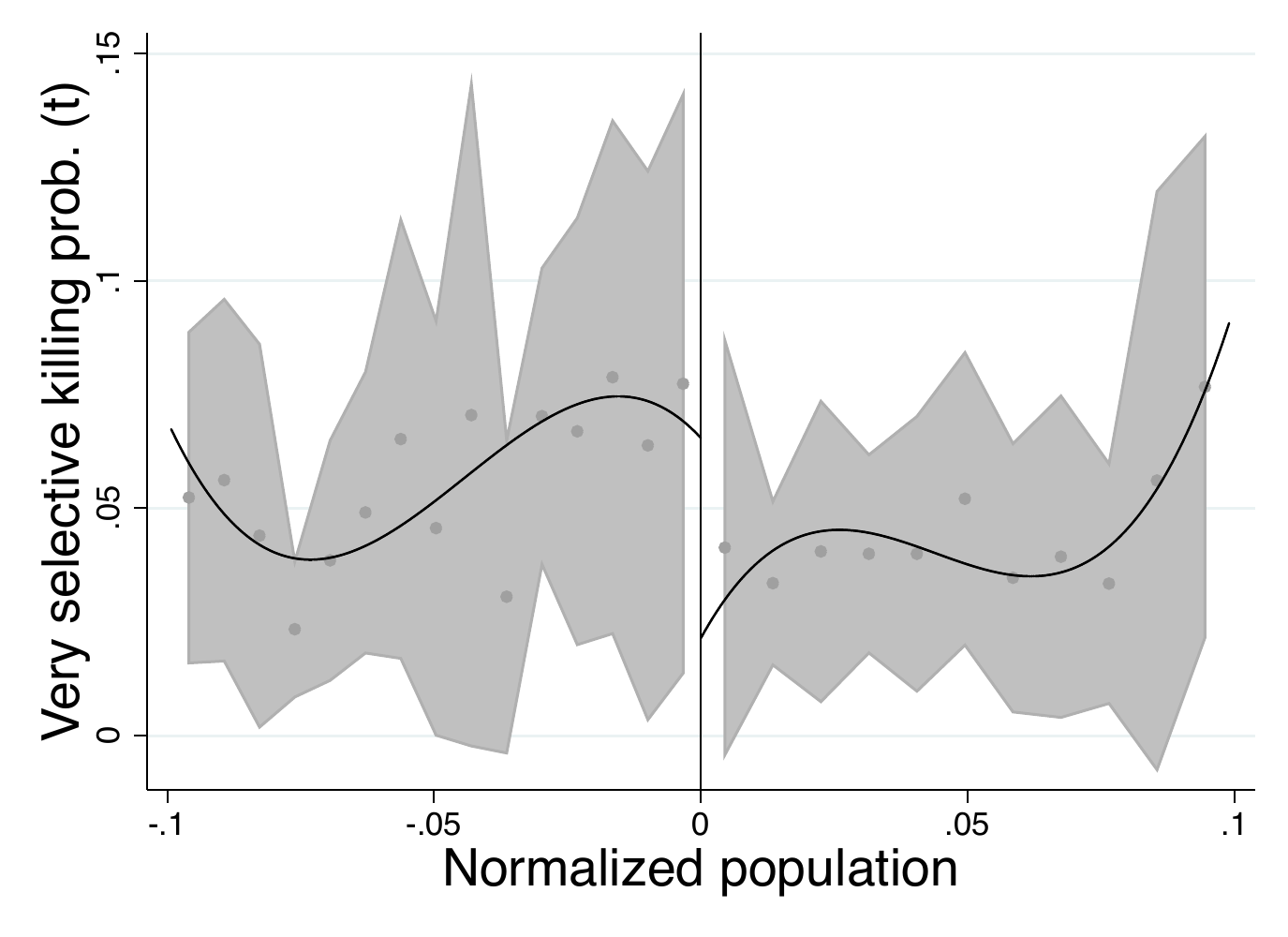}}
\vspace{-0.2cm}
\caption{Selective killing prob. (post-election)}
 \end{subfigure}\\
 \begin{subfigure}[b]{0.45\textwidth}
\resizebox{7cm}{3.5cm}{\includegraphics[width=6cm]{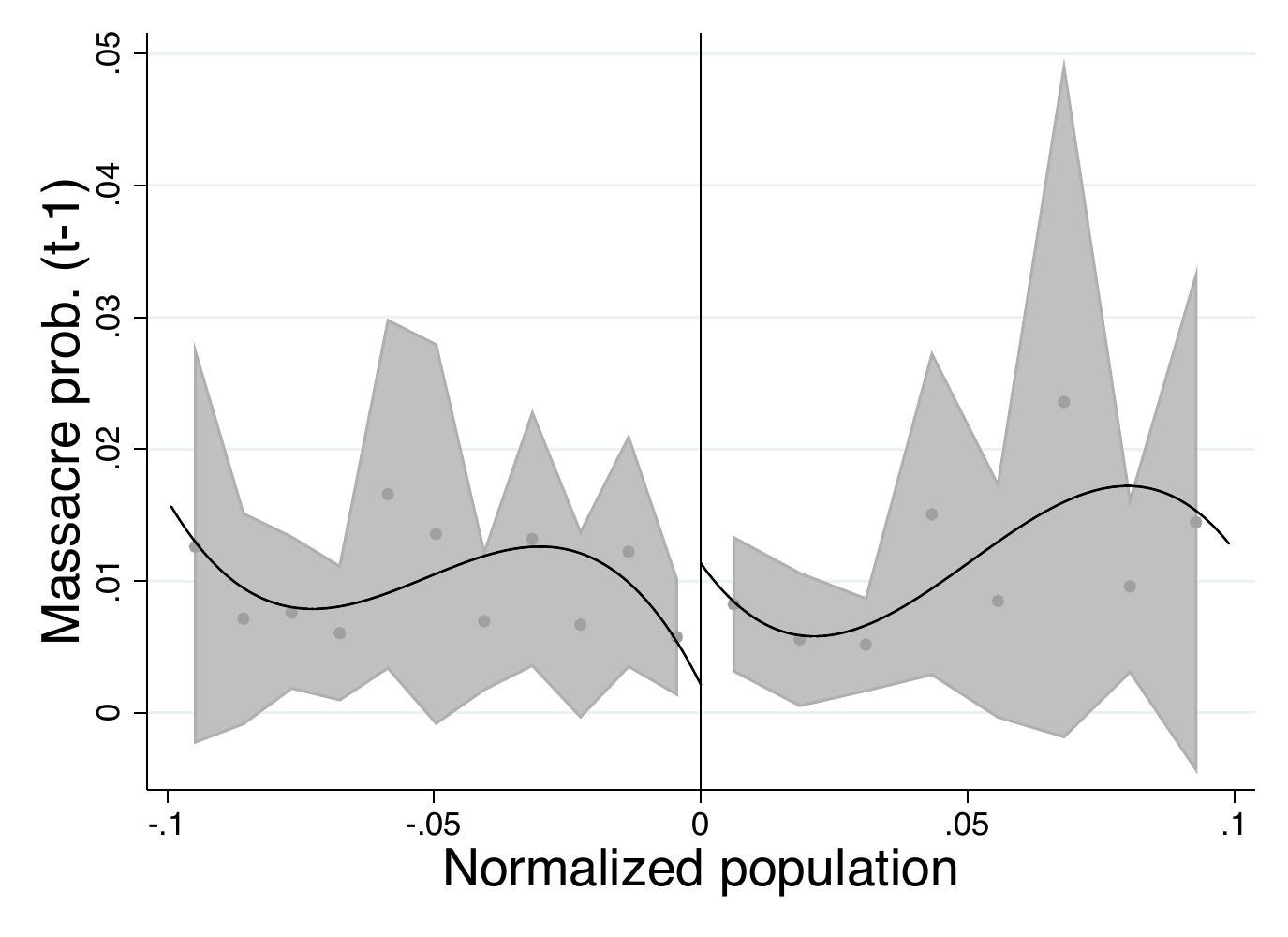}}
\vspace{-0.2cm}
\caption{Massacre prob. (pre-election)}
 \end{subfigure}
\begin{subfigure}[b]{0.45\textwidth}	
\resizebox{7cm}{3.5cm}{\includegraphics[width=6cm]{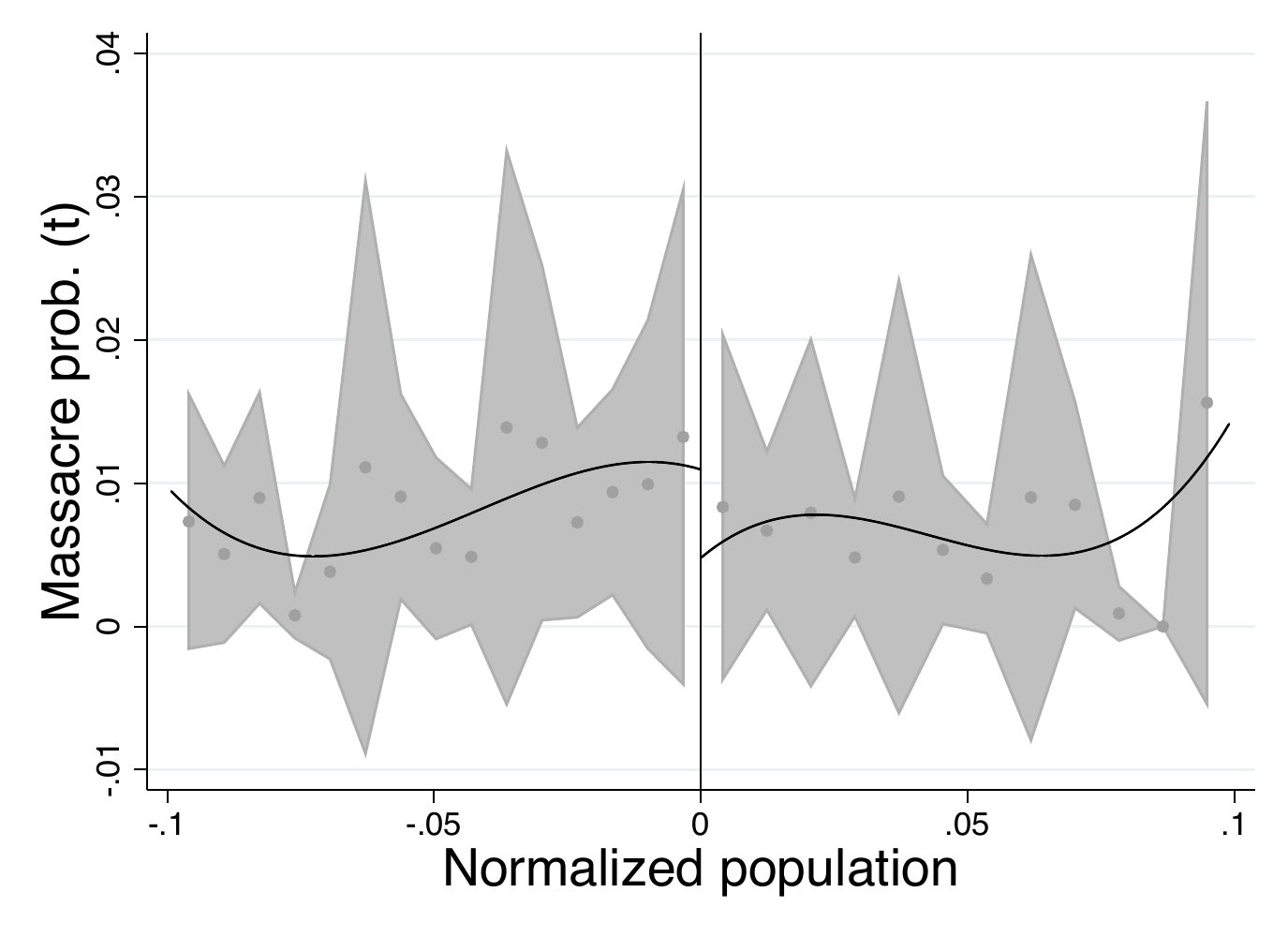}}
\vspace{-0.2cm}
\caption{Massacre prob. (post-election)}
 \end{subfigure}\\
\begin{minipage}{15cm} \scriptsize These figures show conflict-related killings by normalized population size, with a negative value indicating smaller councils.  The bandwidth (h) is  chosen optimally using the algorithm by \cite{CalonicoCattaneoTitiunik2014a} as  implemented in Stata by the command rdplot.ado. No controls or fixed effects are included. The shaded regions represent the 95\% confidence interval.
\end{minipage}
\end{center}
\end{figure}
}


\renewcommand{\arraystretch}{0.8} 
\setlength{\tabcolsep}{5pt}
\begin{table}[H]
\normalsize
\begin{center}
\caption{Effect of council size on conflict-related violence}\label{tableRDconflictbaseline}
\begin{tabular}{lcccc}\hline\hline\addlinespace[0.15cm]
\emph{Dependent variable:} & \multicolumn{3}{c}{Conflict-related homicide}  & \multicolumn{1}{c}{Overall
} \\\cmidrule[0.2pt](l){2-4}
 & \multicolumn{1}{c}{Killing} & \multicolumn{1}{c}{Select. killing}& \multicolumn{1}{c}{Massacre} & \multicolumn{1}{c}{
homicide} \\
\cmidrule[0.2pt](l){2-2}\cmidrule[0.2pt](l){3-3}\cmidrule[0.2pt](l){4-4}\cmidrule[0.2pt](l){5-5}
& (1)& (2)& (3) & (4)\\ \hline \addlinespace[0.15cm]
\multicolumn{1}{l}{\emph{\underline{Panel A}: prob.}} &    \\   \addlinespace[0.15cm]
\hspace{3mm}RD Estimate &      -0.045\sym{***}&      -0.040\sym{**} &      -0.006         &       0.013         \\
            &     (0.016)         &     (0.016)         &     (0.004)         &     (0.022)         \\
Observations&        2355         &        2355         &        2355         &        2325         \\
Eff. Number of obs&         704         &         627         &         573         &         437         \\
BW Loc. Poly. (h)&       0.118         &       0.106         &       0.096         &       0.075         \\
Robust p-value&       0.011         &       0.017         &       0.190         &       0.400         \\

\addlinespace[0.15cm]
\hline \addlinespace[0.15cm]
\multicolumn{1}{l}{\emph{\underline{Panel B}: number}} & &   \\   \addlinespace[0.15cm]
\hspace{3mm}RD Estimate &      -1.099         &      -0.921\sym{*}  &      -0.126         &      -3.580         \\
            &     (0.712)         &     (0.471)         &     (0.291)         &     (4.078)         \\
Observations&        2355         &        2355         &        2355         &        2325         \\
Eff. Number of obs&         544         &         548         &         657         &         569         \\
BW Loc. Poly. (h)&       0.093         &       0.093         &       0.111         &       0.097         \\
Robust p-value&       0.134         &       0.071         &       0.637         &       0.471         \\

\addlinespace[0.15cm]
\hline \addlinespace[0.15cm]
\multicolumn{1}{l}{\emph{\underline{Panel C}: rate}} & &   \\   \addlinespace[0.15cm]
\hspace{3mm}RD Estimate &      -2.829         &      -2.955\sym{*}  &      -0.010         &      -4.488         \\
            &     (2.408)         &     (1.744)         &     (0.982)         &     (9.656)         \\
Observations&        2355         &        2355         &        2355         &        2318         \\
Eff. Number of obs&         558         &         583         &         598         &         589         \\
BW Loc. Poly. (h)&       0.095         &       0.097         &       0.100         &       0.101         \\
Robust p-value&       0.292         &       0.097         &       0.936         &       0.696         \\

 \addlinespace[0.15cm]
\hline     \hline                                                        
\multicolumn{5}{p{12.5cm}}{\scriptsize \textbf{Notes}:  All columns report the RD estimates of having a larger council from Eq. (\ref{rdbaseline}) when the respective characteristic is used as the dependent variable. The dependent variable in panel A is the average over the electoral term of a dummy variable equal to 1 if the particular type of killing is observed in a municipality in a quarter of a year. The dependent variable in panels B and C is the average of the total number  of killings and  killings per 100,000 people, respectively. The bandwidth (h) is chosen optimally using the algorithm by \cite{CalonicoCattaneoTitiunik2014a} as  implemented in Stata by the command rdrobust.ado, and includes  fixed effects for population threshold, electoral term and region. Standard errors clustered by municipality are reported in parentheses. * significant at 10\%, ** significant at 5\%, *** significant at 1\%. }
\end{tabular}
\end{center}
\end{table}


\newgeometry{left=3cm,bottom=2cm,top=3cm}
{ 
\renewcommand{\arraystretch}{0.8} 
\setlength{\tabcolsep}{5pt}
\begin{table}[H]
\normalsize
\begin{center}
\caption{Effect of council size on baseline electoral outcomes  in council elections}\label{electoralbasicRDtab}
\begin{tabularx}{1.1\textwidth}{lYYYY}
\hline\hline\addlinespace[0.15cm]
&  \multicolumn{4}{c}{\emph{Dependent variable:}}   \\
 & \multicolumn{1}{c}{Voter} & \multicolumn{1}{c}{\# of parties}& \multicolumn{1}{c}{\# of parties}& \multicolumn{1}{c}{\# of non-traditional} \\
      & \multicolumn{1}{c}{ turnout} & \multicolumn{1}{c}{ in race}& \multicolumn{1}{c}{on council}& \multicolumn{1}{c}{parties on council} \\\cmidrule[0.2pt](l){2-2}\cmidrule[0.2pt](l){3-3}\cmidrule[0.2pt](l){4-4}\cmidrule[0.2pt](l){5-5}
& (1)& (2)& (3)& (4) \\ \addlinespace[0.15cm] \hline\addlinespace[0.15cm]
RD Estimate &       0.028         &       0.314         &       1.108\sym{***}&       0.959\sym{**} \\
            &     (0.029)         &     (0.710)         &     (0.406)         &     (0.413)         \\
Observations&        1474         &        2355         &        2355         &        2355         \\
Eff. Number of obs&         367         &         494         &         514         &         479         \\
BW Loc. Poly. (h)&       0.099         &       0.082         &       0.086         &       0.080         \\
Robust p-value&       0.353         &       0.717         &       0.018         &       0.036         \\

\addlinespace[0.15cm]
\hline\hline                                               
\multicolumn{5}{p{16.5cm}}{\scriptsize \textbf{Notes}:  All columns report the RD estimates of having a larger council from Eq. (\ref{rdbaseline}) when the respective characteristic is used as the dependent variable. The bandwidth (h) is  chosen optimally using the algorithm by \cite{CalonicoCattaneoTitiunik2014a} as  implemented in Stata by the command rdrobust.ado, and includes  fixed effects for population threshold, electoral term and region. Information on turnout for the 1997 election is unavailable. Standard errors clustered by municipality are reported in parentheses. * significant at 10\%, ** significant at 5\%, *** significant at 1\%. } 
\end{tabularx}
\end{center}
\end{table}
}


\renewcommand{\arraystretch}{0.8} 
\setlength{\tabcolsep}{8pt}
\begin{table}[H]
\normalsize
\begin{center}
\caption{Effect of council size on the success of different parties in council elections}\label{tableRDsuccessparties}
\begin{tabular}{lccccc}
\hline\hline \addlinespace[0.15cm]
     &\multicolumn{4}{c}{\emph{Dependent variable:} Presence on council of a} \\\cmidrule[0.2pt](l){2-5}
  & \multicolumn{1}{c}{Paramilitary-} & \multicolumn{1}{c}{Left-wing}& \multicolumn{1}{c}{Liberal or Con-}& \multicolumn{1}{c}{Paramilitary-linked} \\
      & \multicolumn{1}{c}{linked party} & \multicolumn{1}{c}{party}& \multicolumn{1}{c}{servative party}& \multicolumn{1}{c}{and left-wing party} \\\cmidrule[0.2pt](l){2-2}\cmidrule[0.2pt](l){3-3}\cmidrule[0.2pt](l){4-4}\cmidrule[0.2pt](l){5-5}
& (1)& (2)& (3)& (4)\\ \hline
\addlinespace[0.15cm]                                            
\hspace{3mm}RD Estimate &       0.298\sym{***}&      -0.006         &      -0.064         &       0.196\sym{***}\\
            &     (0.071)         &     (0.072)         &     (0.045)         &     (0.066)         \\
Observations&        2355         &        2355         &        2355         &        2355         \\
Eff. Number of obs&         573         &         639         &         405         &         426         \\
BW Loc. Poly. (h)&       0.096         &       0.107         &       0.069         &       0.073         \\
Robust p-value&       0.000         &       0.898         &       0.112         &       0.003         \\

\addlinespace[0.16cm]
\hline    \hline   
\multicolumn{5}{p{16.1cm}}{\scriptsize \textbf{Notes}:  All columns report the RD estimates of having a larger council from Eq. (\ref{rdbaseline}) when the respective characteristic is used as the dependent variable. The bandwidth (h) is  chosen optimally using the algorithm by \cite{CalonicoCattaneoTitiunik2014a} as  implemented in Stata by the command rdrobust.ado, and includes  fixed effects for population threshold, electoral term and region. Standard errors clustered by municipality are reported in parentheses. * significant at 10\%, ** significant at 5\%, *** significant at 1\%. }  
\end{tabular}
\end{center}
\end{table}


\renewcommand{\arraystretch}{0.8} 
\setlength{\tabcolsep}{6pt}
\begin{table}[H]
\normalsize
\begin{center}
\caption{Effect of council size on armed conflict}\label{tableRDarmedconflicgroup}
\begin{tabularx}{1\textwidth}{lYYY}
\hline\hline \addlinespace[0.15cm]
\emph{Dependent variable:}  &\multicolumn{3}{c}{Presence of violent actions by} \\ 
  & \multicolumn{1}{c}{Any group}& \multicolumn{1}{c}{Guerrillas}& \multicolumn{1}{c}{Paramilitaries} 
      \\\cmidrule[0.2pt](l){2-2}\cmidrule[0.2pt](l){3-3}\cmidrule[0.2pt](l){4-4}
& (1)& (2)& (3) \\ \hline \addlinespace[0.15cm]
\hspace{3mm}RD Estimate &      -0.168\sym{**} &      -0.187\sym{**} &      -0.028         \\
            &     (0.074)         &     (0.079)         &     (0.052)         \\
Observations&        1900         &        1900         &        1900         \\
Eff. Number of obs&         378         &         395         &         545         \\
BW Loc. Poly. (h)&       0.080         &       0.083         &       0.117         \\
Robust p-value&       0.021         &       0.015         &       0.567         \\

\addlinespace[0.15cm]
\hline                \hline                                     
\multicolumn{4}{p{15cm}}{\scriptsize \textbf{Notes}:  All columns report the RD estimates of having a larger council from Eq. (\ref{rdbaseline}) when the respective characteristic is used as the dependent variable. The bandwidth (h) is  chosen optimally using the algorithm by \cite{CalonicoCattaneoTitiunik2014a} as  implemented in Stata by the command rdrobust.ado, and includes  fixed effects for population threshold, electoral term and region. Standard errors clustered by municipality are reported in parentheses. * significant at 10\%, ** significant at 5\%, *** significant at 1\%.} 
\end{tabularx}
\end{center}
\end{table}


\newgeometry{left=1.5cm,bottom=2cm,top=3cm}
{ 
\renewcommand{\arraystretch}{0.8} 
\setlength{\tabcolsep}{5pt}
\begin{table}[H]
\normalsize
\begin{center}
\caption{Effect of council size on conflict-related violence in municipalities with paramilitary-linked parties represented on the local government}\label{tableRDconflictbytypeparty}
\begin{tabular}{lcccccc}
\hline\hline\addlinespace[0.15cm]  
&  \multicolumn{3}{c}{Presence in local government}  &  \multicolumn{3}{c}{No paramilitary-linked}     \\
&  \multicolumn{3}{c}{of a paramilitary-linked party}  &  \multicolumn{3}{c}{party in  local government}      \\\cmidrule[0.2pt](l){2-4}\cmidrule[0.2pt](l){5-7}
Dep. var. is  occurrence of a & \multicolumn{1}{c}{Killing} & \multicolumn{1}{c}{Select. killing}& \multicolumn{1}{c}{Massacre}& \multicolumn{1}{c}{Killing} & \multicolumn{1}{c}{Select. killing}& \multicolumn{1}{c}{Massacre}  \\\cmidrule[0.2pt](l){2-2}\cmidrule[0.2pt](l){3-3}\cmidrule[0.2pt](l){4-4}\cmidrule[0.2pt](l){5-5}\cmidrule[0.2pt](l){6-6}\cmidrule[0.2pt](l){7-7}
& (1)& (2)& (3)& (4)& (5)& (6) \\ \hline \addlinespace[0.15cm]
\underline{Panel A}:  &\multicolumn{6}{c}{\emph{Municipal council}}  \\   \cmidrule[0.2pt](l){2-7}
\hspace{3mm}RD Estimate &      -0.068\sym{**} &      -0.066\sym{**} &      -0.004         &      -0.022         &      -0.011         &      -0.003         \\
            &     (0.032)         &     (0.031)         &     (0.003)         &     (0.027)         &     (0.025)         &     (0.009)         \\
Observations&        1191         &        1191         &        1191         &        1164         &        1164         &        1164         \\
Eff. Number of obs&         264         &         260         &         275         &         307         &         287         &         279         \\
BW Loc. Poly. (h)&       0.083         &       0.082         &       0.088         &       0.109         &       0.101         &       0.098         \\
Robust p-value&       0.037         &       0.041         &       0.231         &       0.580         &       0.802         &       0.730         \\

\addlinespace[0.15cm]
\hline
\addlinespace[0.15cm]
\underline{Panel B}: &\multicolumn{6}{c}{\emph{Mayoralty (entire sample)}}  \\   \cmidrule[0.2pt](l){2-7}
\hspace{3mm}RD Estimate &      -0.061         &      -0.093         &      -0.007         &      -0.045\sym{***}&      -0.040\sym{***}&      -0.005         \\
            &     (0.098)         &     (0.089)         &     (0.005)         &     (0.016)         &     (0.015)         &     (0.004)         \\
Observations&         212         &         212         &         212         &        2143         &        2143         &        2143         \\
Eff. Number of obs&          39         &          44         &          22         &         688         &         648         &         579         \\
BW Loc. Poly. (h)&       0.069         &       0.076         &       0.052         &       0.126         &       0.119         &       0.108         \\
Robust p-value&       0.360         &       0.223         &       0.157         &       0.014         &       0.020         &       0.269         \\

\addlinespace[0.15cm]
\hline
\addlinespace[0.15cm]
\underline{Panel C}:  &\multicolumn{6}{c}{\emph{Mayoralty  (close victory)}}  \\   \cmidrule[0.2pt](l){2-7}
\hspace{3mm}RD Estimate &      -0.151\sym{*}  &      -0.144\sym{*}  &      -0.007\sym{*}  &      -0.024         &      -0.021         &      -0.003         \\
            &     (0.087)         &     (0.086)         &     (0.004)         &     (0.027)         &     (0.028)         &     (0.003)         \\
Observations&          77         &          77         &          77         &          96         &          96         &          96         \\
Eff. Number of obs&          25         &          24         &          20         &          30         &          29         &          28         \\
BW Loc. Poly. (h)&       0.095         &       0.093         &       0.065         &       0.118         &       0.117         &       0.110         \\
Robust p-value&       0.108         &       0.104         &       0.031         &       0.556         &       0.649         &       0.190         \\

\addlinespace[0.15cm]
\hline   \hline                                                          
\multicolumn{7}{p{18cm}}{\scriptsize \textbf{Notes}:  All columns report the RD estimates of having a larger council from Eq. (\ref{rdbaseline}) when the respective characteristic is used as the dependent variable. The bandwidth (h) is  chosen optimally using the algorithm by \cite{CalonicoCattaneoTitiunik2014a} as  implemented in Stata by the command rdrobust.ado, and includes  fixed effects for population threshold, electoral term and region. Standard errors clustered by municipality are reported in parentheses. * significant at 10\%, ** significant at 5\%, *** significant at 1\% } 
\end{tabular}
\end{center}
\end{table}
}


\newgeometry{left=2.5cm,bottom=2cm,top=3cm}
{ 
\renewcommand{\arraystretch}{0.8} 
\setlength{\tabcolsep}{5pt}
\begin{table}[H]
\normalsize
\begin{center}
\caption{Effect of council size on  fiscal outcomes}\label{tableRDfiscaloutcomes}
\begin{tabular}{lcccccc}
\hline\hline\addlinespace[0.15cm] 
 &\multicolumn{6}{c}{\emph{Dependent variable:} Log of } \\  
& \multicolumn{1}{c}{Capital} & \multicolumn{1}{c}{Current}& \multicolumn{1}{c}{Tax}& \multicolumn{1}{c}{Central govt.} & \multicolumn{1}{c}{resource}& \multicolumn{1}{c}{Total} \\
& \multicolumn{1}{c}{spending} & \multicolumn{1}{c}{spending}& \multicolumn{1}{c}{revenue}& \multicolumn{1}{c}{transfers} & \multicolumn{1}{c}{royalties}& \multicolumn{1}{c}{deficit} 
 \\\cmidrule[0.2pt](l){2-2}\cmidrule[0.2pt](l){3-3}\cmidrule[0.2pt](l){4-4}\cmidrule[0.2pt](l){5-5}\cmidrule[0.2pt](l){6-6}\cmidrule[0.2pt](l){7-7}
& (1)& (2)& (3)& (4)& (5)& (6) \\ \hline \addlinespace[0.15cm]
\hspace{3mm}RD Estimate &       0.023         &      -0.041         &       0.085         &       0.082         &       0.134         &      -0.010         \\
            &     (0.105)         &     (0.085)         &     (0.162)         &     (0.079)         &     (0.557)         &     (0.016)         \\
Observations&        2343         &        2341         &        2341         &        2341         &        1769         &        2344         \\
Eff. Number of obs&         575         &         517         &         501         &         674         &         396         &         378         \\
BW Loc. Poly. (h)&       0.097         &       0.088         &       0.084         &       0.114         &       0.091         &       0.064         \\
Robust p-value&       0.971         &       0.593         &       0.762         &       0.358         &       0.923         &       0.364         \\

\addlinespace[0.15cm]
\hline   \hline                                                          
\multicolumn{7}{p{15cm}}{\scriptsize \textbf{Notes}:  All columns report the RD estimates of having a larger council from Eq. (\ref{rdbaseline}) when the respective characteristic is used as the dependent variable. The bandwidth (h) is  chosen optimally using the algorithm by \cite{CalonicoCattaneoTitiunik2014a} as  implemented in Stata by the command rdrobust.ado, and includes  fixed effects for population threshold, electoral term and region. Standard errors clustered by municipality are reported in parentheses. * significant at 10\%, ** significant at 5\%, *** significant at 1\%. } 
\end{tabular}
\end{center}
\end{table}
}


\newgeometry{left=2.5cm,bottom=2cm,top=3cm}
{ 
\renewcommand{\arraystretch}{0.8} 
\begin{table}[H]
\normalsize
\begin{center}
\caption{Effect of council size on the next Senate election}\label{tableNTRDsuccesspartiessenate}
\begin{tabularx}{1.1\textwidth}{lYYYYY}
\hline\hline\addlinespace[0.15cm]  
 &\multicolumn{5}{c}{\emph{Dependent variable:} Success in the next Senate election of}  \\\cmidrule[0.2pt](l){2-6}
 & & &  &   \multicolumn{2}{c}{Paramilitary-linked parties} \\\cmidrule[0.2pt](l){5-6} 
    & \multicolumn{1}{c}{Paramilitary-} & \multicolumn{1}{c}{Left-wing}& \multicolumn{1}{c}{Liberals or} & \multicolumn{1}{c}{Also present} & \multicolumn{1}{c}{Not present}  \\
& \multicolumn{1}{c}{linked parties} & \multicolumn{1}{c}{parties}& \multicolumn{1}{c}{Conservatives}& \multicolumn{1}{c}{in council}& \multicolumn{1}{c}{in council} \\\cmidrule[0.2pt](l){2-2}\cmidrule[0.2pt](l){3-3}\cmidrule[0.2pt](l){4-4} \cmidrule[0.2pt](l){5-5}\cmidrule[0.2pt](l){6-6}  \addlinespace[0.15cm]
& (1)& (2)& (3) & (4) & (5)\\ \hline \addlinespace[0.15cm]
\multicolumn{1}{l}{\emph{\underline{Panel A}}} &  \multicolumn{5}{c}{\emph{ \% of votes in next Senate elections (all elections)}}      \\ \cmidrule[0.2pt](l){2-6}
\addlinespace[0.15cm]
\hspace{3mm}RD Estimate &       0.112         &      -2.559\sym{*}  &       0.981         &       1.238         &      -2.149         \\
            &     (2.033)         &     (1.321)         &     (3.011)         &     (2.314)         &     (3.018)         \\
Observations&        2349         &        2349         &        2349         &        1189         &        1160         \\
Eff. Number of obs&         651         &         595         &         688         &         300         &         277         \\
BW Loc. Poly. (h)&       0.110         &       0.100         &       0.116         &       0.095         &       0.098         \\
Robust p-value&       0.901         &       0.038         &       0.532         &       0.439         &       0.491         \\

\addlinespace[0.15cm]
\hline 
\addlinespace[0.15cm]
\multicolumn{1}{l}{\emph{\underline{Panel B}}} &  \multicolumn{5}{c}{\emph{ \% of votes in next Senate elections (only 2002 and 2006 elections)}}     \\ \cmidrule[0.2pt](l){2-6}
\addlinespace[0.15cm]
\hspace{3mm}RD Estimate &       4.646         &      -2.731         &       0.565         &       6.990\sym{*}  &      -1.063         \\
            &     (3.835)         &     (1.911)         &     (3.755)         &     (4.031)         &     (5.210)         \\
Observations&        1016         &        1016         &        1016         &         498         &         518         \\
Eff. Number of obs&         271         &         235         &         253         &         111         &         137         \\
BW Loc. Poly. (h)&       0.118         &       0.100         &       0.109         &       0.103         &       0.111         \\
Robust p-value&       0.251         &       0.120         &       0.777         &       0.090         &       0.853         \\

\addlinespace[0.15cm]
\hline   \hline                                                          
\multicolumn{6}{p{17cm}}{\scriptsize \textbf{Notes}:  
All columns report the RD estimates of having a larger council from Eq. (\ref{rdbaseline}) when the respective characteristic is used as the dependent variable. The bandwidth (h) is  chosen optimally using the algorithm by \cite{CalonicoCattaneoTitiunik2014a} as  implemented in Stata by the command rdrobust.ado, and includes  fixed effects for population threshold, electoral term and region. Standard errors clustered by municipality are reported in parentheses. * significant at 10\%, ** significant at 5\%, *** significant at 1\%.
 } 
\end{tabularx}
\end{center}
\end{table} 
}


\newpage 
\hbox {} 

\bibliographystyle{ecca}
\bibliography{bibopenness}

\clearpage
\hbox {} 

\appendix
\section{: Supplemental Figures and Tables}\label{appsuppfigtab}

\setcounter{table}{0}
\setcounter{figure}{0}
\renewcommand{\thefigure}{\Alph{section}\arabic{figure}}
\renewcommand{\thetable}{\Alph{section}\arabic{table}}
\small


\begin{figure}[H]
\begin{center}
\caption{Population distribution}\label{popdistribution}
\resizebox{11cm}{8cm}{\includegraphics[width=1cm]{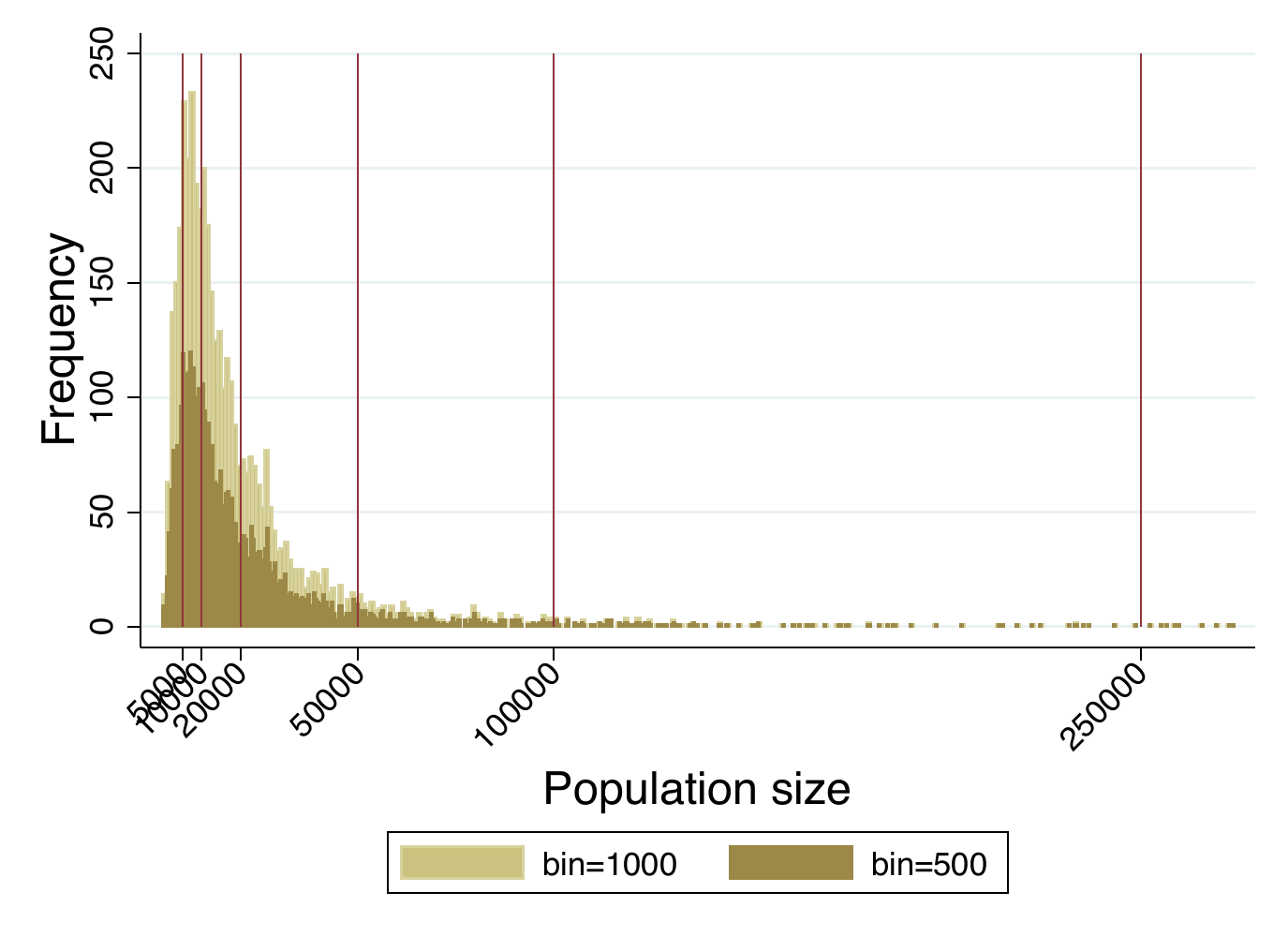}}
\end{center}
\vspace{-1cm}
\begin{center}
\begin{minipage}{12cm} \scriptsize Frequency of municipalities by population size. The vertical lines show the population thresholds used to determine the council size. To facilitate the exposition, it includes municipalities with a population up to 275,000. 
\end{minipage}
\end{center}
\end{figure}


\newpage 
\begin{figure}[H]
\begin{center}
\caption{Manipulation tests by population threshold} \label{McCrarypopthres}
\begin{subfigure}[a]{0.45\textwidth}
\resizebox{7cm}{3.5cm}{\includegraphics[width=\textwidth]{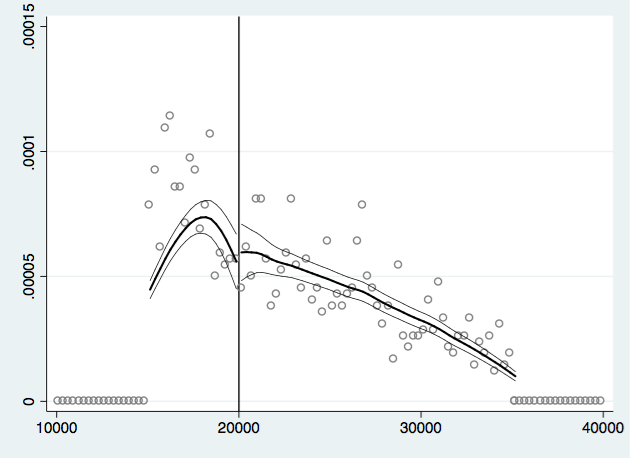}}
\caption{20,000  threshold}
\end{subfigure}
\begin{subfigure}[a]{0.45\textwidth}
\resizebox{7cm}{3.5cm}{\includegraphics[width=\textwidth]{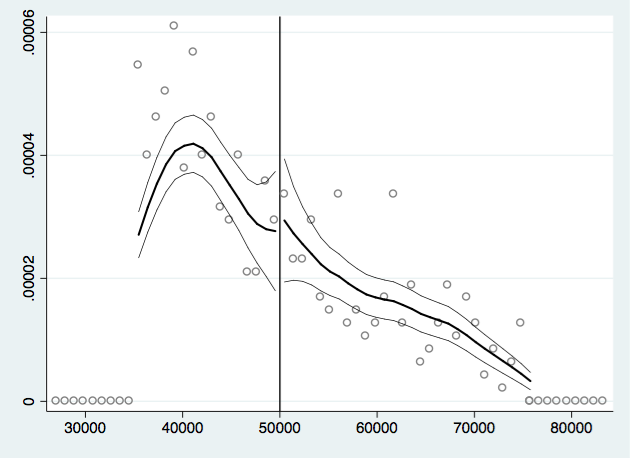}}
\caption{50,000  threshold}
\end{subfigure}\\
\begin{subfigure}[a]{0.45\textwidth}
\resizebox{7cm}{3.5cm}{\includegraphics[width=\textwidth]{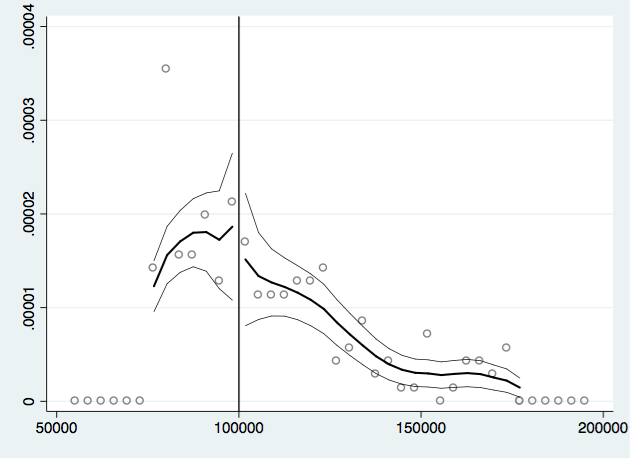}}
\caption{100,000  threshold}
\end{subfigure}
\begin{subfigure}[a]{0.45\textwidth}
\resizebox{7cm}{3.5cm}{\includegraphics[width=\textwidth]{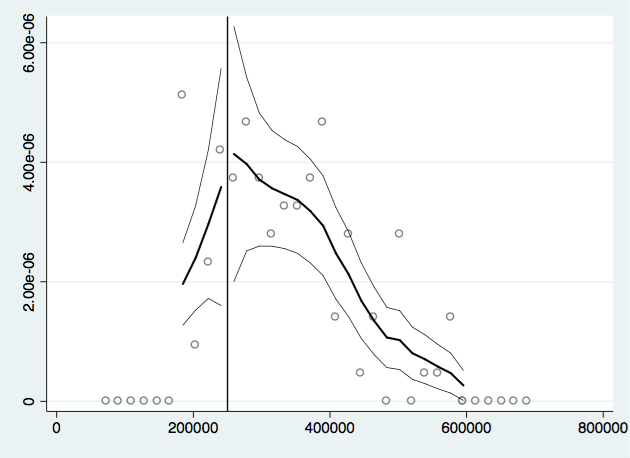}}
\caption{250,000  threshold}
\end{subfigure}\\
\vspace{-0.1cm}
\begin{minipage}{14.5cm} \scriptsize  All figures  pool all years, and show finely-gridded histograms of  the population smoothed using local linear regression, separately on either side of the cutoff  of the density function of the population. Each figure uses data only around the corresponding population threshold. The estimates of the difference in the height at the threshold and  robust manipulation p-values (as implemented in Stata by the command rddensity.ado, \citep[see][]{CattaneoJanssonMa2018a}, are:  (a)  0.092 (s.e. 0.177) and p-value 0.2549; (b) -0.055 (s.e. 0.304) and p-value 0.5612; (c) -0.316 (s.e. 0.441) and p-value 0.6704; (d)  -0.207 (s.e. 0.669) and p-value 0.2358. 
\end{minipage}
\end{center}
\end{figure}

\begin{figure}[H]
\begin{center}
\caption{Manipulation tests by election year}\label{McCraryyear}
\begin{subfigure}[b]{0.45\textwidth}	
\resizebox{7.2cm}{4.5cm}{\includegraphics[width=1cm]{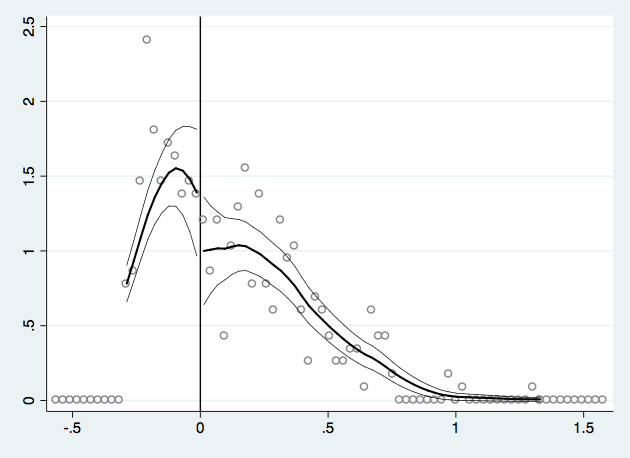}}
\caption{1997}
\end{subfigure}
\begin{subfigure}[b]{0.45\textwidth}	
\resizebox{7.2cm}{4.5cm}{\includegraphics[width=1cm]{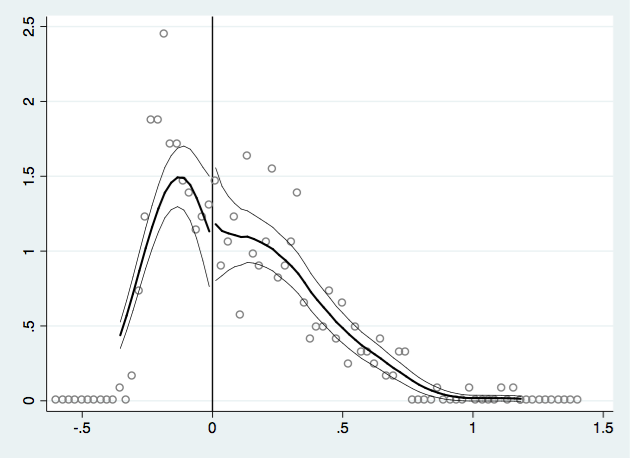}}
\caption{2000}
\end{subfigure}\\
\begin{subfigure}[b]{0.45\textwidth}	
\resizebox{7.2cm}{4.5cm}{\includegraphics[width=1cm]{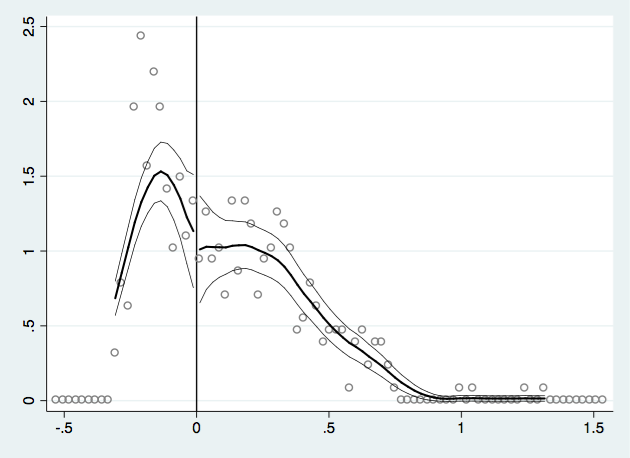}}
\caption{2003}
\end{subfigure}
\begin{subfigure}[b]{0.45\textwidth}	
\resizebox{7.2cm}{4.5cm}{\includegraphics[width=1cm]{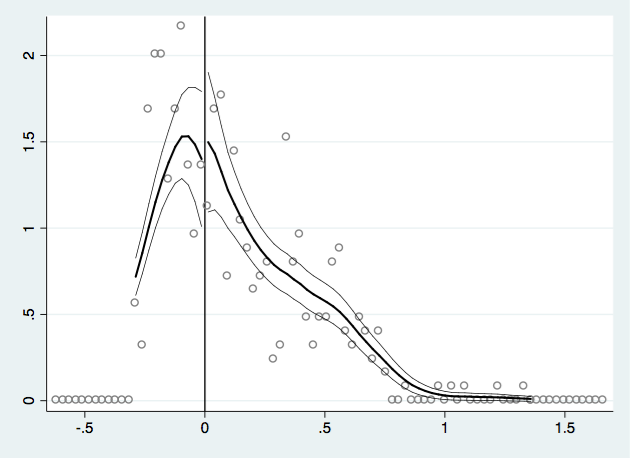}}
\caption{2007}
\end{subfigure}\\
\end{center}
\vspace{-0.8cm}
\begin{center}
\begin{minipage}{14.5cm} \scriptsize All figures show finely-gridded histograms of  the population smoothed using local linear regression, separately on either side of the cutoff  of the density function of the population \cite[][]{McCrary2008, CattaneoJanssonMa2018a}. Each figure uses data only around the corresponding population threshold, and includes municipalities with a population of at least 15,000 inhabitants. The estimates of the difference in the height at the threshold and  robust manipulation p-values (as implemented in Stata by the command rddensity.ado, \citep[see][]{CattaneoJanssonMa2018a}, are:
(a)  -0.299 (s.e. 0.273) and p-value 0.266, (b)  0.112 (s.e. 0.263) and p-value 0.732, (c)  -0.073 (s.e. 0.284) and p-value 0.391, (d)  0.133 (s.e. 0.220) and p-value 0.271.
\end{minipage}
\end{center}
\end{figure}


\newpage 
\begin{figure}[H]
\begin{center}
\caption{Manipulation tests by population threshold: two smallest thresholds} \label{McCrarypopthressmall}
\begin{subfigure}[a]{0.45\textwidth}
\resizebox{7cm}{3.5cm}{\includegraphics[width=\textwidth]{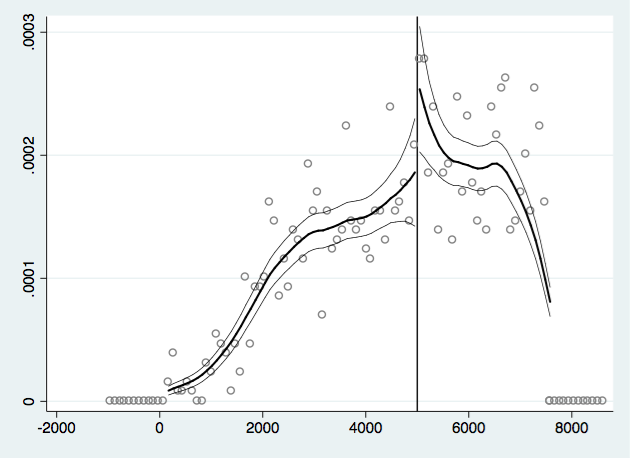}}
\caption{5,000 threshold}
\end{subfigure}
\begin{subfigure}[a]{0.45\textwidth}
\resizebox{7cm}{3.5cm}{\includegraphics[width=\textwidth]{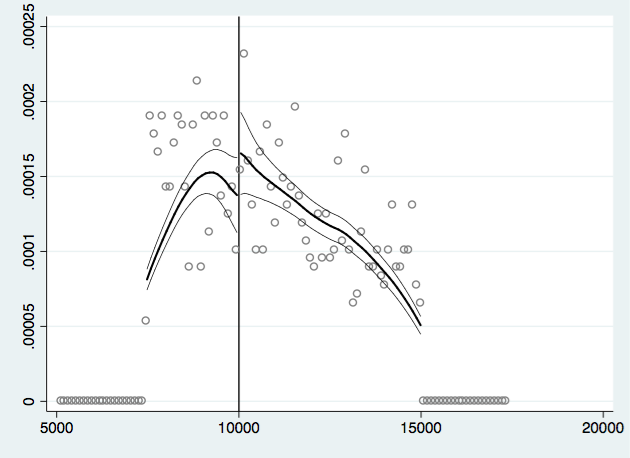}}
\caption{10,000  threshold}
\end{subfigure}\\
\vspace{-0.1cm}
\begin{minipage}{14.5cm} \scriptsize  All figures  pool all years, and show finely-gridded histograms of  the population smoothed using local linear regression, separately on either side of the cutoff  of the density function of the population. Each figure uses data only around the corresponding population threshold. The estimates of the difference in the height at the threshold and  robust manipulation p-values (as implemented in Stata by the command rddensity.ado, \citep[see][]{CattaneoJanssonMa2018a}, are: (a)  0.326 (s.e. 0.190) and p-value 0.0685; (b)  0.350 (s.e. 0.176) and p-value 0.0613. 
\end{minipage}
\end{center}
\end{figure}


\newgeometry{left=2cm,bottom=2cm,top=1.5cm}
{ 
\renewcommand{\arraystretch}{0.5} 
\setlength{\tabcolsep}{2pt}
\begin{table}[H]
\footnotesize
\vspace{-0.4cm}
\caption{Pre-treatment characteristics (20,000 and 50,000 thresholds)}\label{precharacteristics10dpopth3to4}
\vspace{-0.4cm}
\begin{tabular}{lcccccccc}
\hline\hline\addlinespace[0.15cm]
   &\multicolumn{3}{c}{Entire sample} &\multicolumn{3}{c}{10\% population spread}  \\
&&&&& & &RD& SE on \\
   & Obs.  & Mean & St. dev.& Obs.  & Mean& St. dev. &estimate &estimate \\\cmidrule[0.2pt](l){2-4}\cmidrule[0.2pt](l){5-7}\cmidrule[0.2pt](l){8-9}
            &         (1)&         (2)&         (3)&         (4)&         (5)&         (6)&         (7)&         (8)\\
\midrule
\emph{\textbf{Conflict-related violence (pre-election term)}}&            &            &            &            &            &            &            &            \\
Killings (overall)&        1635&       2.288&       4.868&         405&       2.024&       4.010&      -1.119&       0.658\\
Very selective killings&        1635&       1.437&       2.713&         405&       1.315&       2.620&      -0.984&       0.411\\
Massacres   &        1635&       0.851&       2.721&         405&       0.709&       1.977&      -0.152&       0.354\\
Occurrence of a killing (overall)&        1635&       0.069&       0.098&         405&       0.062&       0.096&      -0.032&       0.017\\
Occurrence of a very selective killing&        1635&       0.061&       0.089&         405&       0.056&       0.088&      -0.032&       0.016\\
Occurrence of a massacre&        1635&       0.011&       0.029&         405&       0.010&       0.025&      -0.002&       0.005\\
\emph{\textbf{Armed conflict (pre-election term)}}&            &            &            &            &            &            &            &            \\
Action by guerrilla&        1635&       0.466&       0.403&         405&       0.454&       0.399&      -0.126&       0.085\\
Action by paramilitaries&        1635&       0.183&       0.271&         405&       0.177&       0.257&      -0.060&       0.052\\
Action by Nal. army&        1635&       0.414&       0.379&         405&       0.414&       0.379&      -0.163&       0.077\\
Encounter paramilitaries vs guerrilla&        1635&       0.121&       0.236&         405&       0.103&       0.219&      -0.036&       0.049\\
Encounter Nal. army vs guerrilla&        1635&       0.348&       0.375&         405&       0.338&       0.369&      -0.159&       0.073\\
Encounter Nal. army vs paramilitaries&        1635&       0.091&       0.196&         405&       0.086&       0.187&      -0.039&       0.034\\
Event of massive expulsion&        1635&       0.097&       0.212&         405&       0.102&       0.225&      -0.044&       0.050\\
\emph{\textbf{Crime (pre-election term)}}&            &            &            &            &            &            &            &            \\
Overall homicide rate&        1598&      65.045&      62.271&         391&      66.895&      64.949&      -4.990&      10.596\\
Log of hectares of coca&        1269&       1.172&       2.328&         313&       1.360&       2.506&      -0.612&       0.613\\
Log of hectares of aerial spraying on coca&        1269&       0.804&       2.157&         313&       0.960&       2.323&      -0.823&       0.627\\
\emph{\textbf{Elections (previous election)}}&            &            &            &            &            &            &            &            \\
Turnout rate&         830&       0.566&       0.115&         207&       0.571&       0.114&      -0.044&       0.024\\
Number of parties in council&        1244&       4.796&       1.992&         304&       4.740&       1.977&       0.193&       0.435\\
Council fractionalization&        1629&       0.967&       0.116&         403&       0.980&       0.088&      -0.026&       0.023\\
Liberal party in council&        1244&       0.901&       0.299&         304&       0.862&       0.346&       0.117&       0.078\\
Conservative party in council&        1244&       0.699&       0.459&         304&       0.678&       0.468&       0.054&       0.122\\
Left-wing party in council&        1244&       0.225&       0.418&         304&       0.217&       0.413&      -0.129&       0.087\\
Party with paramilitary links in council&        1244&       0.397&       0.489&         304&       0.444&       0.498&       0.027&       0.121\\
Mayor from Liberal party&        1629&       0.397&       0.489&         403&       0.390&       0.488&      -0.058&       0.112\\
Mayor from Conservative party&        1629&       0.191&       0.393&         403&       0.218&       0.414&       0.124&       0.086\\
Mayor from left-wing party&        1629&       0.022&       0.147&         403&       0.010&       0.099&       0.012&       0.029\\
Mayor from party with paramilitary links&        1629&       0.056&       0.230&         403&       0.067&       0.250&       0.075&       0.058\\
\emph{\textbf{Economy and institutions}}&            &            &            &            &            &            &            &            \\
Municipal category (first year of term)&         835&       5.819&       0.625&         208&       5.889&       0.503&       0.135&       0.141\\
\% unsatisfied basic needs (1993 or 2005)&        1635&      52.506&      21.142&         405&      52.540&      21.515&      -4.719&       4.109\\
Schools per 1000 inhab. (1997)&        1585&      34.058&      19.287&         387&      35.620&      20.070&      -4.586&       5.893\\
Hospitals per 1000 inhab. (1997)&        1585&       3.027&       2.706&         387&       3.145&       2.922&      -0.189&       0.508\\
Bank branches per 1000 inhab. in 1997&        1379&       7.510&       3.964&         331&       7.654&       3.720&       0.176&       0.738\\
Courts per 1000 inhab. (1997)&        1585&      10.095&       8.223&         388&      10.065&       8.143&       1.581&       1.416\\
Police stations per 1000 inhab. (1997)&        1581&       4.560&       2.072&         385&       4.744&       2.083&      -0.491&       0.300\\
\emph{\textbf{Fiscal outcomes (pre-election term)}}&            &            &            &            &            &            &            &            \\
Log current spending per capita&        1613&      -2.508&       0.535&         401&      -2.515&       0.588&       0.085&       0.118\\
Log fixed capital spending per capita&        1612&      -2.255&       0.568&         401&      -2.220&       0.594&      -0.027&       0.106\\
Log other capital spending per capita&        1607&      -2.206&       0.898&         401&      -2.188&       0.878&      -0.172&       0.185\\
Log tax revenue per capita&        1612&      -3.572&       1.134&         400&      -3.598&       1.113&       0.098&       0.189\\
Log royalties per capita&        1064&      -5.086&       2.614&         257&      -5.153&       2.797&      -0.078&       0.739\\
Log transfers per capita&        1591&      -1.803&       0.499&         398&      -1.785&       0.478&       0.053&       0.094\\
Total deficit per capita&        1615&      -0.015&       0.278&         401&      -0.035&       0.502&      -0.105&       0.117\\
\emph{\textbf{Geographic characteristics}}&            &            &            &            &            &            &            &            \\
Surface area (km2)&        1635&    1212.579&    3717.283&         405&    1451.821&    4693.372&    -818.574&     829.630\\
Mean altitude (m)&        1596&    1028.048&    1487.873&         390&     998.369&     918.108&      97.744&     162.835\\
Distance to Bogota (km)&        1635&     350.255&     189.249&         405&     353.554&     189.256&      -2.283&      23.116\\
Distance to the capital of department (km)&        1635&      81.685&      58.385&         405&      79.743&      57.638&      -1.267&      11.366\\
\% Municipalities on the Atlantic coast&        1635&       0.239&       0.426&         405&       0.235&       0.424&       0.015&       0.103\\
\% Municipalities in the eastern region&        1635&       0.179&       0.383&         405&       0.200&       0.400&       0.066&       0.095\\
\% Municipalities in the central region&        1635&       0.175&       0.380&         405&       0.173&       0.379&       0.006&       0.079\\
\% Municipalities on the Pacific coast&        1635&       0.202&       0.402&         405&       0.207&       0.406&      -0.064&       0.097\\
\% Municipalities in Antioquia&        1635&       0.150&       0.358&         405&       0.126&       0.332&      -0.043&       0.078\\
\% Municipalities in the Amazon region&        1635&       0.055&       0.228&         405&       0.059&       0.236&       0.008&       0.048\\

\addlinespace[0.15cm]
\hline\hline\addlinespace[0.1cm]
\multicolumn{9}{p{18cm}}{\scriptsize \textbf{Notes:} The sample in all columns include municipalities with a population up to 75,000 (i.e. it includes municipalities for which the closest population thresholds are 20,000 and 50,000). Pre-election term means that the variable is averaged for the previous election term. The sources of the variables are listed in the note to Table \ref{precharacteristics10}. Column (7) reports the RD estimate from equation (\ref{rdbaseline}) when the respective characteristic is used as the dependent variable, with a data-driven bandwidth chosen optimally using the algorithm by  \cite{CalonicoCattaneoTitiunik2014a} as implemented in Stata by the command rdrobust.ado, and includes  fixed effects for population threshold, electoral term and region.  Column (7) reports the standard errors clustered by municipality.} \\
\end{tabular}
\end{table}
}


\renewcommand{\arraystretch}{0.5} 
\setlength{\tabcolsep}{2pt}
\begin{table}[H]
\footnotesize
\caption{Pre-treatment characteristics (100,000, 250,000 and 1,000,000 thresholds)}\label{precharacteristics10dpopth5to7}
\vspace{-0.4cm}
\begin{tabular}{lcccccccc}
\hline\hline\addlinespace[0.15cm]
   &\multicolumn{3}{c}{Entire sample} &\multicolumn{3}{c}{10\% population spread}  \\
&&&&& & &RD& SE on \\
   & Obs.  & Mean & St. dev.& Obs.  & Mean& St. dev. &estimate &estimate \\\cmidrule[0.2pt](l){2-4}\cmidrule[0.2pt](l){5-7}\cmidrule[0.2pt](l){8-9}
            &         (1)&         (2)&         (3)&         (4)&         (5)&         (6)&         (7)&         (8)\\
\midrule
\emph{\textbf{Conflict-related violence (pre-election term)}}&            &            &            &            &            &            &            &            \\
Killings (overall)&         265&       7.382&      14.055&          63&       6.700&      14.036&       2.707&       3.315\\
Very selective killings&         265&       5.080&       9.006&          63&       4.149&       8.457&       2.669&       2.433\\
Massacres   &         265&       2.303&       5.862&          63&       2.550&       6.155&      -0.536&       1.412\\
Occurrence of a killing (overall)&         265&       0.180&       0.207&          63&       0.151&       0.189&       0.104&       0.077\\
Occurrence of a very selective killing&         265&       0.167&       0.196&          63&       0.136&       0.177&       0.100&       0.074\\
Occurrence of a massacre&         265&       0.031&       0.068&          63&       0.033&       0.073&       0.003&       0.015\\
\emph{\textbf{Armed conflict (pre-election term)}}&            &            &            &            &            &            &            &            \\
Action by guerrilla&         265&       0.567&       0.398&          63&       0.569&       0.411&       0.152&       0.259\\
Action by paramilitaries&         265&       0.352&       0.365&          63&       0.358&       0.346&       0.091&       0.179\\
Action by Nal. army&         265&       0.571&       0.400&          63&       0.616&       0.396&       0.059&       0.211\\
Encounter paramilitaries vs guerrilla&         265&       0.245&       0.338&          63&       0.243&       0.321&       0.155&       0.144\\
Encounter Nal. army vs guerrilla&         265&       0.452&       0.407&          63&       0.475&       0.424&       0.215&       0.256\\
Encounter Nal. army vs paramilitaries&         265&       0.275&       0.349&          63&       0.262&       0.322&       0.244&       0.171\\
Event of massive expulsion&         265&       0.119&       0.235&          63&       0.140&       0.269&      -0.103&       0.100\\
\emph{\textbf{Crime (pre-election term)}}&            &            &            &            &            &            &            &            \\
Overall homicide rate&         263&      60.375&      48.212&          63&      57.075&      60.700&       1.067&      15.722\\
Log of hectares of coca&         207&       0.678&       1.776&          55&       0.757&       1.839&       0.568&       0.456\\
Log of hectares of aerial spraying on coca&         207&       0.254&       1.040&          55&       0.191&       0.913&       0.126&       0.219\\
\emph{\textbf{Elections (previous election)}}&            &            &            &            &            &            &            &            \\
Turnout rate&         138&       0.537&       0.077&          39&       0.553&       0.085&       0.039&       0.036\\
Number of parties in council&         207&       6.749&       2.442&          55&       7.000&       1.944&       0.087&       0.576\\
Council fractionalization&         265&       0.987&       0.069&          63&       0.986&       0.083&      -0.009&       0.010\\
Liberal party in council&         207&       0.990&       0.098&          55&       0.982&       0.135&       0.020&       0.021\\
Conservative party in council&         207&       0.778&       0.417&          55&       0.655&       0.480&       0.455&       0.229\\
Left-wing party in council&         207&       0.391&       0.489&          55&       0.455&       0.503&       0.276&       0.179\\
Party with paramilitary links in council&         207&       0.551&       0.499&          55&       0.545&       0.503&       0.234&       0.277\\
Mayor from Liberal party&         265&       0.400&       0.491&          63&       0.397&       0.493&      -0.091&       0.171\\
Mayor from Conservative party&         265&       0.087&       0.282&          63&       0.048&       0.215&       0.019&       0.045\\
Mayor from left-wing party&         265&       0.064&       0.245&          63&       0.095&       0.296&       0.166&       0.198\\
Mayor from party with paramilitary links&         265&       0.072&       0.258&          63&       0.095&       0.296&      -0.003&       0.195\\
\emph{\textbf{Economy and institutions}}&            &            &            &            &            &            &            &            \\
Municipal category (first year of term)&         137&       3.073&       1.865&          39&       3.308&       1.575&       0.706&       0.667\\
\% unsatisfied basic needs (1993 or 2005)&         265&      32.266&      18.021&          63&      31.057&      17.495&      -0.562&       6.670\\
Schools per 1000 inhab. (1997)&         205&      39.799&      20.204&          50&      48.913&      20.381&      -1.489&       8.783\\
Hospitals per 1000 inhab. (1997)&         205&       1.710&       0.886&          50&       1.588&       0.962&      -0.298&       0.418\\
Bank branches per 1000 inhab. in 1997&         257&       6.999&       3.956&          60&       6.416&       3.550&       1.062&       1.963\\
Courts per 1000 inhab. (1997)&         205&       9.886&       5.026&          50&      10.128&       5.337&       1.985&       2.904\\
Police stations per 1000 inhab. (1997)&         205&       2.673&       1.949&          50&       2.701&       2.129&       0.198&       1.052\\
\emph{\textbf{Fiscal outcomes (pre-election term)}}&            &            &            &            &            &            &            &            \\
Log current spending per capita&         263&      -2.394&       0.555&          62&      -2.423&       0.486&      -0.163&       0.132\\
Log fixed capital spending per capita&         264&      -2.602&       0.594&          62&      -2.657&       0.641&       0.318&       0.258\\
Log other capital spending per capita&         264&      -2.270&       1.024&          62&      -2.188&       0.863&       0.567&       0.348\\
Log tax revenue per capita&         263&      -2.574&       0.837&          62&      -2.485&       0.741&      -0.074&       0.251\\
Log royalties per capita&         203&      -5.847&       2.388&          45&      -5.434&       1.962&       1.600&       0.879\\
Log transfers per capita&         262&      -2.061&       0.649&          62&      -2.145&       0.693&       0.296&       0.268\\
Total deficit per capita&         264&      -0.008&       0.051&          62&      -0.008&       0.040&      -0.000&       0.020\\
\emph{\textbf{Geographic characteristics}}&            &            &            &            &            &            &            &            \\
Surface area (km2)&         265&    1050.407&    1314.433&          63&     995.829&    1476.018&     -85.269&     397.161\\
Mean altitude (m)&         257&     846.459&     860.950&          60&     968.900&     945.462&     177.597&     321.918\\
Distance to Bogota (km)&         265&     362.915&     214.318&          63&     367.744&     225.071&      62.552&      59.693\\
Distance to the capital of department (km)&         265&      48.935&      63.505&          63&      48.311&      57.049&      29.449&      18.872\\
\% Municipalities on the Atlantic coast&         265&       0.272&       0.446&          63&       0.222&       0.419&      -0.135&       0.235\\
\% Municipalities in the eastern region&         265&       0.245&       0.431&          63&       0.381&       0.490&       0.181&       0.257\\
\% Municipalities in the central region&         265&       0.162&       0.369&          63&       0.032&       0.177&       0.039&       0.061\\
\% Municipalities on the Pacific coast&         265&       0.192&       0.395&          63&       0.206&       0.408&      -0.110&       0.149\\
\% Municipalities in Antioquia&         265&       0.109&       0.313&          63&       0.143&       0.353&       0.013&       0.142\\
\% Municipalities in the Amazon region&         265&       0.019&       0.136&          63&       0.016&       0.126&       0.031&       0.042\\

\addlinespace[0.15cm]
\hline\hline\addlinespace[0.1cm]
\multicolumn{9}{p{18cm}}{\scriptsize \textbf{Notes:}  The sample in all columns include municipalities with a population of at least 75,000 (i.e. it includes municipalities for which the closest population thresholds are 100,000, 250,000 and 1,000,000). Pre-election term means that the variable is averaged for the previous election term. The sources of the variables are listed in the note to Table \ref{precharacteristics10}. Column (7) reports the RD estimate from equation (\ref{rdbaseline}) when the respective characteristic is used as the dependent variable, with a data-driven bandwidth chosen optimally using the algorithm by  \cite{CalonicoCattaneoTitiunik2014a} as implemented in Stata by the command rdrobust.ado, and includes  fixed effects for population threshold, electoral term and region.  Column (7) reports the standard errors clustered by municipality.} \\
\end{tabular}
\end{table}


\renewcommand{\arraystretch}{0.5} 
\setlength{\tabcolsep}{2pt}
\begin{table}[H]
\footnotesize
\caption{Pre-treatment characteristics (20,000 threshold)}\label{precharacteristics10popth3}
\vspace{-0.4cm}
\begin{tabular}{lcccccccc}
\hline\hline\addlinespace[0.15cm]
   &\multicolumn{3}{c}{Entire sample} &\multicolumn{3}{c}{10\% population spread}  \\
&&&&& & &RD& SE on \\
   & Obs.  & Mean & St. dev.& Obs.  & Mean& St. dev. &estimate &estimate \\\cmidrule[0.2pt](l){2-4}\cmidrule[0.2pt](l){5-7}\cmidrule[0.2pt](l){8-9}
            &         (1)&         (2)&         (3)&         (4)&         (5)&         (6)&         (7)&         (8)\\
\midrule
\emph{\textbf{Conflict-related violence (pre-election term)}}&            &            &            &            &            &            &            &            \\
Killings (overall)&        1217&       1.912&       4.036&         291&       1.652&       3.302&      -1.890&       0.664\\
Very selective killings&        1217&       1.185&       2.233&         291&       0.994&       1.932&      -1.650&       0.453\\
Massacres   &        1217&       0.727&       2.354&         291&       0.658&       1.991&      -0.266&       0.430\\
Occurrence of a killing (overall)&        1217&       0.059&       0.087&         291&       0.051&       0.082&      -0.063&       0.022\\
Occurrence of a very selective killing&        1217&       0.052&       0.078&         291&       0.045&       0.074&      -0.063&       0.020\\
Occurrence of a massacre&        1217&       0.009&       0.026&         291&       0.009&       0.023&      -0.002&       0.005\\
\emph{\textbf{Armed conflict (pre-election term)}}&            &            &            &            &            &            &            &            \\
Action by guerrilla&        1217&       0.449&       0.399&         291&       0.432&       0.393&      -0.100&       0.100\\
Action by paramilitaries&        1217&       0.164&       0.253&         291&       0.153&       0.227&      -0.052&       0.060\\
Action by Nal. army&        1217&       0.393&       0.373&         291&       0.389&       0.373&      -0.172&       0.083\\
Encounter paramilitaries vs guerrilla&        1217&       0.105&       0.219&         291&       0.082&       0.182&      -0.032&       0.062\\
Encounter Nal. army vs guerrilla&        1217&       0.329&       0.366&         291&       0.317&       0.359&      -0.209&       0.099\\
Encounter Nal. army vs paramilitaries&        1217&       0.078&       0.179&         291&       0.071&       0.164&      -0.060&       0.044\\
Event of massive expulsion&        1217&       0.087&       0.195&         291&       0.092&       0.202&       0.048&       0.050\\
\emph{\textbf{Crime (pre-election term)}}&            &            &            &            &            &            &            &            \\
Overall homicide rate&        1187&      64.302&      62.060&         282&      66.732&      64.749&     -18.812&      14.018\\
Log of hectares of coca&         943&       1.168&       2.295&         227&       1.417&       2.523&       0.124&       0.715\\
Log of hectares of aerial spraying on coca&         943&       0.806&       2.130&         227&       0.986&       2.346&      -0.259&       0.717\\
\emph{\textbf{Elections (previous election)}}&            &            &            &            &            &            &            &            \\
Turnout rate&         618&       0.571&       0.119&         150&       0.572&       0.120&      -0.038&       0.032\\
Number of parties in council&         919&       4.603&       1.929&         219&       4.438&       1.860&       0.187&       0.472\\
Council fractionalization&        1212&       0.964&       0.120&         290&       0.973&       0.102&      -0.027&       0.032\\
Liberal party in council&         919&       0.879&       0.326&         219&       0.831&       0.376&       0.103&       0.145\\
Conservative party in council&         919&       0.700&       0.459&         219&       0.658&       0.476&      -0.037&       0.203\\
Left-wing party in council&         919&       0.209&       0.407&         219&       0.205&       0.405&      -0.202&       0.120\\
Party with paramilitary links in council&         919&       0.370&       0.483&         219&       0.397&       0.490&       0.123&       0.177\\
Mayor from Liberal party&        1212&       0.381&       0.486&         290&       0.369&       0.483&      -0.331&       0.150\\
Mayor from Conservative party&        1212&       0.193&       0.395&         290&       0.214&       0.411&       0.161&       0.130\\
Mayor from left-wing party&        1212&       0.023&       0.150&         290&       0.007&       0.083&       0.036&       0.024\\
Mayor from party with paramilitary links&        1212&       0.059&       0.236&         290&       0.066&       0.248&       0.205&       0.088\\
\emph{\textbf{Economy and institutions}}&            &            &            &            &            &            &            &            \\
Municipal category (first year of term)&         623&       5.934&       0.346&         151&       5.960&       0.303&       0.031&       0.173\\
\% unsatisfied basic needs (1993 or 2005)&        1217&      54.284&      20.616&         291&      54.958&      21.010&      -1.382&       5.738\\
Schools per 1000 inhab. (1997)&        1172&      32.994&      18.975&         275&      33.644&      18.870&      -8.078&       7.810\\
Hospitals per 1000 inhab. (1997)&        1172&       3.298&       3.007&         275&       3.493&       3.334&      -1.493&       0.873\\
Bank branches per 1000 inhab. in 1997&         990&       8.106&       4.046&         223&       8.429&       3.475&      -0.973&       1.128\\
Courts per 1000 inhab. (1997)&        1172&      10.076&       8.129&         276&      10.171&       8.054&       0.455&       1.659\\
Police stations per 1000 inhab. (1997)&        1168&       5.103&       1.859&         273&       5.447&       1.590&      -0.197&       0.383\\
\emph{\textbf{Fiscal outcomes (pre-election term)}}&            &            &            &            &            &            &            &            \\
Log current spending per capita&        1198&      -2.504&       0.524&         287&      -2.502&       0.564&       0.055&       0.143\\
Log fixed capital spending per capita&        1200&      -2.188&       0.566&         287&      -2.121&       0.580&      -0.048&       0.157\\
Log other capital spending per capita&        1196&      -2.158&       0.907&         287&      -2.096&       0.812&      -0.004&       0.206\\
Log tax revenue per capita&        1197&      -3.669&       1.154&         286&      -3.693&       1.123&       0.061&       0.273\\
Log royalties per capita&         757&      -4.922&       2.575&         174&      -4.901&       2.621&      -1.349&       0.966\\
Log transfers per capita&        1178&      -1.728&       0.474&         285&      -1.690&       0.453&       0.098&       0.119\\
Total deficit per capita&        1200&      -0.010&       0.154&         287&      -0.013&       0.149&       0.028&       0.025\\
\emph{\textbf{Geographic characteristics}}&            &            &            &            &            &            &            &            \\
Surface area (km2)&        1217&    1070.192&    3057.046&         291&    1257.260&    3547.250&   -1021.584&    1198.158\\
Mean altitude (m)&        1189&    1006.140&    1120.339&         281&    1018.267&     929.035&      89.235&     232.707\\
Distance to Bogota (km)&        1217&     346.772&     189.067&         291&     351.468&     191.334&      -2.885&      34.905\\
Distance to the capital of department (km)&        1217&      84.123&      60.034&         291&      83.306&      59.920&      -4.183&      15.684\\
\% Municipalities on the Atlantic coast&        1217&       0.233&       0.423&         291&       0.216&       0.413&       0.016&       0.107\\
\% Municipalities in the eastern region&        1217&       0.191&       0.393&         291&       0.213&       0.410&       0.078&       0.143\\
\% Municipalities in the central region&        1217&       0.182&       0.386&         291&       0.179&       0.384&       0.056&       0.114\\
\% Municipalities on the Pacific coast&        1217&       0.210&       0.408&         291&       0.223&       0.417&      -0.191&       0.143\\
\% Municipalities in Antioquia&        1217&       0.137&       0.344&         291&       0.113&       0.318&       0.037&       0.112\\
\% Municipalities in the Amazon region&        1217&       0.047&       0.211&         291&       0.055&       0.228&      -0.068&       0.072\\

\addlinespace[0.15cm]
\hline\hline\addlinespace[0.1cm]
\multicolumn{9}{p{18cm}}{\scriptsize \textbf{Notes:}  The sample in all columns include municipalities with a population between 15,000 and 35,000 (i.e. it includes municipalities for which the closest population threshold is 20,000). Pre-election term means that the variable is averaged for the previous election term. The sources of the variables are listed in the note to Table \ref{precharacteristics10}. Column (7) reports the RD estimate from equation (\ref{rdbaseline}) when the respective characteristic is used as the dependent variable, with a data-driven bandwidth chosen optimally using the algorithm by  \cite{CalonicoCattaneoTitiunik2014a} as implemented in Stata by the command rdrobust.ado, and includes  fixed effects for population threshold, electoral term and region.  Column (7) reports the standard errors clustered by municipality.} \\
\end{tabular}
\end{table}


\renewcommand{\arraystretch}{0.5} 
\setlength{\tabcolsep}{2pt}
\begin{table}[H]
\footnotesize
\caption{Pre-treatment characteristics (50,000 threshold)}\label{precharacteristics10popth4}
\vspace{-0.4cm}
\begin{tabular}{lcccccccc}
\hline\hline\addlinespace[0.15cm]
   &\multicolumn{3}{c}{Entire sample} &\multicolumn{3}{c}{10\% population spread}  \\
&&&&& & &RD& SE on \\
   & Obs.  & Mean & St. dev.& Obs.  & Mean& St. dev. &estimate &estimate \\\cmidrule[0.2pt](l){2-4}\cmidrule[0.2pt](l){5-7}\cmidrule[0.2pt](l){8-9}
            &         (1)&         (2)&         (3)&         (4)&         (5)&         (6)&         (7)&         (8)\\
\midrule
\emph{\textbf{Conflict-related violence (pre-election term)}}&            &            &            &            &            &            &            &            \\
Killings (overall)&         418&       3.381&       6.614&         114&       2.973&       5.314&      -1.183&       1.159\\
Very selective killings&         418&       2.171&       3.684&         114&       2.135&       3.744&      -0.714&       0.798\\
Massacres   &         418&       1.210&       3.562&         114&       0.838&       1.943&      -0.426&       0.460\\
Occurrence of a killing (overall)&         418&       0.096&       0.121&         114&       0.090&       0.120&      -0.001&       0.032\\
Occurrence of a very selective killing&         418&       0.087&       0.112&         114&       0.083&       0.112&      -0.002&       0.030\\
Occurrence of a massacre&         418&       0.015&       0.036&         114&       0.012&       0.029&      -0.005&       0.008\\
\emph{\textbf{Armed conflict (pre-election term)}}&            &            &            &            &            &            &            &            \\
Action by guerrilla&         418&       0.516&       0.409&         114&       0.510&       0.411&      -0.237&       0.131\\
Action by paramilitaries&         418&       0.239&       0.310&         114&       0.238&       0.315&      -0.143&       0.112\\
Action by Nal. army&         418&       0.472&       0.390&         114&       0.480&       0.390&      -0.172&       0.111\\
Encounter paramilitaries vs guerrilla&         418&       0.166&       0.275&         114&       0.156&       0.288&      -0.137&       0.109\\
Encounter Nal. army vs guerrilla&         418&       0.402&       0.393&         114&       0.390&       0.389&      -0.157&       0.114\\
Encounter Nal. army vs paramilitaries&         418&       0.126&       0.236&         114&       0.126&       0.230&      -0.054&       0.066\\
Event of massive expulsion&         418&       0.126&       0.253&         114&       0.126&       0.273&      -0.203&       0.102\\
\emph{\textbf{Crime (pre-election term)}}&            &            &            &            &            &            &            &            \\
Overall homicide rate&         411&      67.191&      62.902&         109&      67.319&      65.761&      -6.892&      11.532\\
Log of hectares of coca&         326&       1.183&       2.426&          86&       1.210&       2.469&      -1.435&       0.860\\
Log of hectares of aerial spraying on coca&         326&       0.798&       2.236&          86&       0.891&       2.274&      -1.048&       0.839\\
\emph{\textbf{Elections (previous election)}}&            &            &            &            &            &            &            &            \\
Turnout rate&         212&       0.552&       0.101&          57&       0.568&       0.098&      -0.040&       0.044\\
Number of parties in council&         325&       5.342&       2.066&          85&       5.518&       2.068&      -0.102&       0.893\\
Council fractionalization&         417&       0.973&       0.104&         113&       0.998&       0.026&      -0.016&       0.015\\
Liberal party in council&         325&       0.963&       0.189&          85&       0.941&       0.237&      -0.005&       0.074\\
Conservative party in council&         325&       0.698&       0.460&          85&       0.729&       0.447&       0.087&       0.155\\
Left-wing party in council&         325&       0.271&       0.445&          85&       0.247&       0.434&      -0.022&       0.151\\
Party with paramilitary links in council&         325&       0.474&       0.500&          85&       0.565&       0.499&      -0.188&       0.154\\
Mayor from Liberal party&         417&       0.444&       0.497&         113&       0.442&       0.499&       0.282&       0.177\\
Mayor from Conservative party&         417&       0.185&       0.388&         113&       0.230&       0.423&       0.015&       0.147\\
Mayor from left-wing party&         417&       0.019&       0.137&         113&       0.018&       0.132&       0.068&       0.060\\
Mayor from party with paramilitary links&         417&       0.046&       0.209&         113&       0.071&       0.258&      -0.047&       0.102\\
\emph{\textbf{Economy and institutions}}&            &            &            &            &            &            &            &            \\
Municipal category (first year of term)&         212&       5.481&       1.019&          57&       5.702&       0.801&       0.473&       0.230\\
\% unsatisfied basic needs (1993 or 2005)&         418&      47.327&      21.815&         114&      46.366&      21.648&     -13.892&       6.296\\
Schools per 1000 inhab. (1997)&         413&      37.075&      19.860&         112&      40.472&      22.099&       3.072&       7.510\\
Hospitals per 1000 inhab. (1997)&         413&       2.256&       1.284&         112&       2.291&       1.103&       0.887&       0.341\\
Bank branches per 1000 inhab. in 1997&         389&       5.991&       3.298&         108&       6.055&       3.714&       1.857&       0.950\\
Courts per 1000 inhab. (1997)&         413&      10.151&       8.493&         112&       9.804&       8.391&       2.080&       2.599\\
Police stations per 1000 inhab. (1997)&         413&       3.025&       1.863&         112&       3.028&       2.153&      -0.581&       0.584\\
\emph{\textbf{Fiscal outcomes (pre-election term)}}&            &            &            &            &            &            &            &            \\
Log current spending per capita&         415&      -2.521&       0.568&         114&      -2.546&       0.647&       0.259&       0.212\\
Log fixed capital spending per capita&         412&      -2.452&       0.527&         114&      -2.467&       0.559&      -0.153&       0.213\\
Log other capital spending per capita&         411&      -2.345&       0.855&         114&      -2.420&       0.990&      -0.291&       0.255\\
Log tax revenue per capita&         415&      -3.291&       1.026&         114&      -3.358&       1.054&       0.400&       0.244\\
Log royalties per capita&         307&      -5.490&       2.672&          83&      -5.683&       3.083&       1.648&       1.166\\
Log transfers per capita&         413&      -2.017&       0.505&         113&      -2.026&       0.456&      -0.178&       0.128\\
Total deficit per capita&         415&      -0.029&       0.483&         114&      -0.091&       0.911&      -0.343&       0.315\\
\emph{\textbf{Geographic characteristics}}&            &            &            &            &            &            &            &            \\
Surface area (km2)&         418&    1627.135&    5163.573&         114&    1948.462&    6790.810&    -662.536&    1458.759\\
Mean altitude (m)&         407&    1092.049&    2240.245&         109&     947.073&     891.467&      48.333&     235.540\\
Distance to Bogota (km)&         418&     360.394&     189.638&         114&     358.879&     184.567&     -13.960&      41.070\\
Distance to the capital of department (km)&         418&      74.589&      52.722&         114&      70.648&      50.468&      -0.456&      17.076\\
\% Municipalities on the Atlantic coast&         418&       0.254&       0.436&         114&       0.281&       0.451&      -0.005&       0.200\\
\% Municipalities in the eastern region&         418&       0.144&       0.351&         114&       0.167&       0.374&      -0.001&       0.135\\
\% Municipalities in the central region&         418&       0.156&       0.363&         114&       0.158&       0.366&      -0.011&       0.147\\
\% Municipalities on the Pacific coast&         418&       0.179&       0.384&         114&       0.167&       0.374&       0.011&       0.157\\
\% Municipalities in Antioquia&         418&       0.189&       0.392&         114&       0.158&       0.366&      -0.149&       0.084\\
\% Municipalities in the Amazon region&         418&       0.079&       0.270&         114&       0.070&       0.257&       0.129&       0.083\\

\addlinespace[0.15cm]
\hline\hline\addlinespace[0.1cm]
\multicolumn{9}{p{18cm}}{\scriptsize \textbf{Notes:} The sample in all columns include municipalities with a population between 35,000 and 75,000 (i.e. it includes municipalities for which the closest population threshold is 50,000). Pre-election term means that the variable is averaged for the previous election term. The sources of the variables are listed in the note to Table \ref{precharacteristics10}. Column (7) reports the RD estimate from equation (\ref{rdbaseline}) when the respective characteristic is used as the dependent variable, with a data-driven bandwidth chosen optimally using the algorithm by  \cite{CalonicoCattaneoTitiunik2014a} as implemented in Stata by the command rdrobust.ado, and includes  fixed effects for population threshold, electoral term and region.  Column (7) reports the standard errors clustered by municipality.} \\
\end{tabular}
\end{table}


\newgeometry{left=3cm,bottom=2cm,top=3cm}
{ 
\renewcommand{\arraystretch}{0.8} 
\setlength{\tabcolsep}{10pt}
\begin{table}[H]
\small
\begin{center}
\caption{Relationship between council size and municipal category}\label{tableRDcategoria}
\begin{tabular}{lccc}\hline\hline\addlinespace[0.15cm]
&\multicolumn{3}{c}{\emph{Dependent variable:} Municipal category for}\\
 & \multicolumn{1}{c}{} & \multicolumn{1}{c}{last}& \multicolumn{1}{c}{next-to-last}  \\
  & \multicolumn{1}{c}{first year} & \multicolumn{1}{c}{year of}& \multicolumn{1}{c}{year of}  \\
  & \multicolumn{1}{c}{of term} & \multicolumn{1}{c}{previous term}& \multicolumn{1}{c}{previous term}  \\
 \cmidrule[0.2pt](l){2-2}\cmidrule[0.2pt](l){3-3}\cmidrule[0.2pt](l){4-4}
& (1)& (2)& (3) \\ \hline \addlinespace[0.15cm]
\hspace{3mm}RD Estimate &       0.158         &       0.133         &      -0.105         \\
            &     (0.186)         &     (0.173)         &     (0.161)         \\
Observations&        1424         &        1427         &        1427         \\
Eff. Number of obs&         444         &         377         &         429         \\
BW Loc. Poly. (h)&       0.118         &       0.100         &       0.115         \\
Robust p-value&       0.418         &       0.431         &       0.603         \\

 \addlinespace[0.15cm]
\hline     \hline                                                        
\multicolumn{4}{p{12cm}}{\scriptsize \textbf{Notes}:  All columns report the RD estimates of having a larger council from Eq. (\ref{rdbaseline}) when the respective characteristic is used as the dependent variable. The bandwidth (h) is chosen optimally using the algorithm by \cite{CalonicoCattaneoTitiunik2014a} as  implemented in Stata by the command rdrobust.ado, and includes  fixed effects for population threshold, electoral term and region. Standard errors clustered by municipality are reported in parentheses. * significant at 10\%, ** significant at 5\%, *** significant at 1\%. }
\end{tabular}
\end{center}
\end{table}
}


\clearpage
\begin{figure}[H]
\begin{center}
\caption{Municipal category and population size}\label{catdistribution}
\vspace{-0.3cm}
\begin{subfigure}[a]{0.45\textwidth}	
\resizebox{7.2cm}{4cm}{\includegraphics[width=1cm]{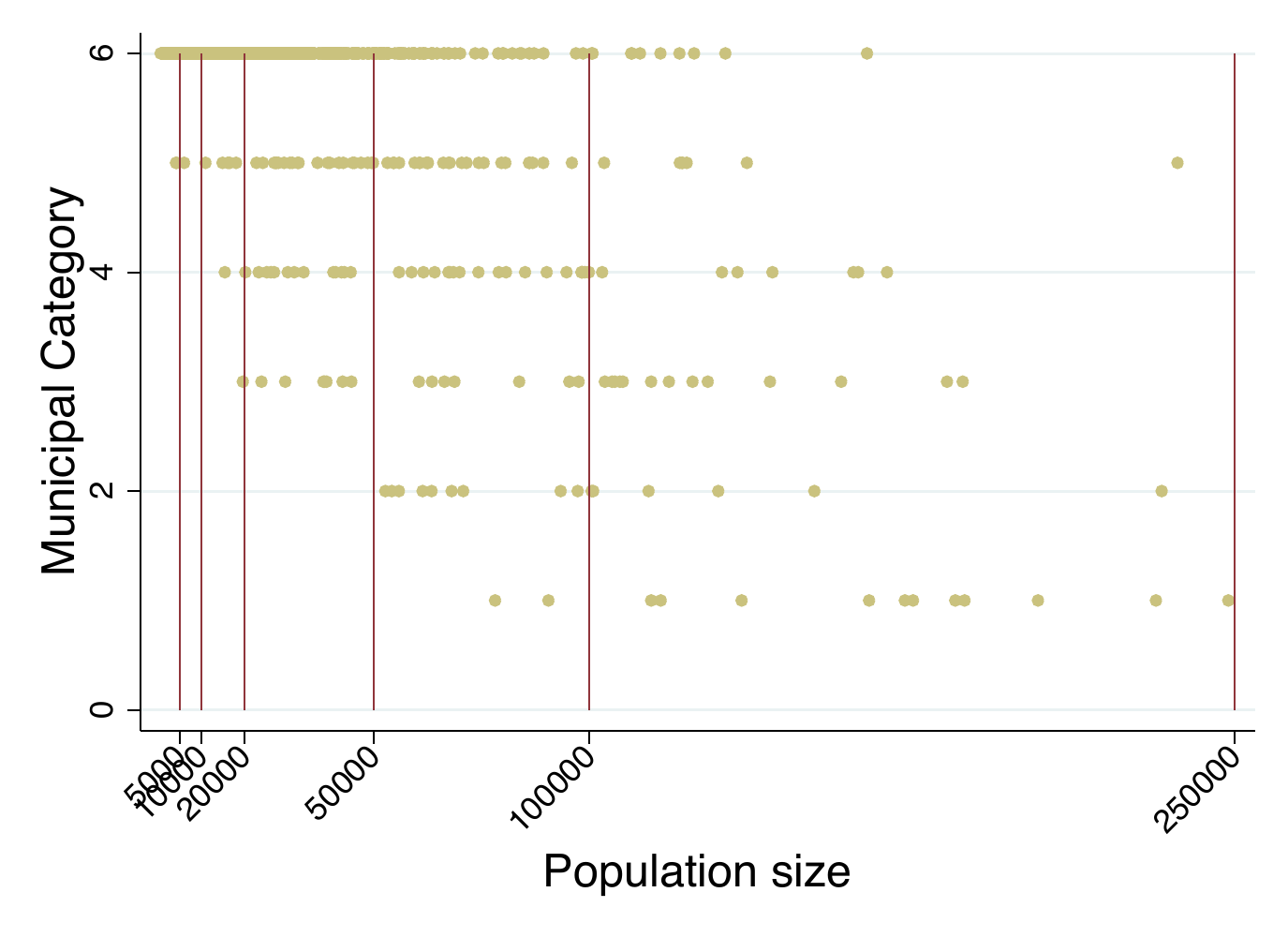}}
\caption{}
\end{subfigure}
\begin{subfigure}[a]{0.45\textwidth}	
\resizebox{7.2cm}{4cm}{\includegraphics[width=1cm]{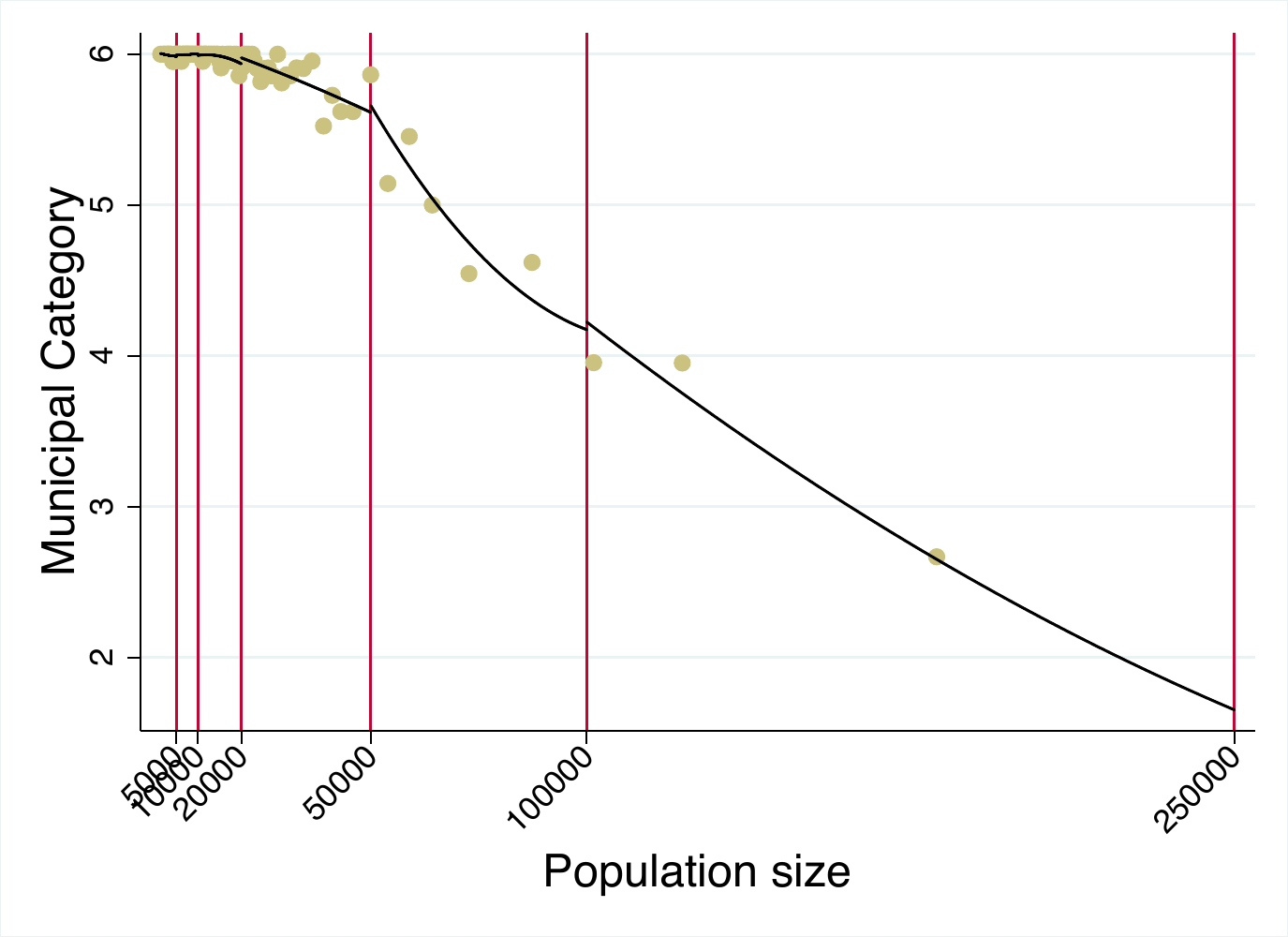}}
\caption{}
\end{subfigure}\\
\end{center}
\vspace{-0.9cm}
\begin{center}
\begin{minipage}{14cm} \scriptsize Panel (a) shows the scatterplot of municipal category by population size; panel (b) shows the scatterplot averaged over 1,000-inhabitant bins plus running-mean smoothing performed separately in each interval between two thresholds. The vertical lines identify the  population thresholds  used in the determination of the council size. To facilitate the exposition it includes municipalities with a population of 250,000 or less. 
\end{minipage}
\end{center}
\end{figure}

 
\begin{figure}[H]
\begin{center}
\caption{Current spending  and population size}\label{currspenddistribution}
\vspace{-0.3cm}
\begin{subfigure}[a]{0.45\textwidth}	
\resizebox{7.2cm}{4cm}{\includegraphics[width=1cm]{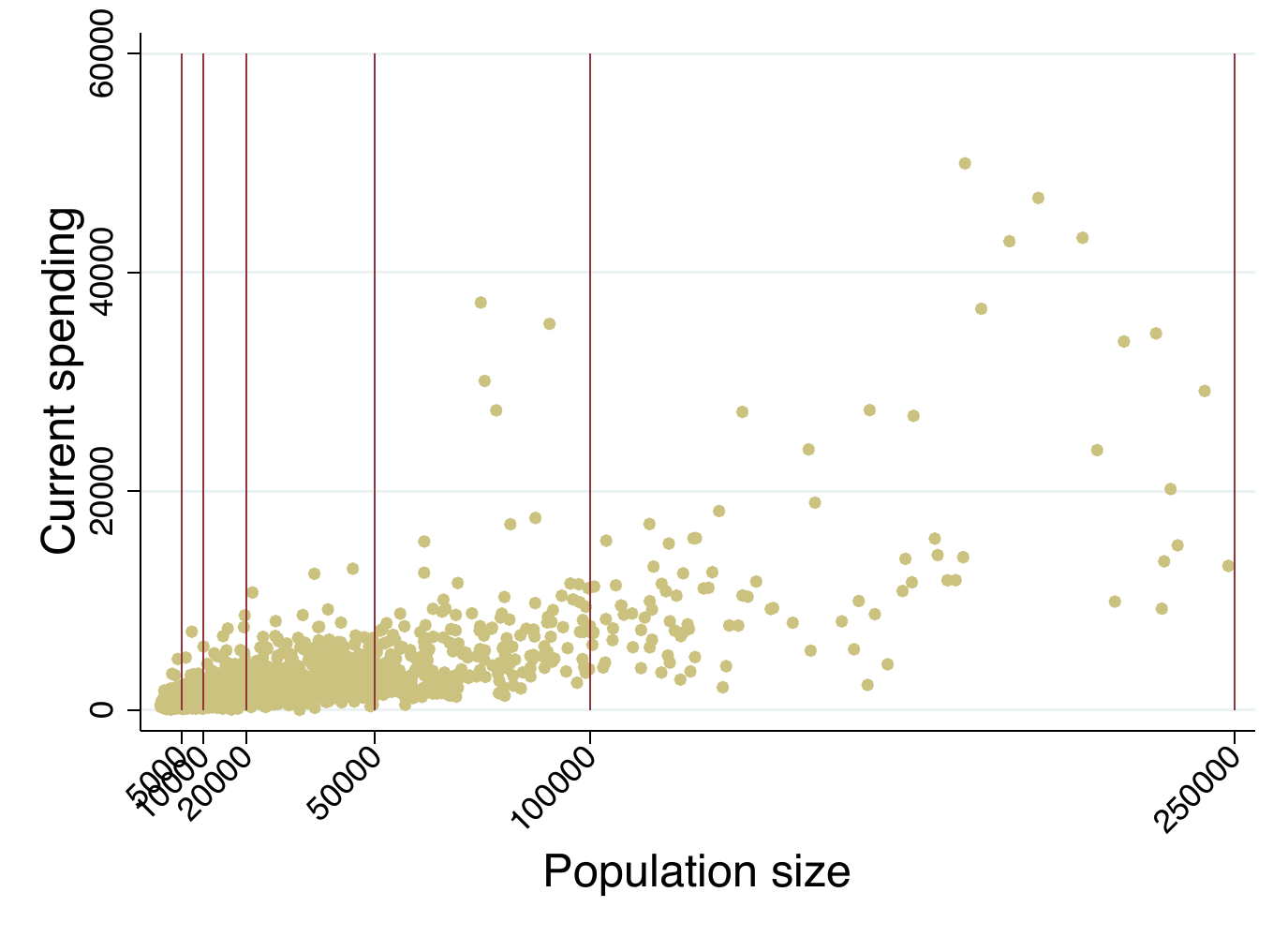}}
\caption{}
\end{subfigure}
\begin{subfigure}[a]{0.45\textwidth}	
\resizebox{7.2cm}{4cm}{\includegraphics[width=1cm]{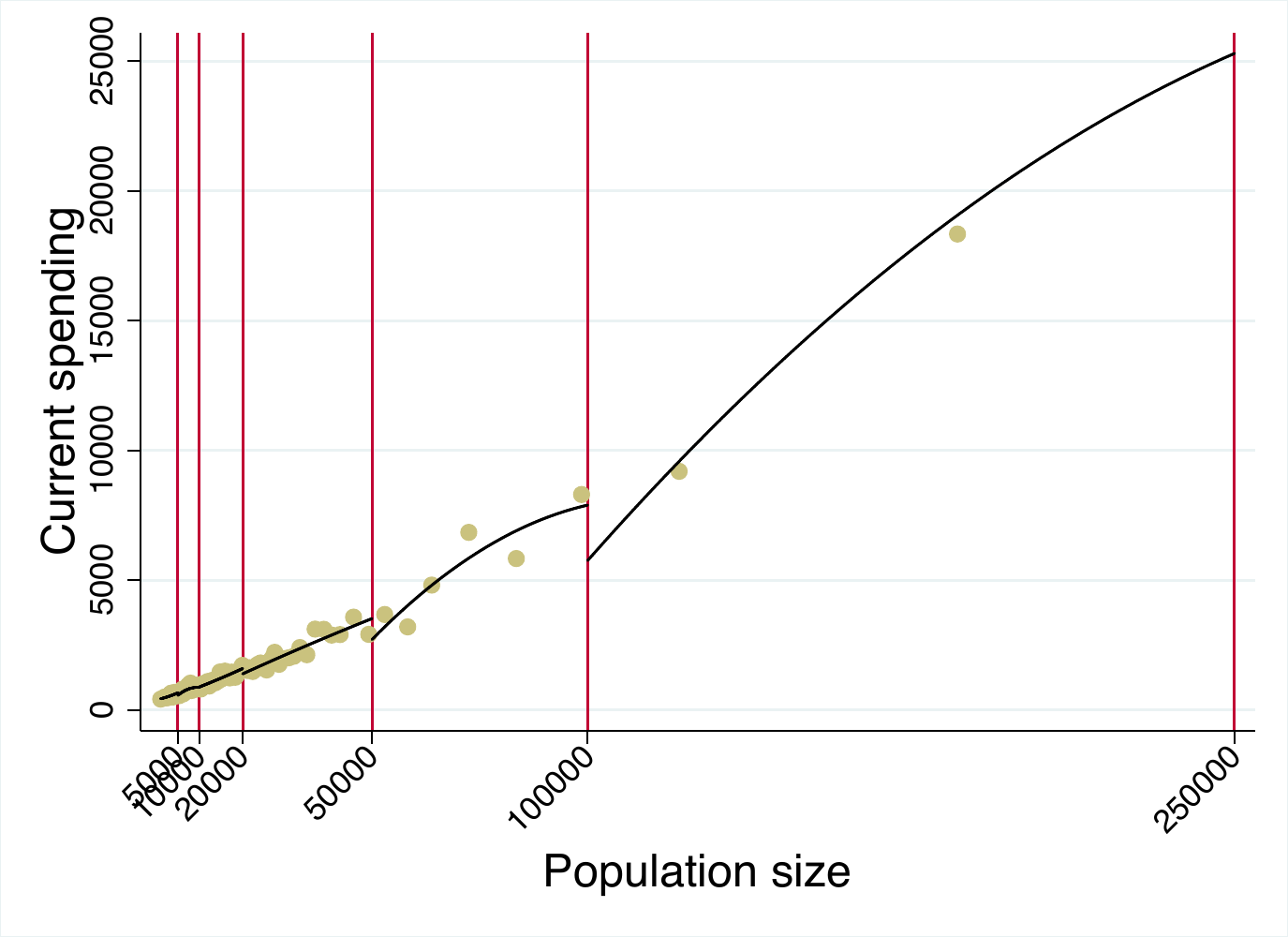}}
\caption{}
\end{subfigure}\\
\end{center}
\vspace{-0.9cm}
\begin{center}
\begin{minipage}{14cm} \scriptsize Panel (a) shows the scatterplot of current spending by population size; panel (b) shows the scatterplot averaged over 1,000-inhabitant bins plus running-mean smoothing performed separately in each interval between two thresholds. The vertical lines identify the  population thresholds  used in the determination of the council size. To facilitate the exposition it includes municipalities with a population of 250,000 or less. 
\end{minipage}
\end{center}
\end{figure}


\newpage

\renewcommand{\arraystretch}{0.8} 
\setlength{\tabcolsep}{5pt}
\begin{table}[H]
\small
\begin{center}
\caption{Effect of council size on baseline electoral outcomes: Robustness to excluding the 20,000 and 100,000 thresholds}\label{electoralbasicRDtabnothresholds3or5}
\begin{tabularx}{1.1\textwidth}{lYYYY}
\hline\hline\addlinespace[0.15cm]
&  \multicolumn{4}{c}{\emph{Dependent variable:}}   \\
 && & & \multicolumn{1}{c}{\# of non-tradi-} \\
 & \multicolumn{1}{c}{Voter} & \multicolumn{1}{c}{\# of parties}& \multicolumn{1}{c}{\# of parties}& \multicolumn{1}{c}{tional parties} \\
      & \multicolumn{1}{c}{ turnout} & \multicolumn{1}{c}{ in race}& \multicolumn{1}{c}{on council}& \multicolumn{1}{c}{on council} \\\cmidrule[0.2pt](l){2-2}\cmidrule[0.2pt](l){3-3}\cmidrule[0.2pt](l){4-4}\cmidrule[0.2pt](l){5-5}
& (1)& (2)& (3)& (4) \\ \addlinespace[0.15cm] \hline\addlinespace[0.15cm]
\multicolumn{1}{l}{\emph{\underline{Panel A}:}} & \multicolumn{4}{c}{Excluding 20,000 threshold}    \\ \cmidrule[0.2pt](l){2-5}  \addlinespace[0.15cm]
RD Estimate &       0.035         &       0.039         &       2.146\sym{***}&       1.750\sym{**} \\
            &     (0.026)         &     (1.218)         &     (0.796)         &     (0.735)         \\
Observations&         532         &         683         &         683         &         683         \\
Eff. Number of obs&         151         &         159         &         110         &         115         \\
BW Loc. Poly. (h)&       0.111         &       0.092         &       0.059         &       0.063         \\
Robust p-value&       0.160         &       0.907         &       0.006         &       0.016         \\

\addlinespace[0.15cm]\hline\addlinespace[0.15cm]
\multicolumn{1}{l}{\emph{\underline{Panel B}:}} &   \multicolumn{4}{c}{Excluding 100,000 threshold}   \\  \cmidrule[0.2pt](l){2-5}  \addlinespace[0.15cm]
RD Estimate &       0.018         &       0.672         &       0.861\sym{*}  &       0.764\sym{*}  \\
            &     (0.032)         &     (0.859)         &     (0.463)         &     (0.459)         \\
Observations&        1350         &        1743         &        1743         &        1743         \\
Eff. Number of obs&         329         &         379         &         412         &         370         \\
BW Loc. Poly. (h)&       0.100         &       0.089         &       0.096         &       0.086         \\
Robust p-value&       0.607         &       0.593         &       0.098         &       0.118         \\

\addlinespace[0.15cm]
\hline\hline                                               
\multicolumn{5}{p{17cm}}{\scriptsize \textbf{Notes}:  All columns report the RD estimates of having a larger council from Eq. (\ref{rdbaseline}) when the respective characteristic is used as the dependent variable. The bandwidth (h) is  chosen optimally using the algorithm by \cite{CalonicoCattaneoTitiunik2014a} as  implemented in Stata by the command rdrobust.ado, and includes  fixed effects for population threshold, electoral term and region. Standard errors clustered by municipality are reported in parentheses. * significant at 10\%, ** significant at 5\%, *** significant at 1\%. } 
\end{tabularx}
\end{center}
\end{table}


\begin{center}
\renewcommand{\arraystretch}{1} 
\setlength{\tabcolsep}{10pt}
\begin{threeparttable}[H]
\small
\caption{Paramilitary-linked parties}\label{tableparamilitaryparties}
\begin{tabular}{|l|c|c|}\hline\hline\addlinespace[0.15cm]
&Political & \% of senators  \\
&party  & prosecuted  or convicted  \\\addlinespace[0.15cm]
\hline\addlinespace[0.15cm]
1&\emph{Colombia Viva}\tnote{1}  & 100\%\\ 
2&\emph{Colombia Democratica}\tnote{2}  & 100\%  \\ 
3&\emph{Convergencia Popular C\'ivica} & 100\%  \\ 
4&\emph{Convergencia Ciudadana}\tnote{3} & 63\%  \\ 
5&\emph{Alas-Equipo Colombia}\tnote{4} & 60\%  \\ 
6&\emph{Cambio Radical} & 47\%  \\ 
7&\emph{Partido de la U} & 37\%  \\ 
8&\emph{Conservador} & 22\%  \\ \addlinespace[0.15cm]
\hline               \hline                                              
\multicolumn{3}{p{13cm}}{\scriptsize \textbf{Notes}: The data on senators prosecuted or convicted because of close ties to paramilitary groups are from \cite{LopezSevillano2008}, \cite{indepa2012} and \cite{MOE2013}, which report information obtained from justice authorities. The baseline definition includes the first five parties listed, where the majority of the leaders were involved in the parapolitics scandal. An alternative definition also includes the \emph{Cambio Radical}.}
\end{tabular}
\begin{tablenotes}\scriptsize
\item[1] Includes the \emph{Integracion Popular}, whose leaders it incorporated \citep{SemanaOCT122007}.
\item[2] Includes the \emph{Mov. Popular Unido}, \emph{Mov. Nacional Progresista} and \emph{Moral} from which it originated, and the  \emph{Mov. de Inclusion y Oportunidades} to which it was renamed \cite[see][]{LopezSevillano2008}.
\item[3] Includes the \emph{Integracion Nacional}, to which it was renamed \cite[see][]{EltiempoNOV092009}.
\item[4] Includes the parties \emph{ALAS} and \emph{Equipo Colombia}, from which it originated \cite[see][]{LopezSevillano2008}.
\end{tablenotes}
\end{threeparttable}
\end{center}


\begin{center}
\renewcommand{\arraystretch}{1} 
\setlength{\tabcolsep}{10pt}
\begin{threeparttable}[H]
\small
\caption{Left-wing  parties}\label{tableleftwingparties}
\begin{tabular}{|c|c|c|c|}
\hline\hline\addlinespace[0.15cm]
\multicolumn{4}{c}{\emph{Political Party}}\\\addlinespace[0.15cm]
\hline\addlinespace[0.15cm]

1&\emph{Union Patriotica} &9&\emph{Frente Social y Politico} \\
2&\emph{Polo Democratico Alternativo} &10&\emph{Movimiento 19 de abril} \\
3&\emph{Polo Democratico Independiente} &11&\emph{Socialdemocrata Colombiano} \\
4&\emph{Partido Comunista} &12&\emph{Independiente Frente de Esperanza} \\
5&\emph{Alianza Nacional Popular ANAPO} &13&\emph{Movimiento Ciudadano} \\
6&\emph{Alianza Democratica M19} &14&\emph{Alternativa Democratica} \\
7&\emph{MOIR} &15&\emph{Unidad Democratica} \\
8&\emph{Frente Social y Politico} &16& \emph{Vamos Ipiales}\\
 \addlinespace[0.15cm]
\hline   \hline                                                          
\multicolumn{4}{p{15.6cm}}{\scriptsize \textbf{Notes}: The classification of left-wing parties mainly follows \cite[see][]{FergussonQuerubinVargas2017},   and adds the \emph{Partido Comunista} and excludes the \emph{Autoridades Indigenas de Colombia}, which may be described as an ethnic party \cite[see][]{Duque2008, Laurent2010}.  An alternative definition also includes the parties \emph{Via Alterna} and  \emph{Socialismo Democratico} \cite[see][]{SemanaFEB022000, SemanaMAY112011}.   \cite{FergussonQuerubinVargas2017} propose a classification of the ideology of 505 different parties that either won or came second in mayoral elections during the period I focus on. Since \cite{FergussonQuerubinVargas2017} focus on parties that are relevant in mayoral elections (while I focus on council elections),  the estimates that use this measure may be less precise than with perfect data. However, it is reasonable to think that if a local party is relevant in mayoral elections, it is also relevant in council elections.}
\end{tabular}
\end{threeparttable}
\end{center}


\renewcommand{\arraystretch}{0.8} 
\setlength{\tabcolsep}{5pt}
\begin{table}[H]
\small
\begin{center}
\caption{Effect of council size on local public goods provision}\label{tableRDpublicgoods}
\begin{tabular}{lccc}
\hline\hline\addlinespace[0.15cm]  
 &\multicolumn{3}{c}{\emph{Dependent variable:}} \\  
  & \multicolumn{1}{c}{Public school} & \multicolumn{1}{c}{Poor with health}& \multicolumn{1}{c}{Access to clean} \\
& \multicolumn{1}{c}{enrollment  (per capita)} & \multicolumn{1}{c}{ insurance (\%)}& \multicolumn{1}{c}{ water  (per capita)}
 \\\cmidrule[0.2pt](l){2-2}\cmidrule[0.2pt](l){3-3}\cmidrule[0.2pt](l){4-4}
& (1)& (2)& (3) \\ \hline \addlinespace[0.15cm]
\hspace{3mm}RD Estimate &       0.013         &       0.024         &      -0.005         \\
            &     (0.010)         &     (0.041)         &     (0.005)         \\
Observations&        1864         &        1900         &        1004         \\
Eff. Number of obs&         506         &         414         &         349         \\
BW Loc. Poly. (h)&       0.111         &       0.088         &       0.137         \\
Robust p-value&       0.213         &       0.840         &       0.298         \\

\addlinespace[0.15cm]
\hline   \hline                                                          
\multicolumn{4}{p{15cm}}{\scriptsize \textbf{Notes}:  All columns report the RD estimates of having a larger council from Eq. (\ref{rdbaseline}) when the respective characteristic is used as the dependent variable. The bandwidth (h) is  chosen optimally using the algorithm by \cite{CalonicoCattaneoTitiunik2014a} as  implemented in Stata by the command rdrobust.ado, and includes  fixed effects for population threshold, electoral term and region. The data on public school enrollment comes from the Ministry of National Education (MEN). Data on access to clean water comes from the Public Services Information System (SUI), and data on the percentage of poor with health insurance comes from the Ministry of Health and Social Protection (MSPS). Standard errors clustered by municipality are reported in parentheses. * significant at 10\%, ** significant at 5\%, *** significant at 1\%.
 } 
\end{tabular}
\end{center}
\end{table}


\renewcommand{\arraystretch}{0.8} 
\setlength{\tabcolsep}{5pt}
\begin{table}[H]
\small
\begin{center}
\caption{Effect of council size on coca cultivation and aerial spraying}\label{tableRDcoca}
\begin{tabularx}{0.7\textwidth}{lYY}
\hline\hline\addlinespace[0.15cm]  
 &\multicolumn{2}{c}{\emph{Dependent variable}: Hectares of  } \\\cmidrule[0.2pt](l){2-3}
& \multicolumn{1}{c}{coca} & \multicolumn{1}{c}{aerial spraying}\\
& \multicolumn{1}{c}{cultivations} & \multicolumn{1}{c}{on coca cult.}
 \\\cmidrule[0.2pt](l){2-2}\cmidrule[0.2pt](l){3-3}
& (1)& (2)  \\ \hline \addlinespace[0.15cm]
\hspace{3mm}RD Estimate &      -0.198         &      -0.127         \\
            &     (0.437)         &     (0.353)         \\
Observations&        1900         &        1900         \\
Eff. Number of obs&         537         &         609         \\
BW Loc. Poly. (h)&       0.115         &       0.129         \\
Robust p-value&       0.688         &       0.799         \\

\addlinespace[0.15cm]
\hline   \hline                                                          
\multicolumn{3}{p{10.5cm}}{\scriptsize \textbf{Notes}: All columns report the RD estimates of having a larger council from Eq. (\ref{rdbaseline}) when the respective characteristic is used as the dependent variable. The bandwidth (h) is  chosen optimally using the algorithm by \cite{CalonicoCattaneoTitiunik2014a} as  implemented in Stata by the command rdrobust.ado, and includes  fixed effects for population threshold, electoral term and region. The data is from the Center of Studies on Economic Development, which compiled and processed information provided by the United Nations Office on Drugs and Crime. The data is available for 2000-2009. Standard errors clustered by municipality are reported in parentheses. * significant at 10\%, ** significant at 5\%, *** significant at 1\%. } 
\end{tabularx}
\end{center}
\end{table}


\newpage
\setcounter{table}{0}
\setcounter{figure}{0}
\newgeometry{left=2cm,right=2cm,bottom=2cm,top=3cm}
{\section{Omitted Robustness Checks}\label{appomittedrobust}
\small

 In this Appendix, I provide additional details about the robustness exercises reported in the text. 
 
 Tables  \ref{tableRDconflictrobustnescontrolcrime} to \ref{tableRDconflictavnpqdcl} show that  the  results in Section \ref{mainresults}  are robust to  a variety of additional specifications.   Table \ref{tableRDconflictrobustnescontrolcrime} includes controls for previous crime rates.  Table \ref{tableRDconflictbaselinyearly} uses yearly averages instead of averages over the electoral period.  Table \ref{tableRDconflictnothresholds3or5} excludes the 20,000 and 100,000 population thresholds (since some key pre-treatment variables may change discontinuously at these thresholds, or other policies may serve as confounding factors). Table \ref{tableRDconflictbaselineextendeperiod} includes the 2011 and 2015 elections (using the data from the NCHM described in the footnote \ref{dataNCHM}).  All the estimates are consistent with those in Table  \ref{tableRDconflictbaseline}, and many of them are more precise.  
 
 Table \ref{tableRDconflictrobustnessratio} examines the robustness of the results in Table \ref{tableRDconflictbaseline} to the use of an alternative dependent variable: the ratio of conflict-related killings to total killings.  Notably, the estimates for selective killings are again negative, statistically significant and more precise than the estimates in Panel C in Table \ref{tableRDconflictbaseline}. Table \ref{tableRDconflictavnpqdcl}  examines whether there is an effect on other types of conflict-related violence. It looks at combatant deaths and unintended civilian casualties. No estimates are statistically different from zero, which provides additional evidence that the effect of larger councils on conflict-related violence is specific to killings that deliberately target civilians.

\renewcommand{\arraystretch}{0.8} 
\setlength{\tabcolsep}{12pt}
\begin{table}[H]
\small
\begin{center}
\caption{Effect of council size on conflict-related killings: Robustness to controlling for previous overall homicide rate}\label{tableRDconflictrobustnescontrolcrime}
\begin{tabular}{lcccccc}\hline\hline\addlinespace[0.15cm]
\emph{Dependent variable:} &\multicolumn{3}{c}{Conflict-related killings}  \\
 &\multicolumn{3}{c}{per 100,000 inhabitants}  \\\cmidrule[0.2pt](l){2-4}
 & \multicolumn{1}{c}{Killing} & \multicolumn{1}{c}{Select. killing}& \multicolumn{1}{c}{Massacre}  \\
\cmidrule[0.2pt](l){2-2}\cmidrule[0.2pt](l){3-3}\cmidrule[0.2pt](l){4-4}
& (1)& (2)& (3) \\ \hline \addlinespace[0.15cm]
\hspace{3mm}RD Estimate &      -2.577         &      -2.790\sym{**} &      -0.020         \\
            &     (1.900)         &     (1.418)         &     (0.928)         \\
Observations&        2318         &        2318         &        2318         \\
Eff. Number of obs&         526         &         495         &         500         \\
BW Loc. Poly. (h)&       0.090         &       0.084         &       0.085         \\
Robust p-value&       0.215         &       0.072         &       0.917         \\

 \addlinespace[0.15cm]
\hline     \hline                                                        
\multicolumn{4}{p{11cm}}{\scriptsize \textbf{Notes}:  All columns report the RD estimates of having a larger council from Eq. (\ref{rdbaseline}) when the respective characteristic is used as the dependent variable.  All columns include fixed effects for population threshold, electoral term, region, and  the overall homicide rate lagged one period. The bandwidth (h) is chosen optimally using the algorithm by \cite{CalonicoCattaneoTitiunik2014a} as  implemented in Stata by the command rdrobust.ado. Standard errors clustered by municipality are reported in parentheses. * significant at 10\%, ** significant at 5\%, *** significant at 1\%. }
\end{tabular}
\end{center}
\end{table}


\renewcommand{\arraystretch}{0.8} 
\setlength{\tabcolsep}{10pt}
\begin{table}[H]
\small
\begin{center}
\caption{Effect of council size on conflict-related violence: Robustness to using yearly averages}\label{tableRDconflictbaselinyearly}
\begin{tabular}{lccc}\hline\hline\addlinespace[0.15cm]
&\multicolumn{3}{c}{\emph{Dep. variable:}  Conflict-related homicide}\\
 & \multicolumn{1}{c}{Killing} & \multicolumn{1}{c}{Select. killing}& \multicolumn{1}{c}{Massacre}  \\\cmidrule[0.2pt](l){2-2}\cmidrule[0.2pt](l){3-3}\cmidrule[0.2pt](l){4-4}
& (1)& (2)& (3) \\ \hline \addlinespace[0.15cm]
\multicolumn{1}{l}{\emph{\underline{Panel A}: Prob.}} &    \\   \addlinespace[0.15cm]
\hspace{3mm}RD Estimate &      -0.047\sym{***}&      -0.043\sym{***}&      -0.005         \\
            &     (0.010)         &     (0.009)         &     (0.003)         \\
Observations&        7128         &        7128         &        7128         \\
Eff. Number of obs&        2066         &        1909         &        1582         \\
BW Loc. Poly. (h)&       0.117         &       0.107         &       0.090         \\
Robust p-value&       0.000         &       0.000         &       0.094         \\

\addlinespace[0.15cm]
\hline \addlinespace[0.15cm]
\multicolumn{1}{l}{\emph{\underline{Panel B}: Number}} & &   \\   \addlinespace[0.15cm]
\hspace{3mm}RD Estimate &      -1.177\sym{**} &      -1.082\sym{***}&      -0.048         \\
            &     (0.493)         &     (0.298)         &     (0.219)         \\
Observations&        7128         &        7128         &        7128         \\
Eff. Number of obs&        1604         &        1552         &        2010         \\
BW Loc. Poly. (h)&       0.091         &       0.087         &       0.113         \\
Robust p-value&       0.015         &       0.001         &       0.800         \\

\addlinespace[0.15cm]
\hline \addlinespace[0.15cm]
\multicolumn{1}{l}{\emph{\underline{Panel C}: Rate}} & &   \\   \addlinespace[0.15cm]
\hspace{3mm}RD Estimate &      -3.526\sym{**} &      -4.068\sym{***}&       0.157         \\
            &     (1.616)         &     (1.075)         &     (0.738)         \\
Observations&        7128         &        7128         &        7128         \\
Eff. Number of obs&        1567         &        1391         &        1552         \\
BW Loc. Poly. (h)&       0.088         &       0.078         &       0.087         \\
Robust p-value&       0.035         &       0.000         &       0.667         \\

 \addlinespace[0.15cm]
\hline     \hline                                                        
\multicolumn{4}{p{11.5cm}}{\scriptsize \textbf{Notes}:  All columns report the RD estimates of having a larger council from Eq. (\ref{rdbaseline}) when the respective characteristic is used as the dependent variable. The dependent variable in panel A is the average over the years following an election of a dummy variable equal to 1 if the particular type of killing is observed in a municipality in a quarter of a year. The dependent variable in panels B and C is the average of the total number  of killings and  killings per 100,000 people, respectively. The bandwidth (h) is chosen optimally using the algorithm by \cite{CalonicoCattaneoTitiunik2014a} as  implemented in Stata by the command rdrobust.ado, and includes  fixed effects for population threshold, year and region. Standard errors clustered by municipality are reported in parentheses. * significant at 10\%, ** significant at 5\%, *** significant at 1\%. }
\end{tabular}
\end{center}
\end{table}


\newgeometry{left=3cm,bottom=2cm,top=3cm}
{ 
\renewcommand{\arraystretch}{0.8} 
\setlength{\tabcolsep}{10pt}
\begin{table}[H]
\small
\begin{center}
\caption{Effect of council size on conflict-related violence: Robustness to excluding the 20,000 and 100,000 thresholds}\label{tableRDconflictnothresholds3or5}
\begin{tabular}{lccccc}\hline\hline\addlinespace[0.15cm]
& \multicolumn{3}{c}{\emph{Dep. variable:}  Prob. of conflict-related:} \\\cmidrule[0.2pt](l){2-4}
 & \multicolumn{1}{c}{Killing} & \multicolumn{1}{c}{Select. killing}& \multicolumn{1}{c}{Massacre}  \\
\cmidrule[0.2pt](l){2-2}\cmidrule[0.2pt](l){3-3}\cmidrule[0.2pt](l){4-4}
& (1)& (2)& (3) \\ \hline \addlinespace[0.15cm]
\multicolumn{1}{l}{\emph{\underline{Panel A}: }} & \multicolumn{3}{c}{Excluding 20,000 threshold}     \\  \cmidrule[0.2pt](l){2-4} \addlinespace[0.15cm]
\hspace{3mm}RD Estimate &      -0.061\sym{*}  &      -0.055\sym{*}  &      -0.015         \\
            &     (0.032)         &     (0.031)         &     (0.010)         \\
Observations&         683         &         683         &         683         \\
Eff. Number of obs&         221         &         203         &         152         \\
BW Loc. Poly. (h)&       0.130         &       0.121         &       0.088         \\
Robust p-value&       0.072         &       0.124         &       0.108         \\

\addlinespace[0.15cm]
\hline \addlinespace[0.15cm]
\multicolumn{1}{l}{\emph{\underline{Panel B}: }} & \multicolumn{3}{c}{Excluding 100,000 threshold}     \\   \cmidrule[0.2pt](l){2-4}\addlinespace[0.15cm]
\hspace{3mm}RD Estimate &      -0.037\sym{**} &      -0.032\sym{*}  &      -0.005         \\
            &     (0.019)         &     (0.017)         &     (0.006)         \\
Observations&        1743         &        1743         &        1743         \\
Eff. Number of obs&         537         &         551         &         411         \\
BW Loc. Poly. (h)&       0.126         &       0.128         &       0.096         \\
Robust p-value&       0.071         &       0.096         &       0.397         \\

 \addlinespace[0.15cm]
\hline     \hline                                                        
\multicolumn{4}{p{12cm}}{\scriptsize \textbf{Notes}:  All columns report the RD estimates of having a larger council from Eq. (\ref{rdbaseline}) when the respective characteristic is used as the dependent variable. The bandwidth (h) is chosen optimally using the algorithm by \cite{CalonicoCattaneoTitiunik2014a} as  implemented in Stata by the command rdrobust.ado, and includes  fixed effects for population threshold, electoral term (or year) and region. Standard errors clustered by municipality are reported in parentheses. * significant at 10\%, ** significant at 5\%, *** significant at 1\%. }
\end{tabular}
\end{center}
\end{table}
}


\renewcommand{\arraystretch}{0.8} 
\setlength{\tabcolsep}{5pt}
\begin{table}[H]
\small
\begin{center}
\caption{Effect of council size on conflict-related violence: NCHM data and  1997-2018 period}\label{tableRDconflictbaselineextendeperiod}
\begin{tabular}{lcccc}\hline\hline\addlinespace[0.15cm]
\emph{Dependent variable:} & \multicolumn{3}{c}{Conflict-related homicide}  & \multicolumn{1}{c}{Overall
} \\\cmidrule[0.2pt](l){2-4}
 & \multicolumn{1}{c}{Killing} & \multicolumn{1}{c}{Select. killing}& \multicolumn{1}{c}{Massacre} & \multicolumn{1}{c}{
homicide} \\
\cmidrule[0.2pt](l){2-2}\cmidrule[0.2pt](l){3-3}\cmidrule[0.2pt](l){4-4}\cmidrule[0.2pt](l){5-5}
& (1)& (2)& (3) & (4)\\ \hline \addlinespace[0.15cm]
\multicolumn{1}{l}{\emph{\underline{Panel A}: prob.}} &    \\   \addlinespace[0.15cm]
\hspace{3mm}RD Estimate &      -0.002         &      -0.070\sym{**} &      -0.005         &       0.012         \\
            &     (0.004)         &     (0.033)         &     (0.005)         &     (0.019)         \\
Observations&        2822         &        2822         &        2822         &        2817         \\
Eff. Number of obs&         855         &         656         &         755         &         563         \\
BW Loc. Poly. (h)&       0.117         &       0.091         &       0.103         &       0.078         \\
Robust p-value&       0.658         &       0.051         &       0.326         &       0.436         \\

\addlinespace[0.15cm]
\hline \addlinespace[0.15cm]
\multicolumn{1}{l}{\emph{\underline{Panel B}: number}} & &   \\   \addlinespace[0.15cm]
\hspace{3mm}RD Estimate &      -2.226         &      -2.156         &      -0.070         &      -3.797         \\
            &     (1.772)         &     (1.718)         &     (0.088)         &     (3.085)         \\
Observations&        2822         &        2822         &        2822         &        2817         \\
Eff. Number of obs&         636         &         636         &         631         &         704         \\
BW Loc. Poly. (h)&       0.088         &       0.087         &       0.087         &       0.096         \\
Robust p-value&       0.239         &       0.243         &       0.402         &       0.242         \\

\addlinespace[0.15cm]
\hline \addlinespace[0.15cm]
\multicolumn{1}{l}{\emph{\underline{Panel C}: rate}} & &   \\   \addlinespace[0.15cm]
\hspace{3mm}RD Estimate &      -1.047         &      -1.020         &      -0.027         &      -4.420         \\
            &     (5.553)         &     (5.402)         &     (0.245)         &     (7.716)         \\
Observations&        2822         &        2822         &        2822         &        2810         \\
Eff. Number of obs&         776         &         776         &         782         &         783         \\
BW Loc. Poly. (h)&       0.106         &       0.106         &       0.107         &       0.109         \\
Robust p-value&       0.863         &       0.863         &       0.961         &       0.563         \\

 \addlinespace[0.15cm]
\hline     \hline                                                        
\multicolumn{5}{p{12cm}}{\scriptsize \textbf{Notes}:  All columns report the RD estimates of having a larger council from Eq. (\ref{rdbaseline}) when the respective characteristic is used as the dependent variable. The dependent variable in panel A is the average over the electoral term of a dummy variable equal to 1 if the particular type of killing is observed in a municipality in a quarter of a year. The dependent variable in panels B and C is the average of the total number  of killings and  killings per 100,000 people, respectively. The bandwidth (h) is chosen optimally using the algorithm by \cite{CalonicoCattaneoTitiunik2014a} as  implemented in Stata by the command rdrobust.ado, and includes  fixed effects for population threshold, electoral term and region. Standard errors clustered by municipality are reported in parentheses. * significant at 10\%, ** significant at 5\%, *** significant at 1\%. }
\end{tabular}
\end{center}
\end{table}


\renewcommand{\arraystretch}{0.8} 
\setlength{\tabcolsep}{10pt}
\begin{table}[H]
\small
\begin{center}
\caption{Effect of council size on conflict-related violence: Robustness to the use of the ratio of conflict-related killings to total killings}\label{tableRDconflictrobustnessratio}
\begin{tabular}{lccc}\hline\hline\addlinespace[0.15cm]
\emph{Dependent variable:} & \multicolumn{3}{c}{Ratio of conflict-related killings}  \\
& \multicolumn{3}{c}{to total killings}  \\
\cmidrule[0.2pt](l){2-4}
 & \multicolumn{1}{c}{Killing} & \multicolumn{1}{c}{Select. killing}& \multicolumn{1}{c}{Massacre}  \\
\cmidrule[0.2pt](l){2-2}\cmidrule[0.2pt](l){3-3}\cmidrule[0.2pt](l){4-4}
& (1)& (2)& (3) \\ \hline \addlinespace[0.15cm]
\hspace{3mm}RD Estimate &      -0.033         &      -0.054\sym{***}&       0.013         \\
            &     (0.028)         &     (0.019)         &     (0.023)         \\
Observations&        2311         &        2311         &        2311         \\
Eff. Number of obs&         504         &         392         &         763         \\
BW Loc. Poly. (h)&       0.087         &       0.069         &       0.130         \\
Robust p-value&       0.126         &       0.005         &       0.616         \\

 \addlinespace[0.15cm]
\hline     \hline                                                        
\multicolumn{4}{p{10.4cm}}{\scriptsize \textbf{Notes}:  All columns report the RD estimates of having a larger council from Eq. (\ref{rdbaseline}) when the respective characteristic is used as the dependent variable.  All columns include fixed effects for population threshold, electoral term and region. The bandwidth (h) is chosen optimally using the algorithm by \cite{CalonicoCattaneoTitiunik2014a} as  implemented in Stata by the command rdrobust.ado. Standard errors clustered by municipality are reported in parentheses. * significant at 10\%, ** significant at 5\%, *** significant at 1\%. }
\end{tabular}
\end{center}
\end{table}


\renewcommand{\arraystretch}{0.8} 
\setlength{\tabcolsep}{5pt}
\begin{table}[H]
\small
\begin{center}
\caption{Effect of council size on other types of conflict-related violence}\label{tableRDconflictavnpqdcl}
\begin{tabular}{lcccccc}\hline\hline\addlinespace[0.15cm]
&\multicolumn{6}{c}{\emph{Dep. variable:}}\\
&\multicolumn{3}{c}{Combatant deaths} &\multicolumn{3}{c}{Unintended civilian casualties}   \\\cmidrule[0.2pt](l){2-4}\cmidrule[0.2pt](l){5-7}
 & \multicolumn{1}{c}{Occurrence} & \multicolumn{1}{c}{Number}& \multicolumn{1}{c}{Rate}& \multicolumn{1}{c}{Occurrence} & \multicolumn{1}{c}{Number}& \multicolumn{1}{c}{Rate}   \\\cmidrule[0.2pt](l){2-2}\cmidrule[0.2pt](l){3-3}\cmidrule[0.2pt](l){4-4}\cmidrule[0.2pt](l){5-5}\cmidrule[0.2pt](l){6-6}\cmidrule[0.2pt](l){7-7}
& (1)& (2)& (3) & (4) & (5) & (6)\\ \hline \addlinespace[0.15cm]
\hspace{3mm}RD Estimate &       0.001         &       0.058         &       0.180         &       0.000         &       0.043         &       0.052         \\
            &     (0.002)         &     (0.067)         &     (0.194)         &     (0.003)         &     (0.067)         &     (0.206)         \\
Observations&        1900         &        1900         &        1900         &        1900         &        1900         &        1900         \\
Eff. Number of obs&         624         &         491         &         406         &         626         &         501         &         355         \\
BW Loc. Poly. (h)&       0.131         &       0.105         &       0.086         &       0.132         &       0.107         &       0.075         \\
Robust p-value&       0.400         &       0.377         &       0.341         &       0.960         &       0.481         &       0.605         \\

 \addlinespace[0.15cm]
\hline     \hline                                                        
\multicolumn{7}{p{15.2cm}}{\scriptsize \textbf{Notes}:  All columns report the RD estimates of having a larger council from Eq. (\ref{rdbaseline}) when the respective characteristic is used as the dependent variable. The bandwidth (h) is chosen optimally using the algorithm by \cite{CalonicoCattaneoTitiunik2014a} as  implemented in Stata by the command rdrobust.ado, and includes  fixed effects for population threshold, electoral term (or year) and region. Standard errors clustered by municipality are reported in parentheses. * significant at 10\%, ** significant at 5\%, *** significant at 1\%. }
\end{tabular}
\end{center}
\end{table}


 Tables  \ref{electoralbasicRDtabthreshold} to \ref{tableRDarmedconflicgroupyearly} provide suplemental robustness checks to the results in Section \ref{politicalopenness}.  Table \ref{electoralbasicRDtabthreshold} repeats the analysis in Table \ref{electoralbasicRDtab} distinguishing between two groups of municipalities according to their closest population thresholds; the estimates show that the effect is slightly greater for the group with a population closest to the larger thresholds. Table \ref{tableRDsuccesspartiesalt1} shows that the estimates in Table \ref{tableRDsuccessparties} are robust to the use of  the alternative definition of paramilitary-linked and left-wing parties mentioned in the notes of Tables \ref{tableparamilitaryparties} and \ref{tableleftwingparties}. The estimates are similar to those in Table  \ref{tableRDsuccessparties}. Table \ref{tableRDarmedconflicgroupyearly} shows that the results in Table \ref{tableRDarmedconflicgroup} are robust to using yearly averages instead of averages over the electoral period.

 Table  \ref{tableRDpartyofmayor} and Figure \ref{McCraryparamayorpopthresl} provide suplemental robustness checks to the results in Section \ref{paramilitarylinkedrepresentationpoliticalviolence}.  Table \ref{tableRDpartyofmayor} shows that a larger council does not affect the probability of electing a mayor from any group I have already defined. Figure \ref{McCraryparamayorpopthresl} shows that a systematic manipulation of the electoral results around the zero paramilitary margin of victory threshold within normalized population windows of 10\% or  20\% is unlikely.


\renewcommand{\arraystretch}{0.8} 
\setlength{\tabcolsep}{5pt}
\begin{table}[H]
\small
\begin{center}
\caption{Effect of council size on number of parties on council (by group of thresholds)}\label{electoralbasicRDtabthreshold}
\begin{tabularx}{0.8\textwidth}{lYYYY}
\hline\hline\addlinespace[0.15cm]
&  \multicolumn{2}{c}{\emph{Dependent variable:}}   \\
   & \multicolumn{1}{c}{\# of parties}& \multicolumn{1}{c}{\# of non-traditional} \\
      &  \multicolumn{1}{c}{on council}& \multicolumn{1}{c}{parties on council} \\\cmidrule[0.2pt](l){2-2}\cmidrule[0.2pt](l){3-3}
& (1)& (2)\\\addlinespace[0.15cm] \hline\addlinespace[0.15cm]
\multicolumn{1}{l}{\emph{\underline{Panel A}:}} &    \multicolumn{2}{c}{20,000 and 50,000 thresholds} \\ \cmidrule[0.2pt](l){2-3}  \addlinespace[0.15cm]
\hspace{3mm}RD Estimate &       0.967\sym{*}  &       0.864\sym{*}  \\
            &     (0.494)         &     (0.498)         \\
Observations&        1635         &        1635         \\
Eff. Number of obs&         364         &         328         \\
BW Loc. Poly. (h)&       0.090         &       0.080         \\
Robust p-value&       0.078         &       0.104         \\

\addlinespace[0.15cm]\hline\addlinespace[0.15cm]
\multicolumn{1}{l}{\emph{\underline{Panel B}:}} & \multicolumn{2}{c}{100,000, 250,000 and 1,000,000 thresholds}   \\  \cmidrule[0.2pt](l){2-3}  \addlinespace[0.15cm]
\hspace{3mm}RD Estimate &       1.076\sym{**} &       0.834\sym{*}  \\
            &     (0.506)         &     (0.477)         \\
Observations&         265         &         265         \\
Eff. Number of obs&          48         &          51         \\
BW Loc. Poly. (h)&       0.076         &       0.083         \\
Robust p-value&       0.072         &       0.212         \\

\addlinespace[0.15cm]
\hline    \hline                                                 
\multicolumn{3}{p{12.5cm}}{\scriptsize \textbf{Notes}:  All columns report the RD estimates of having a larger council from Eq. (\ref{rdbaseline}) when the respective characteristic is used as the dependent variable. The bandwidth (h) is  chosen optimally using the algorithm by \cite{CalonicoCattaneoTitiunik2014a} as  implemented in Stata by the command rdrobust.ado, and includes  fixed effects for population threshold, electoral term and region. Standard errors clustered by municipality are reported in parentheses. * significant at 10\%, ** significant at 5\%, *** significant at 1\%. } 
\end{tabularx}
\end{center}
\end{table}


\renewcommand{\arraystretch}{0.8} 
\setlength{\tabcolsep}{10pt}
\begin{table}[H]
\small
\begin{center}
\caption{Effect of council size on success of different parties: Robustness to using an alternative definition of paramilitary-linked and left-wing party}\label{tableRDsuccesspartiesalt1}
\begin{tabular}{lccc}
\hline\hline \addlinespace[0.15cm]
     &\multicolumn{3}{c}{\emph{Dependent variable:} Presence on council of a} \\\cmidrule[0.2pt](l){2-4}
  & \multicolumn{1}{c}{Paramilitary-} & \multicolumn{1}{c}{Left-wing} & \multicolumn{1}{c}{Paramilitary-linked} \\
      & \multicolumn{1}{c}{linked party} & \multicolumn{1}{c}{party}& \multicolumn{1}{c}{and a left-wing party} \\\cmidrule[0.2pt](l){2-2}\cmidrule[0.2pt](l){3-3}\cmidrule[0.2pt](l){4-4}
& (1)& (2)& (3) \\ \hline
\addlinespace[0.15cm]                                            
\hspace{3mm}RD Estimate &       0.241\sym{***}&      -0.015         &       0.196\sym{***}\\
            &     (0.078)         &     (0.075)         &     (0.074)         \\
Observations&        1900         &        1900         &        1900         \\
Eff. Number of obs&         391         &         580         &         357         \\
BW Loc. Poly. (h)&       0.083         &       0.125         &       0.076         \\
Robust p-value&       0.005         &       0.937         &       0.009         \\

\addlinespace[0.15cm]
\hline    \hline   
\multicolumn{4}{p{13.8cm}}{\scriptsize \textbf{Notes}:  All columns report the RD estimates of having a larger council from Eq. (\ref{rdbaseline}) when the respective characteristic is used as the dependent variable. The bandwidth (h) is  chosen optimally using the algorithm by \cite{CalonicoCattaneoTitiunik2014a} as  implemented in Stata by the command rdrobust.ado, and includes  fixed effects for population threshold, electoral term and region. Standard errors clustered by municipality are reported in parentheses. * significant at 10\%, ** significant at 5\%, *** significant at 1\%. }  
\end{tabular}
\end{center}
\end{table}


\renewcommand{\arraystretch}{0.8} 
\setlength{\tabcolsep}{5pt}
\begin{table}[H]
\small
\begin{center}
\caption{Effect of council size on armed conflict: Robustness to using yearly averages}\label{tableRDarmedconflicgroupyearly}
\begin{tabularx}{1\textwidth}{lYYY}
\hline\hline \addlinespace[0.15cm]
 &\multicolumn{3}{c}{\emph{Dependent variable:} Presence of violent actions by} \\ 
  & \multicolumn{1}{c}{Any group}& \multicolumn{1}{c}{Guerrillas}& \multicolumn{1}{c}{Paramilitaries}  
      \\\cmidrule[0.2pt](l){2-2}\cmidrule[0.2pt](l){3-3}\cmidrule[0.2pt](l){4-4}
& (1)& (2)& (3) \\ \hline \addlinespace[0.15cm]
\hspace{3mm}RD Estimate &      -0.152\sym{***}&      -0.154\sym{***}&      -0.035         \\
            &     (0.047)         &     (0.052)         &     (0.036)         \\
Observations&        5765         &        5765         &        5765         \\
Eff. Number of obs&         827         &         879         &        1215         \\
BW Loc. Poly. (h)&       0.058         &       0.062         &       0.086         \\
Robust p-value&       0.001         &       0.003         &       0.246         \\

\addlinespace[0.15cm]
\hline                \hline                                     
\multicolumn{4}{p{15cm}}{\scriptsize \textbf{Notes}:  All columns report the RD estimates of having a larger council from Eq. (\ref{rdbaseline}) when the respective characteristic is used as the dependent variable. The bandwidth (h) is  chosen optimally using the algorithm by \cite{CalonicoCattaneoTitiunik2014a} as  implemented in Stata by the command rdrobust.ado, and includes  fixed effects for population threshold, year and region. Standard errors clustered by municipality are reported in parentheses. * significant at 10\%, ** significant at 5\%, *** significant at 1\% } 
\end{tabularx}
\end{center}
\end{table}


\newgeometry{left=2.5cm,bottom=2cm,top=3cm}
{ 
\renewcommand{\arraystretch}{0.8} 
\setlength{\tabcolsep}{4pt}
\begin{table}[H]
\small
\begin{center}
\caption{Effect of council size on mayor's party}\label{tableRDpartyofmayor}
\begin{tabularx}{0.9\textwidth}{lYYYY}  
\hline\hline \addlinespace[0.15cm]
      & \multicolumn{3}{c}{\emph{Dependent variable:} Mayor from a} \\\cmidrule[0.2pt](l){2-4}
  & \multicolumn{1}{c}{Paramilitary-} & \multicolumn{1}{c}{Left-wing}& \multicolumn{1}{c}{Liberal or Con-}\\
      & \multicolumn{1}{c}{ linked party} & \multicolumn{1}{c}{party}& \multicolumn{1}{c}{servative party} \\\cmidrule[0.2pt](l){2-2}\cmidrule[0.2pt](l){3-3}\cmidrule[0.2pt](l){4-4}
& (1)& (2)& (3) \\ \hline     
\addlinespace[0.15cm]                                            
\hspace{3mm}RD Estimate &       0.033         &      -0.050         &      -0.134         \\
            &     (0.040)         &     (0.047)         &     (0.093)         \\
Observations&        1900         &        1900         &        1900         \\
Eff. Number of obs&         343         &         357         &         525         \\
BW Loc. Poly. (h)&       0.073         &       0.076         &       0.113         \\
Robust p-value&       0.691         &       0.287         &       0.167         \\

\addlinespace[0.15cm]
\hline    \hline   
\multicolumn{4}{p{14cm}}{\scriptsize \textbf{Notes}:  All columns report the RD estimates of having a larger council from Eq. (\ref{rdbaseline}) when the respective characteristic is used as the dependent variable. The bandwidth (h) is  chosen optimally using the algorithm by \cite{CalonicoCattaneoTitiunik2014a} as  implemented in Stata by the command rdrobust.ado, and includes  fixed effects for population threshold, electoral term and region. Standard errors clustered by municipality are reported in parentheses. * significant at 10\%, ** significant at 5\%, *** significant at 1\%.} 
\end{tabularx}
\end{center}
\end{table}
}


\begin{figure}[H]
\begin{center}
\caption{Manipulation tests for margin of victory of paramilitary-linked parties in small windows for normalized population} \label{McCraryparamayorpopthresl}
\vspace{-0.3cm}
\begin{subfigure}[b]{0.45\textwidth}	
\resizebox{7.2cm}{4.5cm}{\includegraphics[width=1cm]{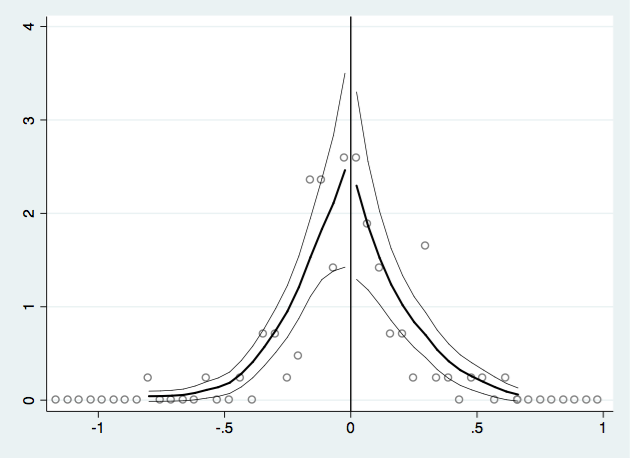}}
\caption{Window for normalized pop.: 10\%}
\end{subfigure}
\begin{subfigure}[b]{0.45\textwidth}	
\resizebox{7.2cm}{4.5cm}{\includegraphics[width=1cm]{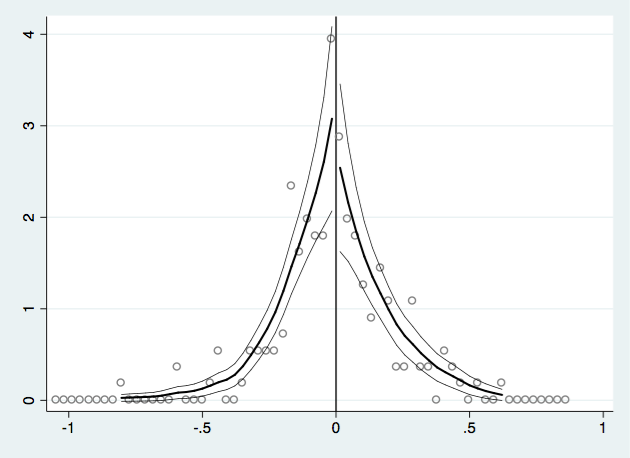}}
\caption{Window for normalized pop.: 20\%}
\end{subfigure}\\
\vspace{-0.0cm}
\begin{minipage}{14cm} \scriptsize  All figures  pool all years, and show finely-gridded histograms of the margin of victory of paramilitary-linked parties using local linear regression, separately on either side of the cutoff  of the density function of the margin of victory \cite[see][]{McCrary2008, CattaneoJanssonMa2018a}. Each figure uses data only around the corresponding population threshold.  The estimates of the difference in the height at the threshold and  robust manipulation p-values (as implemented in Stata by the command rddensity.ado, \citep[see][]{CattaneoJanssonMa2018a}, are: (a)  -0.034 (s.e. 0.346) and  p-value=0.990; (b)  -0.182 (s.e.  0.279) and  p-value=0.860.
\end{minipage}
\end{center}
\end{figure}

\newpage
\setcounter{table}{0}
\setcounter{figure}{0}
\newgeometry{left=2cm,right=2cm,bottom=2cm,top=3cm}
{\section{Robustness to excluding the creation of municipalities}\label{appcreated}
\small
 In this appendix, I explore the robustness of the main results to excluding events in which a municipality was either created or gave part of its population to a new municipality. During the period considered, no municipality disappeared. As discussed in footnote \ref{footnotesplitorcreated}, when a municipality is created, the data used in the estimation of the population of the municipalities involved in the event may be systematically manipulated. 

Between 1997 and 2011, in municipalities with a population of at least 15,000 (i.e. 564 municipalities), there were 2,355 local elections. Simultaneously to these elections (i.e. within a two-year window) there were 32 events in which a municipality was either created or lost part of its territory. These events are reported in Table \ref{splitorcreated}.  

\begin{center}
\renewcommand{\arraystretch}{0.7} 
\setlength{\tabcolsep}{10pt}
\begin{threeparttable}[H]
\small
\caption{Municipalities created or that lost part of their territory between 1997-2011}\label{splitorcreated}
\begin{tabular}{|c|c|c|c|c|}\hline\hline\addlinespace[0.15cm]
&  & Type &Year&Year of\\
ID & Municipality (Name/Department)  & of event & of event&closest election\\
\addlinespace[0.15cm]
\hline\addlinespace[0.15cm]
13600 &Rio Viejo (Bol\'ivar)  & Loss of terr. & 2007 & 2007  \\
19100 & Bol\'iva (Cauca)  & Loss of terr. & 1999& 2000  \\
19698 & Santander de Quilichao (Cauca) & Loss of terr. & 1998& 1997 \\
23466 & Montel\'ibano (C\'ordoba)  & Loss of terr. & 2007&  2007 \\
23670 & San Andr\'es de Sotavento (C\'ordoba) & Loss of terr. & 2007& 2007  \\
27001 & Quibd\'o (Choc\'o)& Loss of terr. & 1999&  2000 \\
27073 &   Bagad\'o (Choc\'o) & Loss of terr. & 2000& 2000  \\
27205 & Condoto (Choc\'o) & Loss of terr. & 2000 &  2000 \\
27361 &  Istmina (Choc\'o) & Loss of terr. & 1999 &  2000 \\
27615 &  Riosucio (Choc\'o) & Loss of terr. & 2000 &  2000 \\
27787 & Tad\'o (Choc\'o) & Loss of terr. & 2000 & 2000  \\
44078 &  Barrancas (La Guajira) & Loss of terr. & 1999&  2000 \\
44430 &  Maicao (La Guajira) & Loss of terr. & 2000 &  2000 \\
47030 &  Algarrobo (Magdalena) & creation & 1999& 2000 \\
47058 & Ariguan\'i  (Magdalena)& Loss of terr. & 1999& 2000 \\
47170 & Chivolo (Magdalena)& Loss of terr. & 1999&  2000\\
47189 & Ci\'enaga (Magdalena)& Loss of terr. & 1999& 2000 \\
47205 &  Concordia (Magdalena)& creation & 1999&  2000 \\
47258 &  El Pi\~n\'on (Magdalena)& Loss of terr. & 2000&  2000 \\
47288 &  Fundaci\'on (Magdalena)& Loss of terr. & 1999& 2000  \\
47460 & Nueva Granada (Magdalena)& creation & 2000& 2000 \\
47551 &  Pivijay (Magdalena)& Loss of terr. & 1999& 2000 \\
47555 &  Plato (Magdalena)& Loss of terr. & 2000 & 2000 \\
47707 &  Santa Ana (Magdalena)& Loss of terr. & 2000&  2000 \\
47798 &  Tenerife (Magdalena)& Loss of terr. & 2000&  2000 \\
47980 &  Zona Bananera (Magdalena)& creation & 1999& 2000  \\
52001 &  Pasto (Nari\~no) & Loss of terr. & 1999&  2000 \\
52260 &  El Tambo (Nari\~no) & Loss of terr. & 1998&  1997 \\
70215 & Corozal (Sucre) & Loss of terr. & 1998&  1997 \\
70678 &  San Benito Abad (Sucre)& Loss of terr. & 1998& 1997  \\
70742 & Sinc\'e (Sucre)& Loss of terr. & 1998& 1997  \\
70820 & Santiago de Tol\'u (Sucre) & Loss of terr. & 2002& 2003  \\
\addlinespace[0.15cm]
\hline               \hline          
\addlinespace[0.15cm]                                    
\multicolumn{5}{p{16cm}}{\scriptsize \textbf{Notes}: This table lists all the events between 1997 and 2011 in which a Colombian municipality with a population of at least 15,000 was created or split (during this period no municipality disappeared). The table uses data from \cite{Chavarro2013}.} 
\end{tabular}
\end{threeparttable}
\end{center}


\newpage
Figure \ref{McCrarypopthresC1}  and Table \ref{RDconflictbaselineC} exclude from the baseline sample the events listed in Table \ref{splitorcreated}. Figure \ref{McCrarypopthresC1} reports the \citeauthor{McCrary2008}'s and \citeauthor{CattaneoJanssonMa2018a}'s tests for each population threshold and pooled thresholds. The figure provides evidence against the existence of systematic manipulation of the population at the thresholds. Table \ref{RDconflictbaselineC} repeats the analysis in Table \ref{tableRDconflictbaseline}, which include the main results of the paper. All the estimates in Table \ref{RDconflictbaselineC} are virtually the same as those in Table \ref{tableRDconflictbaseline}. This  shows that the results of Table \ref{tableRDconflictbaseline} are robust to the existence of events in which the population of a municipality changes due to the municipality's creation or splitting. 


\begin{figure}[H]
\begin{center}
\caption{Manipulation tests excluding events in which a municipality was created} \label{McCrarypopthresC1}
\begin{subfigure}[a]{0.45\textwidth}
\resizebox{7cm}{4cm}{\includegraphics[width=\textwidth]{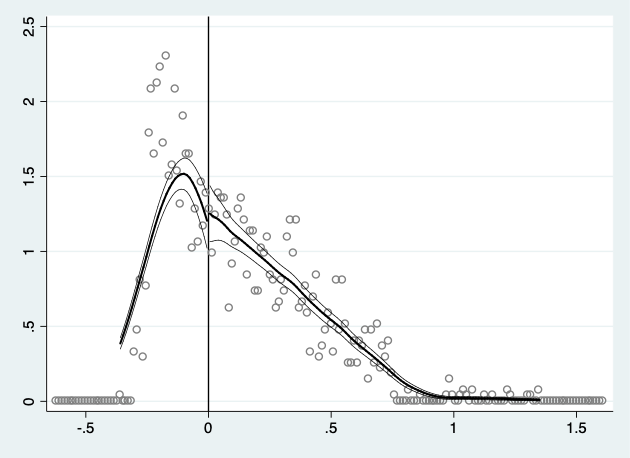}}
\caption{Pooled thresholds}
\end{subfigure}\\
\begin{subfigure}[a]{0.45\textwidth}
\resizebox{7cm}{3.4cm}{\includegraphics[width=\textwidth]{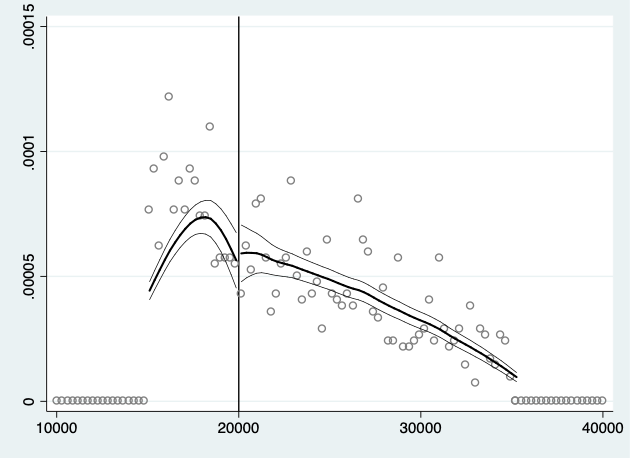}}
\caption{20,000  threshold}
\end{subfigure}
\begin{subfigure}[a]{0.45\textwidth}
\resizebox{7cm}{3.4cm}{\includegraphics[width=\textwidth]{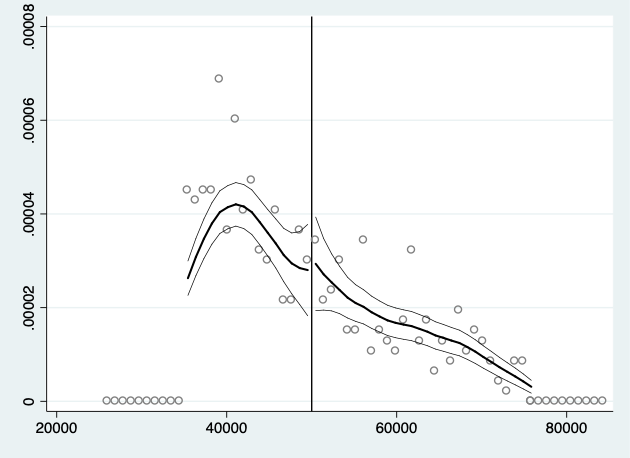}}
\caption{50,000  threshold}
\end{subfigure}\\
\begin{subfigure}[a]{0.45\textwidth}
\resizebox{7cm}{3.4cm}{\includegraphics[width=\textwidth]{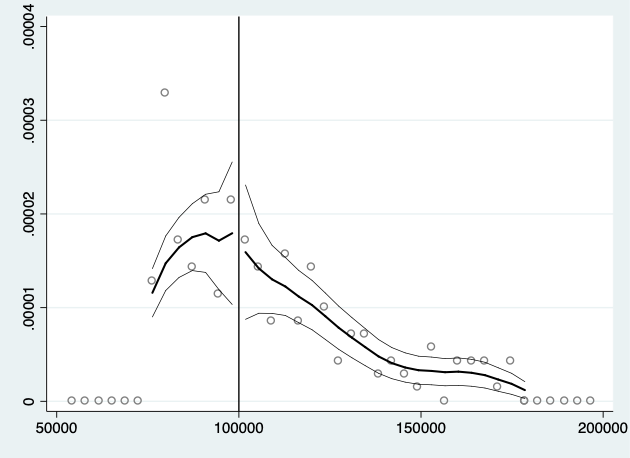}}
\caption{100,000  threshold}
\end{subfigure}
\begin{subfigure}[a]{0.45\textwidth}
\resizebox{7cm}{3.4cm}{\includegraphics[width=\textwidth]{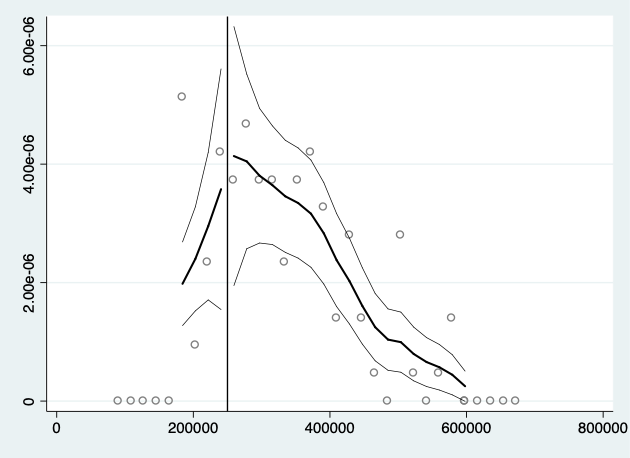}}
\caption{250,000  threshold}
\end{subfigure}\\
\vspace{-0.1cm}
\begin{minipage}{16cm} \scriptsize  All figures  pool all years and exclude the events listed in Table \ref{splitorcreated}. The figures show finely-gridded histograms of  the population smoothed using local linear regression, separately on either side of the cutoff  of the density function of the population. Figures in panels (b) to (e) use data only around the corresponding population threshold. The estimates of the difference in the height at the threshold and  robust manipulation p-values (as implemented in Stata by the command rddensity.ado, \citep[see][]{CattaneoJanssonMa2018a}, are: (a) 0.070 (s.e.  0.115) and p-value 0.369;  (b)  0.084 (s.e. 0.150) and p-value 0.161; (c) 0.0823 (s.e. 0.279) and p-value 0.674; (d) -0.118 (s.e. 0.374) and p-value 0.862; (e)  0.072 (s.e. 0.497) and p-value 0.066. 
\end{minipage}
\end{center}
\end{figure}


\renewcommand{\arraystretch}{0.8} 
\setlength{\tabcolsep}{5pt}
\begin{table}[H]
\small
\begin{center}
\caption{Effect of council size on conflict-related violence: Robustness to excluding events in which a municipality was created}\label{RDconflictbaselineC}
\begin{tabular}{lcccc}\hline\hline\addlinespace[0.15cm]
\emph{Dependent variable:} & \multicolumn{3}{c}{Conflict-related homicide}  & \multicolumn{1}{c}{Overall
} \\\cmidrule[0.2pt](l){2-4}
 & \multicolumn{1}{c}{Killing} & \multicolumn{1}{c}{Select. killing}& \multicolumn{1}{c}{Massacre} & \multicolumn{1}{c}{
homicide} \\
\cmidrule[0.2pt](l){2-2}\cmidrule[0.2pt](l){3-3}\cmidrule[0.2pt](l){4-4}\cmidrule[0.2pt](l){5-5}
& (1)& (2)& (3) & (4)\\ \hline \addlinespace[0.15cm]
\multicolumn{1}{l}{\emph{\underline{Panel A}: prob.}} &    \\   \addlinespace[0.15cm]
\hspace{3mm}RD Estimate &      -0.044\sym{***}&      -0.040\sym{***}&      -0.006         &       0.012         \\
            &     (0.016)         &     (0.015)         &     (0.004)         &     (0.021)         \\
Observations&        2323         &        2323         &        2323         &        2294         \\
Eff. Number of obs&         722         &         639         &         588         &         437         \\
BW Loc. Poly. (h)&       0.122         &       0.109         &       0.099         &       0.076         \\
Robust p-value&       0.013         &       0.018         &       0.182         &       0.419         \\

\addlinespace[0.15cm]
\hline \addlinespace[0.15cm]
\multicolumn{1}{l}{\emph{\underline{Panel B}: number}} & &   \\   \addlinespace[0.15cm]
\hspace{3mm}RD Estimate &      -1.092         &      -0.923\sym{*}  &      -0.131         &      -3.679         \\
            &     (0.713)         &     (0.471)         &     (0.293)         &     (4.078)         \\
Observations&        2323         &        2323         &        2323         &        2294         \\
Eff. Number of obs&         570         &         554         &         663         &         601         \\
BW Loc. Poly. (h)&       0.096         &       0.095         &       0.113         &       0.103         \\
Robust p-value&       0.131         &       0.068         &       0.657         &       0.420         \\

\addlinespace[0.15cm]
\hline \addlinespace[0.15cm]
\multicolumn{1}{l}{\emph{\underline{Panel C}: rate}} & &   \\   \addlinespace[0.15cm]
\hspace{3mm}RD Estimate &      -3.153         &      -3.123\sym{*}  &      -0.091         &      -5.383         \\
            &     (2.406)         &     (1.750)         &     (1.003)         &     (9.571)         \\
Observations&        2323         &        2323         &        2323         &        2287         \\
Eff. Number of obs&         592         &         584         &         621         &         623         \\
BW Loc. Poly. (h)&       0.100         &       0.099         &       0.105         &       0.108         \\
Robust p-value&       0.204         &       0.082         &       0.946         &       0.641         \\

 \addlinespace[0.15cm]
\hline     \hline                                                        
\multicolumn{5}{p{12.5cm}}{\scriptsize \textbf{Notes}:  All columns report the RD estimates of having a larger council from Eq. (\ref{rdbaseline}) when the respective characteristic is used as the dependent variable. All columns exclude the events listed in Table \ref{splitorcreated}. The dependent variable in panel A is the average over the electoral term of a dummy variable equal to 1 if the particular type of killing is observed in a municipality in a quarter of a year. The dependent variable in panels B and C is the average of the total number  of killings and  killings per 100,000 people, respectively. The bandwidth (h) is chosen optimally using the algorithm by \cite{CalonicoCattaneoTitiunik2014a} as  implemented in Stata by the command rdrobust.ado, and includes  fixed effects for population threshold, electoral term and region. Standard errors clustered by municipality are reported in parentheses. * significant at 10\%, ** significant at 5\%, *** significant at 1\%. }
\end{tabular}
\end{center}
\end{table}

\newpage
\setcounter{table}{0}
\setcounter{figure}{0}
\newgeometry{left=2cm,right=2cm,bottom=2cm,top=3cm}
{\section{Additional Evidence for the Preferred Mechanism}\label{appmechanism}
\small
To further examine the plausibility of my deterrence hypothesis, this appendix contains several additional exercises based on certain key consequences. Specifically, I examine whether the impact of a larger municipal council on selective violence is affected by the degree to which the municipality was contested militarily in the past.  I hypothesize that the effect of an exogenous increase in the influence of politicians with paramilitary links (as a consequence of a larger municipal council) on selective killings is amplified in areas that have been more military contested.

I use two proxies for the degree to which municipalities were contested.  First, I look at past occurrences of violence by both guerrillas and paramilitaries. Specifically, I compute the proportion of years in which both a guerrilla and paramilitary group perpetrated a violent act in the municipality (using CERAC data), and define a dummy variable equal to 1 for those municipalities with a proportion higher than the median (which I interpret as highly contested), and zero for those with proportion below the median (which I interpret as relatively uncontested). Columns (1) and (2) of Panel A in Table \ref{tableRDhighlydisputedhighforcdispl} examine how the results in Section \ref{mainresults} are affected when we distinguish between highly contested municipalities (column (1)) and relatively uncontested municipalities (column (2)). Consistent with the deterrence hypothesis, the table shows that the effect of a larger council size on conflict-related killings is concentrated in highly contested municipalities.\footnote{Table \ref{tableRDhighlydisputedhighforcdisplPARAM} uses a parametric specification to explore robustness to using the extensive margin. The table reports estimates for two bandwidths (0.10) and (0.15), which are close to \citeauthor{CalonicoCattaneoTitiunik2014b}'s endogenously chosen bandwidths in the baseline specifications. The table also reports results for specifications with linear and quadratic polynomials. Columns (1) and (6) examine whether the results in Table \ref{tableRDconflictbaseline} are robust to using a variety of parametric specifications. The estimates in these columns are virtually the same as those in Table \ref{tableRDconflictbaseline}. Columns (2) and (7) in Table \ref{tableRDhighlydisputedhighforcdisplPARAM}   introduce an interaction term between the dummy for a larger council size and the dummy previously described for  highly contested municipalities. Columns (3) and (8) replace the dummy with the intensive margin variable (i.e. share of years the municipality has been contested). The estimates of the interaction term are always negative and statistically significant, which confirms the results in columns (1) and (2) of panel A in Table \ref{tableRDhighlydisputedhighforcdispl}.\label{parametricontested}}\footnote{Columns (1) and (2) of panel B in Table \ref{tableRDhighlydisputedhighforcdispl} examine whether the effect of a larger council on paramilitary-linked representation is concentrated in highly contested municipalities. The estimates show this is not the case. These results, combined with those in panel A, imply that even though paramilitary groups may have had greater influence in municipalities with larger councils that were not contested in the past, this influence only reduces guerrilla attacks in municipalities that were contested militarily in the past.}

A second proxy is mass forced displacement. By 2005, approximately 7\% of the Colombian population had been forcibly displaced. All armed groups deliberately triggered the forced migration of civilians, which increased the resources of these armed groups and hampered their enemies' fighting capacity \cite[see][]{IbanezVelez2008}. Paramilitary groups instigated half of all forced migrations, while guerrilla groups and the simultaneous presence of two armed groups were responsible for 20\% and 22\% of such migrations, respectively \cite[see][]{IbanezVelez2008}. I construct a dummy variable that equals 1 if a municipality experienced a mass forced migration in a given year. I compute the average for the period in which data exists, and define a dummy variable equal to 1 for municipalities with a proportion higher than this average (which I interpret as highly contested), and 0 for those with a proportion below the average (which I interpreted as relatively uncontested). Approximately 28.7\% of observations are classified as highly contested.  Columns (3) and (4) of Panel A in Table \ref{tableRDhighlydisputedhighforcdispl} show results that are also consistent with the deterrence hypothesis: the effect of a large council on conflict-related killings is concentrated in municipalities with significant previous exposure to forced migration.\footnote{Columns (4), (5), (9) and (10) in Table \ref{tableRDhighlydisputedhighforcdisplPARAM} in Web Appendix \ref{appsuppfigtab} explore robustness to using the intensive margin and a parametric specification (see footnote \ref{parametricontested}). As is the case for contested municipalities, the estimates in these columns are consistent with those in columns (3) and (4) of Panel A in Table \ref{tableRDhighlydisputedhighforcdispl}.}\footnote{Columns (3) and (4) of Panel B in Table \ref{tableRDhighlydisputedhighforcdispl} examine whether the effect of a larger council on the representation of paramilitary-linked parties is concentrated in municipalities with frequent past forced displacement. The estimates show this is not the case.} 
}


\renewcommand{\arraystretch}{0.8} 
\setlength{\tabcolsep}{10pt}
\begin{table}[H]
\small
\begin{center}
\caption{Effect of council size on conflict-related violence in highly disputed municipalities}\label{tableRDhighlydisputedhighforcdispl}
\begin{tabular}{lcccc}
\hline\hline\addlinespace[0.15cm]  
& (1)& (2)& (3) & (4)\\  \addlinespace[0.15cm] \hline \addlinespace[0.15cm]
\underline{Panel A}:  &\multicolumn{4}{c}{\emph{Dependent variable:} Occurrence of a killing}     \\\cmidrule[0.2pt](l){2-5}   \addlinespace[0.15cm]
&  \multicolumn{1}{c}{Frequent}  &  \multicolumn{1}{c}{Infrequent}  &  \multicolumn{1}{c}{Frequent}  &  \multicolumn{1}{c}{Infrequent}   \\
&  \multicolumn{1}{c}{past violent}  &  \multicolumn{1}{c}{past violent }  &  \multicolumn{1}{c}{past forced}  &  \multicolumn{1}{c}{ past forced}   \\
&  \multicolumn{1}{c}{actions}  &  \multicolumn{1}{c}{actions} &  \multicolumn{1}{c}{displacement}  &  \multicolumn{1}{c}{displacement} \\\cmidrule[0.2pt](l){2-2}\cmidrule[0.2pt](l){3-3} \cmidrule[0.2pt](l){4-4} \cmidrule[0.2pt](l){5-5} 
 \addlinespace[0.15cm]
\hspace{3mm}RD Estimate &      -0.064\sym{***}&      -0.018         &      -0.087\sym{***}&      -0.018         \\
            &     (0.024)         &     (0.013)         &     (0.030)         &     (0.019)         \\
Observations&        1303         &        1052         &         770         &        1585         \\
Eff. Number of obs&         345         &         360         &         164         &         413         \\
BW Loc. Poly. (h)&       0.101         &       0.133         &       0.083         &       0.104         \\
Robust p-value&       0.010         &       0.210         &       0.003         &       0.521         \\

\addlinespace[0.15cm]\hline \addlinespace[0.15cm]
\underline{Panel B}:  &\multicolumn{4}{c}{\emph{Dependent variable:} Presence on council of a paramilitary-linked party}     \\\cmidrule[0.2pt](l){2-5}   \addlinespace[0.15cm]
&  \multicolumn{1}{c}{Frequent}  &  \multicolumn{1}{c}{Infrequent}  &  \multicolumn{1}{c}{Frequent}  &  \multicolumn{1}{c}{Infrequent}   \\
&  \multicolumn{1}{c}{past violent}  &  \multicolumn{1}{c}{past violent }  &  \multicolumn{1}{c}{past forced}  &  \multicolumn{1}{c}{ past forced}   \\
&  \multicolumn{1}{c}{actions}  &  \multicolumn{1}{c}{actions} &  \multicolumn{1}{c}{displacement}  &  \multicolumn{1}{c}{displacement} \\\cmidrule[0.2pt](l){2-2}\cmidrule[0.2pt](l){3-3} \cmidrule[0.2pt](l){4-4} \cmidrule[0.2pt](l){5-5} 
 \addlinespace[0.15cm]
\hspace{3mm}RD Estimate &       0.304\sym{***}&       0.380\sym{***}&       0.490\sym{***}&       0.213\sym{**} \\
            &     (0.085)         &     (0.115)         &     (0.120)         &     (0.090)         \\
Observations&        1303         &        1052         &         770         &        1585         \\
Eff. Number of obs&         367         &         200         &         178         &         408         \\
BW Loc. Poly. (h)&       0.110         &       0.075         &       0.089         &       0.103         \\
Robust p-value&       0.005         &       0.001         &       0.000         &       0.041         \\

\addlinespace[0.15cm]
\hline   \hline                                                          
\multicolumn{5}{p{16cm}}{\scriptsize \textbf{Notes}:  All columns report the RD estimates of having a larger council from Eq. (\ref{rdbaseline}) when the respective characteristic is used as the dependent variable. The bandwidth (h) is  chosen optimally using the algorithm by \cite{CalonicoCattaneoTitiunik2014a} as  implemented in Stata by the command rdrobust.ado, and includes  fixed effects for population threshold, electoral term and region. Standard errors clustered by municipality are reported in parentheses. * significant at 10\%, ** significant at 5\%, *** significant at 1\% } 
\end{tabular}
\end{center}
\end{table}


\clearpage
\newgeometry{left=2.5cm,right=1.5cm,bottom=2cm,top=2cm}
{ 
 \begin{landscape}
\renewcommand{\arraystretch}{0.6} 
\setlength{\tabcolsep}{4pt}
\begin{table}[H]
\small
\caption{Effect of council size on conflict-related violence in highly disputed municipalities: Robustness to using a parametric specification}\label{tableRDhighlydisputedhighforcdisplPARAM}
\begin{tabular}{lcccccccccc}
\hline\hline\addlinespace[0.15cm]  
 &\multicolumn{10}{c}{\emph{Dep. var.: Occurrence of a killing}}     \\\cmidrule[0.2pt](l){2-11}   \addlinespace[0.15cm]
& (1)& (2)& (3) & (4)& (5)& (6)& (7)& (8)& (9)& (10)\\  \addlinespace[0.15cm] \hline \addlinespace[0.15cm]
& \multicolumn{5}{c}{h=0.10}  & \multicolumn{5}{c}{h=0.15}  \\\cmidrule[0.2pt](l){2-6}\cmidrule[0.2pt](l){7-11}  
 \addlinespace[0.15cm]
Bigger council size &      -0.040\sym{**} &      -0.000         &      -0.009         &      -0.015         &      -0.015         &      -0.051\sym{**} &       0.004         &      -0.006         &      -0.013         &      -0.013         \\
                    &     (0.017)         &     (0.015)         &     (0.015)         &     (0.020)         &     (0.018)         &     (0.021)         &     (0.017)         &     (0.018)         &     (0.025)         &     (0.023)         \\
Frenquent past violent actions&                     &       0.093\sym{***}&                     &                     &                     &                     &       0.112\sym{***}&                     &                     &                     \\
                    &                     &     (0.027)         &                     &                     &                     &                     &     (0.034)         &                     &                     &                     \\
\shortstack[l]{Bigger council size $\times$ Frenquent \\  \hspace{0.1cm}  past violent actions}&                     &      -0.076\sym{**} &                     &                     &                     &                     &      -0.112\sym{***}&                     &                     &                     \\
                    &                     &     (0.032)         &                     &                     &                     &                     &     (0.041)         &                     &                     &                     \\
Share of years with violent actions&                     &                     &       0.366\sym{***}&                     &                     &                     &                     &       0.420\sym{***}&                     &                     \\
                    &                     &                     &     (0.120)         &                     &                     &                     &                     &     (0.151)         &                     &                     \\
\shortstack[l]{Bigger council size $\times$ Share of \\ \hspace{0.1cm}  of years with violent actions}&                     &                     &      -0.398\sym{***}&                     &                     &                     &                     &      -0.523\sym{***}&                     &                     \\
                    &                     &                     &     (0.143)         &                     &                     &                     &                     &     (0.174)         &                     &                     \\
Frenquent past forced displacement&                     &                     &                     &       0.066\sym{**} &                     &                     &                     &                     &       0.085\sym{**} &                     \\
                    &                     &                     &                     &     (0.032)         &                     &                     &                     &                     &     (0.041)         &                     \\
\shortstack[l]{Bigger council size $\times$ Frenquent \\ \hspace{0.1cm}  past forced displacement}&                     &                     &                     &      -0.079\sym{*}  &                     &                     &                     &                     &      -0.121\sym{**} &                     \\
                    &                     &                     &                     &     (0.040)         &                     &                     &                     &                     &     (0.048)         &                     \\
Share of years forced displacement&                     &                     &                     &                     &       0.145\sym{***}&                     &                     &                     &                     &       0.176\sym{***}\\
                    &                     &                     &                     &                     &     (0.043)         &                     &                     &                     &                     &     (0.065)         \\
\shortstack[l]{Bigger council size $\times$ Share of \\ \hspace{0.1cm}  of years with forced displacement}&                     &                     &                     &                     &      -0.233\sym{**} &                     &                     &                     &                     &      -0.352\sym{***}\\
                    &                     &                     &                     &                     &     (0.099)         &                     &                     &                     &                     &     (0.105)         \\
R-squared           &       0.352         &       0.403         &       0.447         &       0.390         &       0.431         &       0.361         &       0.403         &       0.428         &       0.390         &       0.419         \\
Observations        &         468         &         468         &         468         &         468         &         468         &         732         &         732         &         732         &         732         &         732         \\

\addlinespace[0.15cm]
polynomial & linear &linear& linear & linear& linear&quad.& quad. &quad.&quad.&quad.\\
\addlinespace[0.15cm]
\hline   \hline                                                          
\multicolumn{11}{p{23.5cm}}{\scriptsize \textbf{Notes}:  Columns (1) and (6) report the coefficient for having a bigger council size $D_{mt}$ from an equation of the form $$Y_{rmt+1}=\alpha+\beta  D_{rmt}+f(\tilde{N}_{rmt}\cdot \boldsymbol\gamma)+\delta_t+\zeta_T+\eta_r+\epsilon_{mt}$$ where $Y_{rmt+1}$ is the respective characteristic in municipality  $m$ i region $r$ in the term  immediately following election $t$, $\tilde{N}_{rmt}$ is  the normalized population, $D_{rmt}$ is an indicator variable taking the value of 1 if the municipality has a bigger council size,  $f(\cdot)$ is a first or second-order polynomial in $\tilde{N}_{rmt}$, $\delta_t $  is fixed effects by year, $\zeta_T$ threshold fixed effects, $\eta_r$ region fixed effects, and $\epsilon_{mt}$ the error term clustered at the municipal level. Columns (2) to (5) and (7) to (10) include interactions  between the dummy for a bigger council and the respective pre-treatment characteristic. The bandwidth h in columns (1) to (5) is 10 percent points and in columns (6) to (10) is 15 percent points.  * significant at 10\%, ** significant at 5\%, *** significant at 1\%.} 
\end{tabular}
\end{table}
\end{landscape}
}

\end{document}